%
%
%
%

\input harvmac
\newcount\yearltd\yearltd=\year\advance\yearltd by 0

\noblackbox 
\input epsf

\def\tilde{\widetilde}

\newcount\figno
\figno=0

\def\fig#1#2#3{
\par\begingroup\parindent=0pt\leftskip=1cm\rightskip=1cm\parindent=0pt
\baselineskip=11pt \global\advance\figno by 1 \midinsert
\epsfxsize=#3 \centerline{\epsfbox{#2}} \vskip 12pt {\bf Figure\
\the\figno: } #1\par
\endinsert\endgroup\par
}
\def\figlabel#1{\xdef#1{\the\figno}}
\def\encadremath#1{\vbox{\hrule\hbox{\vrule\kern8pt\vbox{\kern8pt
\hbox{$\displaystyle #1$}\kern8pt} \kern8pt\vrule}\hrule}}

%
%

\def\twovertfig#1#2#3#4{
\par\begingroup\parindent=0pt\leftskip=1cm\rightskip=1cm\parindent=0pt
\baselineskip=11pt \global\advance\figno by 1 \midinsert
\epsfxsize=#2
\centerline{\epsfbox{#3}}
\epsfxsize=#2
\centerline{\epsfbox{#4}} \vskip 12pt
{\bf Figure\ \the\figno: } #1\par
\endinsert\endgroup\par
}

\def\twosidefig#1#2#3#4{
\par\begingroup\parindent=0pt\leftskip=1cm\rightskip=1cm\parindent=0pt
\baselineskip=11pt \global\advance\figno by 1 \midinsert
\centerline{
\epsfxsize=#2
\epsfbox{#3}
\epsfxsize=#2
\epsfbox{#4}
}
\vskip 12pt
{\bf Figure\ \the\figno: } #1\par
\endinsert\endgroup\par
}

\def\quadfig#1#2#3#4#5#6{
\par\begingroup\parindent=0pt\leftskip=1cm\rightskip=1cm\parindent=0pt
\baselineskip=11pt \global\advance\figno by 1 \midinsert
\centerline{
\epsfxsize=#2
\epsfbox{#3}
\epsfxsize=#2
\epsfbox{#4}
}
\centerline{
\epsfxsize=#2
\epsfbox{#5}
\epsfxsize=#2
\epsfbox{#6}
}
\vskip 12pt
{\bf Figure\ \the\figno: } #1\par
\endinsert\endgroup\par
}

\def\quadvertfig#1#2#3#4#5#6{
\par\begingroup\parindent=0pt\leftskip=1cm\rightskip=1cm\parindent=0pt
\baselineskip=11pt \global\advance\figno by 1 \midinsert
\centerline{
\epsfxsize=#2
\epsfbox{#3}
}
\centerline{
\epsfxsize=#2
\epsfbox{#4}
}
\centerline{
\epsfxsize=#2
\epsfbox{#5}
}
\centerline{
\epsfxsize=#2
\epsfbox{#6}
}
\vskip 12pt
{\bf Figure\ \the\figno: } #1\par
\endinsert\endgroup\par
}

%
%

\def\half{{\textstyle{1\over2}}}

\def\apm{{\alpha^{\prime}}}

\def\half{{1\over 2}}

 \def\a{{\alpha}}
 
 \def\frac#1#2{{#1\over #2}}

 \def\CO{{\cal O}}

 \def\CN{{\cal N}}

 \def\apm{{\a^{\prime}}}

\def\IR{\relax{\rm I\kern-.18em R}}
\def\IZ{\relax\ifmmode\hbox{Z\kern-.4em Z}\else{Z\kern-.4em Z}\fi}

\lref\sphere{ O.~Aharony, J.~Marsano, S.~Minwalla, K.~Papadodimas
and M.~Van Raamsdonk,
 ``The Hagedorn / deconfinement phase transition in weakly coupled large N
gauge theories,''
Adv.\ Theor.\ Math.\ Phys.\  {\bf 8}, 603 (2004)
[arXiv:hep-th/0310285].
}

\lref\Sundborg{ B.~Sundborg, ``The Hagedorn transition,
deconfinement and N = 4 SYM theory,'' Nucl.\ Phys.\ B {\bf 573}, 349
(2000) [arXiv:hep-th/9908001].
}

\lref\hawkingpage{ S.~W.~Hawking and D.~N.~Page, ``Thermodynamics Of
Black Holes In Anti-De Sitter Space,'' Commun.\ Math.\ Phys.\  {\bf
87}, 577 (1983).
}

\lref\witteno{ E.~Witten, ``Anti-de Sitter space and holography,''
Adv.\ Theor.\ Math.\ Phys.\  {\bf 2}, 253 (1998)
[arXiv:hep-th/9802150].
}

\lref\wittent{ E.~Witten, ``Anti-de Sitter space, thermal phase
transition, and confinement in  gauge theories,'' Adv.\ Theor.\
Math.\ Phys.\  {\bf 2}, 505 (1998) [arXiv:hep-th/9803131].
}

\lref\MaldacenaRE{
  J.~M.~Maldacena,
  ``The large N limit of superconformal field theories and supergravity,''
  Adv.\ Theor.\ Math.\ Phys.\  {\bf 2}, 231 (1998)
  [Int.\ J.\ Theor.\ Phys.\  {\bf 38}, 1113 (1999)]
  [arXiv:hep-th/9711200].
}

\lref\adsrev{ O.~Aharony, S.~S.~Gubser, J.~M.~Maldacena, H.~Ooguri
and Y.~Oz, ``Large N field theories, string theory and gravity,''
Phys.\ Rept.\  {\bf 323}, 183 (2000) [arXiv:hep-th/9905111].
}

\lref\susskind{ L.~Susskind, ``Matrix theory black holes and the
Gross Witten transition,'' arXiv:hep-th/9805115.
}

\lref\creutz{ M.~Creutz, ``On Invariant Integration Over SU(N),''
J.\ Math.\ Phys.\  {\bf 19}, 2043 (1978).
}

\lref\gw{ D.~J.~Gross
and E.~Witten, ``Possible Third Order Phase Transition In The Large
N Lattice Gauge Theory,'' Phys.\ Rev.\ D {\bf 21}, 446 (1980).
}

\lref\AmbjornXW{ J.~Ambjorn and S.~Wolfram,
``Properties Of The Vacuum. 1. Mechanical And Thermodynamic,''
Annals Phys.\  {\bf 147}, 1 (1983).
}

\lref\martineco{ M.~Li, E.~J.~Martinec and V.~Sahakian,
``Black holes and the SYM phase diagram,''
Phys.\ Rev.\ D {\bf 59}, 044035 (1999) [arXiv:hep-th/9809061].
}

\lref\martinect{ E.~J.~Martinec and V.~Sahakian,
``Black holes and the SYM phase diagram. II,''
Phys.\ Rev.\ D {\bf 59}, 124005 (1999) [arXiv:hep-th/9810224].
}

\lref\wit{ E.~Witten,
``Anti-de Sitter space, thermal phase transition, and confinement in  gauge
theories,''
Adv.\ Theor.\ Math.\ Phys.\  {\bf 2}, 505 (1998)
[arXiv:hep-th/9803131].
}

\lref\polstr{ J.~Polchinski and M.~J.~Strassler,
``The string dual of a confining four-dimensional gauge theory,''
arXiv:hep-th/0003136.
}

\lref\klestr{ I.~R.~Klebanov and M.~J.~Strassler,
``Supergravity and a confining gauge theory: Duality cascades and
$\chi$SB-resolution of naked singularities,''
JHEP {\bf 0008}, 052 (2000) [arXiv:hep-th/0007191].
}

\lref\kletsy{ I.~R.~Klebanov and A.~A.~Tseytlin,
``Gravity duals of supersymmetric SU(N) x SU(N+M) gauge theories,''
Nucl.\ Phys.\ B {\bf 578}, 123 (2000) [arXiv:hep-th/0002159].
}

\lref\buchel{ A.~Buchel, C.~P.~Herzog, I.~R.~Klebanov, L.~A.~Pando
Zayas and A.~A.~Tseytlin,
``Non-extremal gravity duals for fractional D3-branes on the conifold,''
JHEP {\bf 0104}, 033 (2001) [arXiv:hep-th/0102105].
}

\lref\GubserEG{
  S.~S.~Gubser, A.~A.~Tseytlin and M.~S.~Volkov,
  ``Non-Abelian 4-d black holes, wrapped 5-branes, and their dual
  descriptions,''
  JHEP {\bf 0109}, 017 (2001)
  [arXiv:hep-th/0108205].
}

\lref\BuchelQI{
  A.~Buchel and A.~R.~Frey,
  ``Comments on supergravity dual of pure N = 1 super Yang Mills theory  with
  unbroken chiral symmetry,''
  Phys.\ Rev.\ D {\bf 64}, 064007 (2001)
  [arXiv:hep-th/0103022].
}

\lref\GubserRI{
  S.~S.~Gubser, C.~P.~Herzog, I.~R.~Klebanov and A.~A.~Tseytlin,
  ``Restoration of chiral symmetry: A supergravity perspective,''
  JHEP {\bf 0105}, 028 (2001)
  [arXiv:hep-th/0102172].
}

\lref\BuchelCH{
  A.~Buchel,
  ``Finite temperature resolution of the Klebanov-Tseytlin singularity,''
  Nucl.\ Phys.\ B {\bf 600}, 219 (2001)
  [arXiv:hep-th/0011146].
}

\lref\malnun{ J.~M.~Maldacena and C.~Nunez,
``Towards the large N limit of pure N = 1 super Yang Mills,''
Phys.\ Rev.\ Lett.\  {\bf 86}, 588 (2001) [arXiv:hep-th/0008001].
}

\lref\loc{S. Minwalla, `Black Holes in Gauge Theories' (talks at the
Post-Strings workshop at CERN, July 2004, and at the Tata Institute,
September 2004.)}

\lref\nastase{
K.~Kang and H.~Nastase,
``High energy QCD from Planckian scattering in AdS and the Froissart bound,''
arXiv:hep-th/0410173; ~~~
K.~Kang and H.~Nastase,
``Heisenberg saturation of the Froissart bound from AdS-CFT,''
arXiv:hep-th/0501038; ~~~
H.~Nastase,
``The soft pomeron from AdS-CFT,''
arXiv:hep-th/0501039; ~~~
H.~Nastase,
``The RHIC fireball as a dual black hole,''
arXiv:hep-th/0501068.}

\lref\giddings{ S.~B.~Giddings,
``High energy QCD scattering, the shape of gravity on an IR brane, and  the
Froissart bound,''
Phys.\ Rev.\ D {\bf 67}, 126001 (2003) [arXiv:hep-th/0203004];
~~D.~M.~Eardley and S.~B.~Giddings,
  Phys.\ Rev.\ D {\bf 66}, 044011 (2002)
  [arXiv:gr-qc/0201034].
}

\lref\GiddingsAY{
  S.~B.~Giddings and E.~Katz,
  ``Effective theories and black hole production in warped
  compactifications,''
  J.\ Math.\ Phys.\  {\bf 42}, 3082 (2001)
  [arXiv:hep-th/0009176].
}

\lref\ScherkTA{
  J.~Scherk and J.~H.~Schwarz,
  ``Spontaneous Breaking Of Supersymmetry Through Dimensional Reduction,''
  Phys.\ Lett.\ B {\bf 82}, 60 (1979).
}

\lref\SusskindWS{
  L.~Susskind,
  ``Some speculations about black hole entropy in string theory,''
  arXiv:hep-th/9309145.
}

\lref\HorowitzBJ{
  G.~T.~Horowitz and H.~Ooguri,
  ``Spectrum of large N gauge theory from supergravity,''
  Phys.\ Rev.\ Lett.\  {\bf 80}, 4116 (1998)
  [arXiv:hep-th/9802116].
}

\lref\HorowitzNW{
  G.~T.~Horowitz and J.~Polchinski,
  ``A correspondence principle for black holes and strings,''
  Phys.\ Rev.\ D {\bf 55}, 6189 (1997)
  [arXiv:hep-th/9612146].
}

\lref\ItzhakiDD{
  N.~Itzhaki, J.~M.~Maldacena, J.~Sonnenschein and S.~Yankielowicz,
  ``Supergravity and the large N limit of theories with sixteen
  supercharges,''
  Phys.\ Rev.\ D {\bf 58}, 046004 (1998)
  [arXiv:hep-th/9802042].
}

\lref\PolicastroYC{
  G.~Policastro, D.~T.~Son and A.~O.~Starinets,
  ``The shear viscosity of strongly coupled N = 4 supersymmetric Yang-Mills
  plasma,''
  Phys.\ Rev.\ Lett.\  {\bf 87}, 081601 (2001)
  [arXiv:hep-th/0104066]; ~~~
  G.~Policastro, D.~T.~Son and A.~O.~Starinets,
  ``From AdS/CFT correspondence to hydrodynamics,''
  JHEP {\bf 0209}, 043 (2002)
  [arXiv:hep-th/0205052]; ~~~
  E.~V.~Shuryak and I.~Zahed,
  ``Rethinking the properties of the quark gluon plasma at T approx. T(c),''
  Phys.\ Rev.\ C {\bf 70}, 021901 (2004)
  [arXiv:hep-ph/0307267]; ~~~
  A.~Buchel and J.~T.~Liu,
  ``Universality of the shear viscosity in supergravity,''
  Phys.\ Rev.\ Lett.\  {\bf 93}, 090602 (2004)
  [arXiv:hep-th/0311175].
}

\lref\deepin{ J.~Polchinski and M.~J.~Strassler,
``Deep inelastic scattering and gauge/string duality,''
JHEP {\bf 0305}, 012 (2003) [arXiv:hep-th/0209211].
}

\lref\heisenberg{W.~Heisenberg,
``Production of mesons as a shock wave problem,''
Zeit. Phys. {\bf 133}, 65 (1952).
}

\lref\GregoryVY{
  R.~Gregory and R.~Laflamme,
  ``Black strings and p-branes are unstable,''
  Phys.\ Rev.\ Lett.\  {\bf 70}, 2837 (1993)
  [arXiv:hep-th/9301052].
}

\lref\GregoryBJ{
  R.~Gregory and R.~Laflamme,
  ``The instability of charged black strings and p-branes,''
  Nucl.\ Phys.\ B {\bf 428}, 399 (1994)
  [arXiv:hep-th/9404071].
}

\lref\PeetWN{
  A.~W.~Peet and J.~Polchinski,
  ``UV/IR relations in AdS dynamics,''
  Phys.\ Rev.\ D {\bf 59}, 065011 (1999)
  [arXiv:hep-th/9809022].
}

\lref\EmparanWA{
  R.~Emparan, G.~T.~Horowitz and R.~C.~Myers,
  ``Exact description of black holes on branes,''
  JHEP {\bf 0001}, 007 (2000)
  [arXiv:hep-th/9911043].
}

\lref\kraus{ P.~Kraus, F.~Larsen and R.~Siebelink, ``The
gravitational action in asymptotically AdS and flat spacetimes,''
Nucl.\ Phys.\ B {\bf 563}, 259 (1999) [arXiv:hep-th/9906127].
}

\lref\KharzeevIZ{
  D.~Kharzeev and K.~Tuchin,
  ``From color glass condensate to quark gluon plasma through the event
  horizon,''
  arXiv:hep-ph/0501234.
}

\lref\teper{
B.~Lucini, M.~Teper and U.~Wenger,
``The deconfinement transition in SU(N) gauge theories,''
Phys.\ Lett.\ B {\bf 545}, 197 (2002)
[arXiv:hep-lat/0206029]; ~~~
B.~Lucini, M.~Teper and U.~Wenger,
``The high temperature phase transition in SU(N) gauge theories,''
JHEP {\bf 0401}, 061 (2004)
[arXiv:hep-lat/0307017]; ~~~
  B.~Lucini, M.~Teper and U.~Wenger,
  ``Properties of the deconfining phase transition in SU(N) gauge theories,''
  JHEP {\bf 0502}, 033 (2005)
  [arXiv:hep-lat/0502003].
}

\lref\GyulassyZY{
  M.~Gyulassy and L.~McLerran,
  ``New forms of QCD matter discovered at RHIC,''
  Nucl.\ Phys.\ A {\bf 750}, 30 (2005)
  [arXiv:nucl-th/0405013].
}

\lref\LudlamRH{
  T.~Ludlam and L.~McLerran,
  ``What have we learned from the Relativistic Heavy Ion Collider?,''
  Phys.\ Today {\bf 56N10}, 48 (2003).
}

\lref\HeinzAX{
  U.~W.~Heinz,
  ``From SPS to RHIC: Breaking the barrier to the quark-gluon plasma,''
  AIP Conf.\ Proc.\  {\bf 602}, 281 (2001)
  [arXiv:hep-ph/0109006].
}

\lref\BraunMunzingerIP{
  P.~Braun-Munzinger, D.~Magestro, K.~Redlich and J.~Stachel,
  ``Hadron production in Au Au collisions at RHIC,''
  Phys.\ Lett.\ B {\bf 518}, 41 (2001)
  [arXiv:hep-ph/0105229].
}

\lref\bjorken{
  J.~D.~Bjorken,
  ``Highly Relativistic Nucleus-Nucleus Collisions: The Central Rapidity
  Region,''
  Phys.\ Rev.\ D {\bf 27}, 140 (1983).
}

\lref\pisarski{
  R.~D.~Pisarski,
  ``Finite Temperature QCD At Large N,''
  Phys.\ Rev.\ D {\bf 29}, 1222 (1984).
}

\lref\tHooftJZ{
G.~'t Hooft,
``A planar diagram theory for strong interactions,''
Nucl.\ Phys.\ B {\bf 72}, 461 (1974).
}

\lref\FerraraGV{
  S.~Ferrara, A.~Kehagias, H.~Partouche and A.~Zaffaroni,
  ``AdS(6) interpretation of 5D superconformal field theories,''
  Phys.\ Lett.\ B {\bf 431}, 57 (1998)
  [arXiv:hep-th/9804006].
}

\lref\DouglasYP{
  M.~R.~Douglas, D.~Kabat, P.~Pouliot and S.~H.~Shenker,
  ``D-branes and short distances in string theory,''
  Nucl.\ Phys.\ B {\bf 485}, 85 (1997)
  [arXiv:hep-th/9608024].
}

\lref\kro{ H.~Heiselberg and A.~D.~Jackson,
``Signatures of QCD matter at RHIC,''
arXiv:nucl-th/9809013; ~~~
H.~Heiselberg and A.~D.~Jackson,
``Anomalous multiplicity fluctuations from phase transitions in heavy ion
collisions,''
Phys.\ Rev.\ C {\bf 63}, 064904 (2001) [arXiv:nucl-th/0006021].
}

\lref\krt{ I.~N.~Mishustin,
``Non-equilibrium phase transition in rapidly expanding {QCD} matter,''
Phys.\ Rev.\ Lett.\  {\bf 82}, 4779 (1999) [arXiv:hep-ph/9811307].
}

\lref\krwi{ M.~G.~Alford, K.~Rajagopal and F.~Wilczek,
``QCD at finite baryon density: Nucleon droplets and color
superconductivity,''
Phys.\ Lett.\ B {\bf 422}, 247 (1998) [arXiv:hep-ph/9711395].
}

\lref\krnwi{ J.~Berges and K.~Rajagopal, ``Color superconductivity
and chiral symmetry restoration at nonzero  baryon density and
temperature,'' Nucl.\ Phys.\ B {\bf 538}, 215 (1999)
[arXiv:hep-ph/9804233].
}

\lref\HubenyXN{
  V.~E.~Hubeny and M.~Rangamani,
  ``Unstable horizons,''
  JHEP {\bf 0205}, 027 (2002)
  [arXiv:hep-th/0202189].
}

\lref\HorowitzMaeda{
   G.~T.~Horowitz and K.~Maeda,
   ``Fate of the black string instability,''
    Phys.\ Rev. \ Lett. {\bf 87}, 131301 (2001) [arXiv:hep-th/0105111].
}

\lref\krit{
  K.~Rajagopal,
  ``Mapping the QCD phase diagram,''
  Nucl.\ Phys.\ A {\bf 661}, 150 (1999)
  [arXiv:hep-ph/9908360].
}

\lref\firstord{
O.~Aharony, J.~Marsano, S.~Minwalla, K.~Papadodimas and M.~Van Raamsdonk,
  ``A first order deconfinement transition in large N Yang-Mills theory on a
  small $S^3$,''
  Phys.\ Rev.\ D {\bf 71}, 125018 (2005)
  [arXiv:hep-th/0502149].
}

\lref\henningson{ M.~Henningson and K.~Skenderis, ``The
holographic Weyl anomaly,'' JHEP {\bf 9807}, 023 (1998)
[arXiv:hep-th/9806087].
}

\lref\fefferman{
  C.~Fefferman and C.R.~Graham,
  ``Conformal invariants,''
  in {\it Elie Cartan et les Mathematiques d'Aujourd'hui}, Asterisque (1985) 95.
}

\lref\hollands{
  S.~Hollands, A.~Ishibashi and D.~Marolf,
  ``Counter-term charges generate bulk symmetries,''
  arXiv:hep-th/0503105.
}

\lref\gubsermitra{
  S.~Gubser and I.~Mitra,
  ``Instability of charged black holes in anti-de Sitter space,''
   arXiv:hep-th/0009126.
}

\lref\BrownBR{
  J.~D.~Brown and J.~W.~York,
  ``Quasilocal energy and conserved charges derived from the gravitational
  action,''
  Phys.\ Rev.\ D {\bf 47}, 1407 (1993).
}

\lref\BalasubramanianRE{
  V.~Balasubramanian and P.~Kraus,
  ``A stress tensor for anti-de Sitter gravity,''
  Commun.\ Math.\ Phys.\  {\bf 208}, 413 (1999)
  [arXiv:hep-th/9902121].
}

\lref\weyl{
  H.~Weyl, ``Zur Gravitationstheorie,''
  Ann. Phys.\ (Leipzig) {\bf 54}, 117 (1917).
}

\lref\reall{
  R.~Emparan and H.~Reall,
  ``Generalized Weyl solutions,''
  Phys.\ Rev.\ D {\bf 65}, 084025 (2002)
  [arXiv:hep-th/0110258].
}

\lref\wisemanA{
  T.~Wiseman,
  ``Relativistic stars in Randall-Sundrum gravity,''
  Phys.\ Rev.\ D {\bf 65}, 124007 (2002)
  [arXiv:hep-th/0111057].
}

\lref\wisemanB{
  T.~Wiseman,
  ``Static axisymmetric vacuum solutions and non-uniform black strings,''
  Class. \ Quant.\ Grav. \ {\bf 20}, 1137 (2003)
  [arXiv:hep-th/0209051].
}

\lref\kudoh{
  H.~Kudoh, T.~Tanaka and T.~Nakamura,
  ``Small localized black holes in braneworld: Formulation and numerical method,''
  Phys.\ Rev.\ D {\bf 68}, 024035 (2003)
  [arXiv:gr-qc/0301089].
}

\lref\kleihaus{
  B.~Kleihaus and J.~Kunz,
  ``Static black hole solutions with axial symmetry,''
  Phys.\ Rev.\ Lett.\  {\bf 79}, 1595 (1997) [arXiv:gr-qc/9704060].
}

\lref\horowitz{ R.~Emparan, G.~T.~Horowitz and R.~C.~Myers,
  ``Exact description of black holes on branes,''
  JHEP {\bf 01}, 007 (2000) [arXiv:hep-th/9911043].
}

\lref\plebanski{ J.~F.~Plebanski and M.~Demianski,
  ``Rotating, charged, and uniformly accelerating mass in general relativity,''
  Annals Phys.\  {\bf 98}, 98 (1976).
}

\lref\zakout{
  I.~Zakout,
  ``The C-metric black hole near the IR-brane in the AdS(4) space,''
  arXiv:hep-th/0210063.
}

\lref\GubserQJ{
  S.~S.~Gubser, C.~P.~Herzog and I.~R.~Klebanov,
  ``Symmetry breaking and axionic strings in the warped deformed conifold,''
  JHEP {\bf 0409}, 036 (2004)
  [arXiv:hep-th/0405282].
}

\lref\HorowitzMyers{
  G.~T.~Horowitz and R.~C.~Myers,
  ``The AdS/CFT correspondence and a new positive energy conjecture for  general relativity,''
     Phys.\ Rev.\ D {\bf 59}, 026005 (1999) [arXiv:hep-th/9808079].
}

\lref\glg{ O.~Aharony, J.~Marsano, S.~Minwalla and T.~Wiseman,
``Black hole - black string phase transitions in thermal 1+1
dimensional supersymmetric Yang-Mills theory on a circle,'' Class.\
Quant.\ Grav.\  {\bf 21}, 5169 (2004) [arXiv:hep-th/0406210].
}

\lref\CharGregory{
     C.~Charmousis and R.~Gregory,
     ``Axisymmetric metrics in arbitrary dimensions,''
     Class. \ Quant.\ Grav. \ {\bf 21}, 527 (2004) [arXiv:gr-qc/0306069].
}

\lref\EmparanWN{
  R.~Emparan and H.~S.~Reall,
  ``A rotating black ring in five dimensions,''
  Phys.\ Rev.\ Lett.\  {\bf 88}, 101101 (2002)
  [arXiv:hep-th/0110260].
}

\lref\spenta{
L.~Alvarez-Gaume, C.~Gomez, H.~Liu and S.~Wadia,
``Finite temperature effective action, AdS(5) black holes, and 1/N expansion,''
Phys.\ Rev.\ D {\bf 71}, 124023 (2005) [arXiv:hep-th/0502227].
}

\lref\Creminelli{
  P.~Creminelli, A.~Nicolis and R.~Rattazzi,
  ``Holography and the electroweak phase transition,''
  JHEP {\bf 03}, 051 (2002) [arXiv:hep-th/0107141].
}

%
%

\lref\skena{
  M.~Henningson and K.~Skenderis,
  ``Holography and the Weyl anomaly,''
  Fortsch.\ Phys. {\bf 48}, 125 (2000) [arXiv:hep-th/9812032].
}

\lref\skenb{
  S.~de~Haro, S.~Solodukhin and K.~Skenderis,
  ``Holographic reconstruction of spacetime and renormalization in the  AdS/CFT correspondence,''
  Commun.\ Math.\ Phys.\  {\bf 217}, 595 (2001) [arXiv:hep-th/0002230].
}

\lref\skenc{
  K.~Skenderis,
  ``Asymptotically anti-de Sitter spacetimes and their stress energy tensor,''
  Int.\ J.\ Mod.\ Phys. {\bf A16}, 740 (2001)
  [arXiv:hep-th/0010138].
}

\lref\skend{
  I.~Papadimitriou and K.~Skenderis,
  ``Thermodynamics of asymptotically locally AdS spacetimes,''
  arXiv:hep-th/0010138.
}

\lref\schnitzer{
  H.~Schnitzer,
  ``Confinement / deconfinement transition of large N gauge theories with  N(f) fundamentals: N(f)/N finite,''
  Nucl. \ Phys. {\bf B695}, 267 (2004) [arXiv:hep-th/0402219].
}

\def\my_Title#1#2{\nopagenumbers\abstractfont\hsize=\hstitle\rightline{#1}%
\vskip .5in\centerline{\titlefont
#2}\abstractfont\vskip.5in\pageno=0}

\my_Title {\vbox{\baselineskip12pt \hbox{WIS/18/05-JUL-DPP}
\hbox{HUTP-05/A0035} \hbox{\tt hep-th/0507219}}}
{\vbox{\centerline{Plasma-Balls in Large $N$ Gauge
Theories}\vskip7pt \centerline { and Localized Black Holes }}}

\centerline{Ofer Aharony$^{a,}$\foot{E-mail :
{\tt Ofer.Aharony@weizmann.ac.il}.},
Shiraz Minwalla$^{b,c,}$\foot{E-mail : {\tt
minwalla@theory.tifr.res.in}.} and Toby Wiseman$^{c,}$\foot{E-mail :
{\tt twiseman@fas.harvard.edu}.}}

\medskip

\centerline{\sl $^{a}$Department of Particle Physics, Weizmann Institute of
Science, Rehovot 76100, Israel}
\centerline{\sl $^{b}$Department of Theoretical Physics, Tata Institute
of Fundamental Research,}
\centerline{\sl Homi Bhabha Rd, Mumbai 400005, India}
\centerline{\sl $^{c}$Jefferson Physical Laboratory, Harvard University,
Cambridge, MA 02138, USA}


\vskip 0.3cm

\noindent We argue for the existence of plasma-balls -- meta-stable,
nearly homogeneous lumps of gluon plasma at just above the
deconfinement energy density -- in a class of large $N$ confining
gauge theories that undergo first order deconfinement transitions.
Plasma-balls decay over a time scale of order $N^2$ by thermally
radiating hadrons at the deconfinement temperature. In gauge
theories that have a dual description that is well approximated by a
theory of gravity in a warped geometry, we propose that plasma-balls
map to a family of classically stable finite energy black holes
localized in the IR. We present a conjecture for the qualitative
nature of large mass black holes in such backgrounds, and
numerically construct these black holes in a particular class of
warped geometries. These black holes have novel properties; in
particular their temperature approaches a nonzero constant value at
large mass. Black holes dual to plasma-balls shrink as they decay by
Hawking radiation; towards the end of this process they resemble ten
dimensional Schwarzschild black holes, which we propose are dual to
small plasma-balls. Our work may find practical applications in the
study of the physics of localized black holes from a dual viewpoint.

\Date{July 2005}



\centerline{\bf Contents}\nobreak\medskip{\baselineskip=12pt
 \parskip=0pt\catcode`\@=11  

\noindent {1.} {Introduction and Summary} \leaderfill{2} \par 
\noindent \quad{1.1.} {Meta-stable Bubbles of Plasma in Confining Large $N$ Gauge Theories} \leaderfill{2} \par 
\noindent \quad{1.2.} {Black Holes in the IR} \leaderfill{3} \par 
\noindent \quad{1.3.} {Plasma-Balls as Black Holes} \leaderfill{5} \par 
\noindent {2.} {The Plasma-ball as a Stable Lump of Plasma Fluid} \leaderfill{7} \par 
\noindent {3.} {Hadronization of Plasma-Balls at Large $N$} \leaderfill{10} \par 
\noindent \quad{3.1.} {Counting Powers of $N$} \leaderfill{11} \par 
\noindent \quad{3.2.} {k=0: Gluon Mean Free Time} \leaderfill{12} \par 
\noindent \quad{3.3.} {k=1: Glueball Production Rate} \leaderfill{12} \par 
\noindent \quad{3.4.} {k=2: Friction} \leaderfill{13} \par 
\noindent {4.} {Localized Black Holes in Warped Backgrounds} \leaderfill{13} \par 
\noindent {5.} {Localized Black Holes as Plasma-balls} \leaderfill{16} \par 
\noindent {6.} {Plasma-Ball Dynamics from Black Holes} \leaderfill{18} \par 
\noindent \quad{6.1.} {Plasma-Ball Hadronization as Hawking Radiation} \leaderfill{18} \par 
\noindent \quad{6.2.} {Plasma-Ball Production in Hadron-Hadron Collisions} \leaderfill{18} \par 
\noindent {7.} {Black Hole Physics from Gluon Plasmas} \leaderfill{20} \par 
\noindent \quad{7.1.} {The Fate of Small Schwarzschild Black Holes} \leaderfill{21} \par 
\noindent \quad{7.2.} {Information Conservation in Hawking Radiation} \leaderfill{21} \par 
\noindent \quad{7.3.} {How Black is a Black Hole ?} \leaderfill{22} \par 
\noindent {8.} {Numerical Solutions for Domain Walls in some Specific Backgrounds} \leaderfill{23} \par 
\noindent \quad{8.1.} {Scherk-Schwarz Compactifications of Anti-De Sitter Space} \leaderfill{23} \par 
\noindent \quad{8.2.} {The Domain Wall: Asymptotic Behaviors} \leaderfill{25} \par 
\noindent Appendix {A.} {Thermodynamics of Large $N$ Gauge Theories} \leaderfill{34} \par 
\noindent \quad{\hbox {A.}1.} {Local Stability of Homogeneous Configurations} \leaderfill{34} \par 
\noindent \quad{\hbox {A.}2.} {Density of States in Confining Gauge Theories} \leaderfill{35} \par 
\noindent \quad{\hbox {A.}3.} {Phase Separation in Confining Large $N$ gauge theories} \leaderfill{37} \par 
\noindent Appendix {B.} {Properties of Plasma Dynamics} \leaderfill{38} \par 
\noindent \quad{\hbox {B.}1.} {The Surface Tension of the Plasma-Ball} \leaderfill{38} \par 
\noindent \quad{\hbox {B.}2.} {Dynamical Evolution of Lumps of Plasma} \leaderfill{39} \par 
\noindent \quad{\hbox {B.}3.} {Small Plasma-Balls} \leaderfill{40} \par 
\noindent \quad{\hbox {B.}4.} {Plasma Lumps in Theories with Second Order Deconfinement Transitions} \leaderfill{40} \par 
\noindent \quad{\hbox {B.}5.} {Plasma-Balls in the Real World ?} \leaderfill{40} \par 
\noindent Appendix {C.} {Counting Powers of $N$ at Lowest Order in Perturbation Theory} \leaderfill{41} \par 
\noindent \quad{\hbox {C.}1.} {$k=0$ : Mean Free Path} \leaderfill{41} \par 
\noindent \quad{\hbox {C.}2.} {$k=1$ : Rate of Glueball Production} \leaderfill{42} \par 
\noindent Appendix {D.} {The Final Decay of Small Plasma-Balls} \leaderfill{43} \par 
\noindent Appendix {E.} {The Boundary Stress Tensor for Compactifications of $AdS_{d+2}$} \leaderfill{44} \par 
\noindent Appendix {F.} {Numerical Construction of the Domain Wall Solution} \leaderfill{44} \par 
\noindent \quad{\hbox {F.}1.} {Equations and Constraints} \leaderfill{44} \par 
\noindent \quad{\hbox {F.}2.} {The IR Boundary Conditions} \leaderfill{46} \par 
\noindent \quad{\hbox {F.}3.} {The UV Boundary Conditions} \leaderfill{47} \par 
\noindent \quad{\hbox {F.}4.} {Solving the Constraint System } \leaderfill{51} \par 
\noindent \quad{\hbox {F.}5.} {Summary} \leaderfill{52} \par 
\noindent Appendix {G.} {Numerical Details} \leaderfill{52} \par 
\noindent References \leaderfill{56} \par
\catcode`\@=12 \bigbreak\bigskip}

\newsec{Introduction and Summary}

In this paper we study the spectrum of localized excitations in
confining large $N$ gauge theories with energies of order $N^2$, the
spectrum of static black holes in warped geometries with an effective
infrared wall, and the connection between these two spectra, using
generalizations of the AdS/CFT correspondence.  In the rest of this
section we will first introduce each of these questions and then
explain the relationship between them. Readers who are only interested
in gauge theory aspects may read sections 1.1, 2 and 3 and appendices
A, B, C, which are independent of the rest of the paper.

\subsec{Meta-stable Bubbles of Plasma in Confining Large $N$ Gauge Theories}

It has long been believed \tHooftJZ\ that confining large $N$ $SU(N)$
gauge theories
are dual to weakly coupled string theories with a string coupling
$g_s \propto 1/N$. The string duals possess
an exponential tower of long lived excitations (perturbative string states)
which are identified with
glueballs. However, at least some perturbative string theories also
possess other long lived excitations -- black holes -- at energies of order
$1/g_s^2$. In this paper we argue that such
configurations have analogues in a class of confining large $N$ gauge
theories (for earlier related works see \refs{\giddings,\nastase}).

Consider a confining large $N$ gauge theory, with mass gap $\Lambda_{gap}$,
whose thermal deconfinement phase transition is of first order (e.g. pure Yang
Mills theory at large $N$ \refs{\teper, \firstord}). In this paper we
argue that, provided the theory in question obeys one additional
condition (see section 2), it hosts long lived excitations at all
masses $m \gg N^2 \Lambda_{gap}$.
We call these meta-stable configurations plasma-balls.
Plasma-balls are spherical, approximately homogeneous lumps of
deconfined plasma fluid, whose energy density is just above
the critical density (which is the energy density at the deconfinement phase
transition temperature $T_d$; this energy density is of
order $N^2$ in the large $N$ limit). These lumps are static because
the pressure of the plasma in such theories
vanishes at the critical energy density.

Of course a plasma-ball is not completely stable; eventually it
decays into a gas of hadrons. In section 3 we study the $N$
dependence of the processes that contribute to this decay, and
conclude that the hadronization of the plasma-ball is very slow in
the large $N$ limit. Even though the density of the plasma-ball is
of order $N^2$, and the number of gluon-gluon interactions per
unit time within the plasma-ball is of order $N^2$, glueballs are
produced from these collisions at a rate that is independent of
$N$. Intuitively, only  $1/N^2$ of the gluon-gluon collisions --
collisions between a gluon and its color anti-partner -- can form
a singlet state that can escape from the plasma.

Large $N$ counting is insufficient to determine the precise rate of
radiation of glueballs as a function of (for instance) their mass;
however the thermal nature of the plasma-ball suggests that these
rates are controlled by the Boltzmann factor at the temperature of
the plasma-ball. We conjecture that this is indeed the case.
In a class of examples described below, the dual gravitational
description of plasma-balls confirms this conjecture.

We have already explained that the rate of radiation of any given
glueball from the surface of a plasma-ball is independent of $N$.
Using the conjecture of the previous paragraph and the expectation
that the density of glueball species at high energies is
proportional to $\exp(E/T_H)$ where $T_H$ is the Hagedorn
temperature, the rate of loss of energy integrated over all glueball
species is proportional to $\int \exp\left( \left( {1 \over T_H}-{1
\over T_d}\right) E \right) dE $. As the deconfinement temperature
is strictly smaller than the Hagedorn temperature in theories that
undergo first order deconfinement transitions \refs{\pisarski,
\sphere}, this is a finite number of order $N^0$. As a consequence,
the lifetime of a plasma-ball of radius $R$ ($R\gg 1/\Lambda_{gap}$)
is of order  $N^2 R$.

Note that QCD at zero chemical potential (for baryon number) does
not undergo a first order phase transition as a function of
temperature. Thus, we do not expect plasma-ball-like
meta-stable configurations to be created in experiments like
RHIC\foot{Dynamical plasma configurations which are dual to dynamical black
holes could perhaps be related to RHIC, as suggested by Nastase and
collaborators \nastase. In this context note that the decay products
of the RHIC fireball have a distribution which is close to
thermal, at approximately
the deconfinement crossover temperature.}. See appendix B.5 for further
discussion.

\subsec{Black Holes in the IR}

We start this subsection by characterizing the class of warped geometries
whose  black holes we study.
Consider any solution to Einstein's equations coupled to appropriate
matter fields, whose metric can be put in the form
\eqn\backmet{ds^2=\alpha' L^2 \left( W^2(u) dx_{\mu}^2 + d s_{int}^2
\right),}
where $\mu = 0,1,2, \cdots, p$, $\alpha' L^2$ is a constant with units
of length squared, and $ds_{int}^2$ is the metric
on an internal manifold, one of whose coordinates is $u$, and whose
constant $u$ slices are compact. The variable $u$ in \backmet\ has
the range $u_0<u<\infty$ and the function $W(u)$ increases
monotonically from a positive nonzero value at $u_0$ to infinity at
$u=\infty$. We assume that \backmet\ is smooth and without
boundaries everywhere, including at the IR wall $u=u_0$ \foot{This
is possible because a $k$-cycle of the internal manifold shrinks to
zero size at $u=u_0$, so that, locally, $u-u_0$ may simply be
thought of as the radial coordinate of an $\IR^{k+1}$ component of
the geometry.}. We further assume that the space in question has well
defined thermodynamics (in particular it admits the definition of an
ADM mass),
and that the spectrum of fluctuations about \backmet\ is gapped.
Examples include the confining backgrounds constructed in \refs{
\wit,\klestr,\malnun}. \foot{The background of \klestr\ actually does
not have a mass gap \GubserQJ, but we expect that our analysis should
still apply to it.}

We are interested in backgrounds \backmet\ which have an
additional property; they host black brane solutions of finite
energy density \foot{More precisely, there exist smooth gravitational
solutions with horizons, asymptoting to \backmet, which
preserve translational invariance in $\IR^p$ and
have a finite energy density above that of the background \backmet.}
at all energy densities larger than a critical value $\rho_e$.
Moreover, these branes are required to be stable (and have positive
specific heat) above $\rho_s > \rho_e$, and to have negative free
energy density above $\rho_c>\rho_s$. Let $T_d$ denote the
temperature of the black brane of energy density $\rho_c$.
Assuming that there are no other phases of the theory,
it follows that the thermodynamics in this background
is dominated by a graviton gas about \backmet\
for $T<T_d$, and by the black brane for $T>T_d$; the system undergoes
a first order phase transition at $T=T_d$.

In this paper we conjecture that spaces that obey all these
properties possess a one parameter set of spherically symmetric (in
$p$ dimensions) black hole solutions\foot{Black holes in this
context were first discussed in \giddings.} labeled by their mass,
and that, in the large mass limit, these black holes have the
following properties:
\item{1.} Their volume in $p$ dimensions is proportional to their mass, i.e.
the radius of the black hole in $p$ dimensions scales with mass
like $(m / \rho_c)^{1 / p}$.
\item{2.} In the interior (meaning away from the edges in the
$\IR^p$ spatial directions, not the black hole interior behind the horizon)
these black hole
solutions approximate the black brane at energy density $\rho_c$.
\item{3.} In the vicinity of their edge the solutions reduce to
a domain wall in the $\IR^p$ directions, that interpolates between the
black brane at the critical energy density $\rho_c$ at one end, and
the background \backmet\ at the other end.

We emphasize that the black holes described in the previous paragraph
have thermodynamical properties that are qualitatively
different from black holes in flat space; in particular their
temperature tends to a finite value $T_d$ in the limit of infinite mass.

In section 4 we use intuitive arguments to motivate the conjecture
described above. In addition, as concrete evidence for our
conjecture, in section 8 we present a numerical construction of the
domain wall about a particularly simple background of the form
\backmet.  The background \wit\ we study is a solution to Einstein's
equations with a negative cosmological constant but no other matter
fields, and asymptotes to a Scherk-Schwarz circle compactification
of Lorentzian $AdS_{d+2}$, or (at finite temperature in Euclidean
space) to a torus compactification of Euclidean $AdS_{d+2}$.  An
infinite series of translationally invariant (in $d-1$ dimensions)
solutions with these asymptotics is known. The topology of each of
these solutions is $\IR^{d}$ times the solid torus; the distinct
solutions are labeled by which cycle of the boundary torus is
`filled in' in the solution. When the boundary torus is rectangular,
the thermodynamics is dominated by the solution that fills the
smaller of the $(1,0)$ and $(0,1)$ cycles. A phase transition
between these two solutions occurs when the boundary torus is a
square. In section 8 we give an overview of the numerical
construction of the domain wall solution that interpolates between
these two solutions at the phase transition temperature, and compute
the (positive) surface tension of this domain wall. More details of
our construction may be found in appendices F and G, and an example
of the program used to generate the solutions may be downloaded from
{\tt http://schwinger.harvard.edu/$\sim$wiseman/IRblackholes/} .

\subsec{Plasma-Balls as Black Holes}

Dual string descriptions of large $N$ $SU(N)$ gauge theories are
expected to exist on general grounds from 't Hooft's analysis of the
large $N$ limit \tHooftJZ. The dual string coupling is always
proportional to $1/N$, so the interactions between strings are small
in the large $N$ limit. Unfortunately the string duals to most
familiar confining gauge theories (such as pure $SU(N)$ Yang-Mills
theory) have yet to be discovered; furthermore, the relevant dual
dynamics may be argued to be controlled by a strongly coupled
worldsheet theory that does not admit a truncation to the lowest
(gravitational) sector.
In several examples \refs{\wit, \polstr, \klestr, \malnun}, however,
it turns out to be possible to modify standard confining gauge
theories by coupling them to a judiciously chosen set of additional
degrees of freedom. The energy scale of `new physics' is a free
parameter, which is most conveniently labeled by the value of the 't
Hooft coupling $\lambda\equiv g^2_{YM}N$ at the scale of new
physics. In the limit $\lambda \to 0$ the scale of new physics is
much higher than the mass gap, and, at low energies compared to the
scale of the new physics, we recover the (strongly coupled) dynamics
of the original confining theory. In the opposite limit of large
$\lambda$, the new degrees of freedom give large anomalous
dimensions to all operators that create glueballs dual to string
oscillators. As a consequence such glueballs have masses that are
parametrically higher than the graviton, and the low energy dual
dynamics reduces to a theory of gravity (usually ten dimensional
supergravity) in a warped background. In specific examples
\refs{\wit, \polstr, \klestr, \malnun}  this gravitational dual may
explicitly be identified using a generalization of the AdS/CFT
correspondence.  In each case the resulting gravitational theory
lives in a warped geometry of the form \backmet\ (for a $p+1$-dimensional
confining gauge theory) that obeys all the conditions of the
previous subsection, and so, according to our conjecture, hosts
localized black holes. Since the `black brane nucleation' phase
transitions in these geometries may be identified with deconfinement
transitions in the dual field theory \wit, it follows that these
localized black holes are dual to plasma-balls (see \refs{\giddings,
\loc, \nastase} for related remarks).

\twosidefig{The plasma-ball and its decay via hadronization (left), and
the localized black hole and its decay via Hawking radiation (right).}
{4truein}{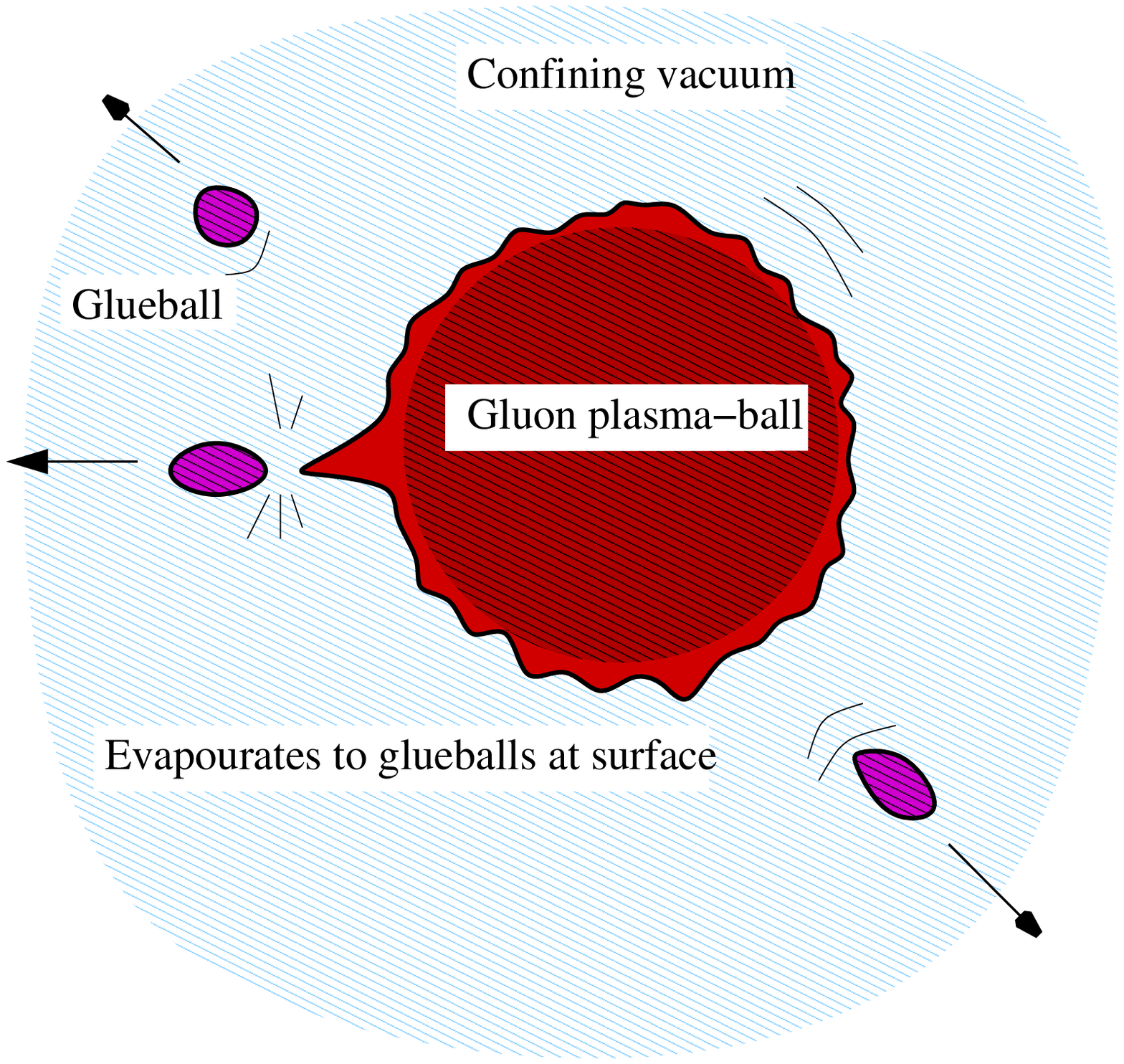}{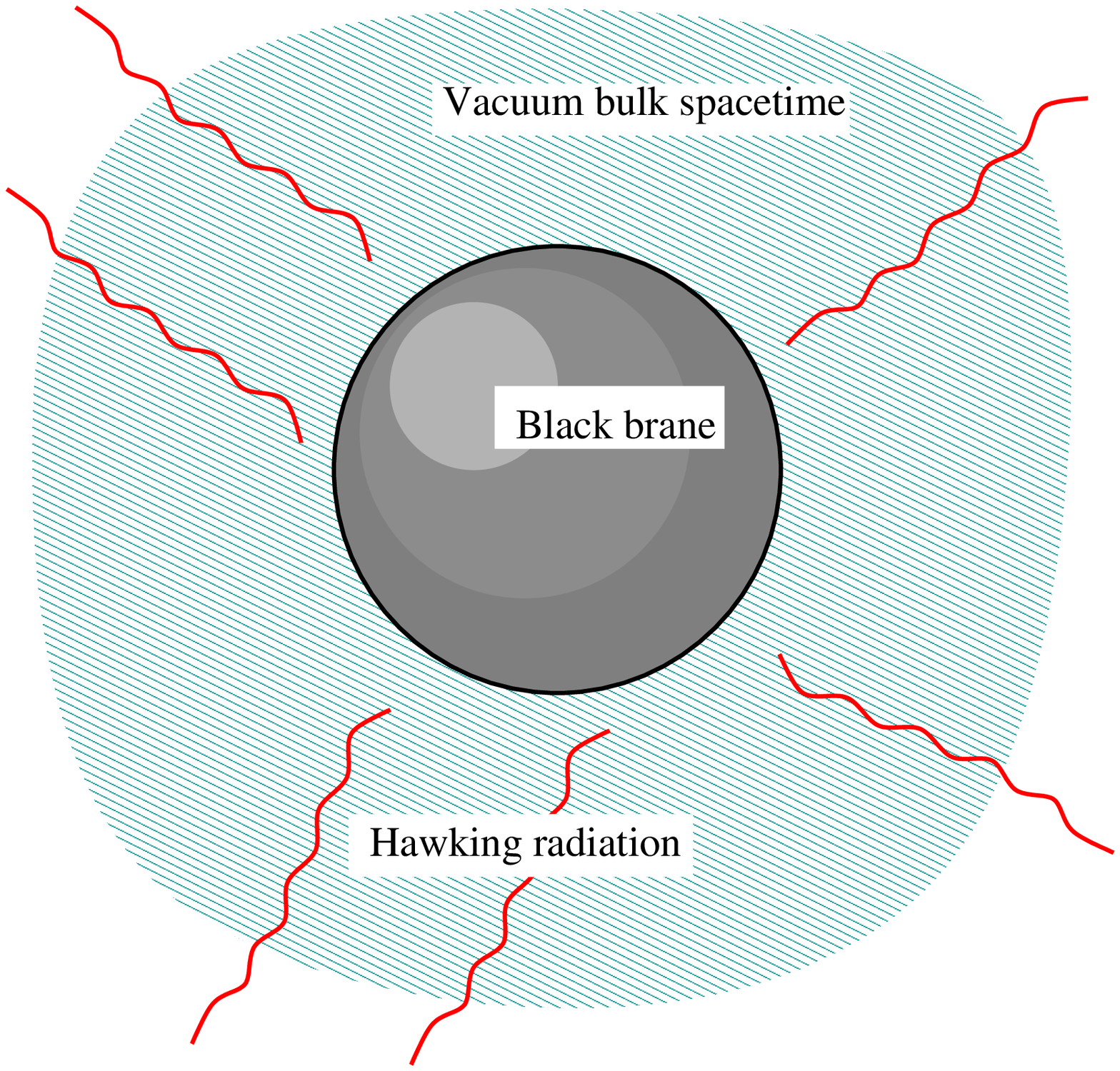}\figlabel{\balls}

The decay of the plasma-ball by hadronization maps to the decay of
its dual black hole by Hawking radiation. It follows that, at least
at large $\lambda$, the decay of plasma-balls is a thermal process
at the temperature $T_d$, in agreement with the guess described in
\S1.1.

Even in warped backgrounds, high energy graviton-graviton collisions
are likely to be dominated by black hole formation \refs{\giddings,
\nastase}. It follows that, in confining theories at large $\lambda$,
high energy hadron-hadron collisions are dominated by plasma-ball
production. Of course, plasma-ball formation does not dominate high
energy glueball-glueball collisions at small $\lambda$. In Section 6
we trace these different behaviors to qualitative differences in
the hadron parton distribution functions at small and large $\lambda$
\deepin.

The duality with plasma-balls could yield interesting lessons for
black hole physics. As we have explained above, at large $\lambda$ the
production of a plasma-ball in high energy hadron-hadron collisions,
and its subsequent decay by hadronization, is dual to the production
of a black hole in high energy graviton-graviton scattering, and its
subsequent decay by Hawking radiation. As the first of these processes
is manifestly unitary, its dual must be as well. It is possible that
this connection could be pursued further to draw more detailed lessons
about the nature of the Hawking radiation process.

As a second application, we argue in section 7 that one of the
most striking features of black hole physics -- the universally
absorptive nature of black holes -- is crucially tied to a feature
of the hadronic parton distribution function at large $\lambda$
that is absent at small $\lambda$. It follows that while black
hole like configurations continue to exist in at least some warped
geometries with curvatures of order the string scale (those that
are dual to gauge theories at small $\lambda$), they have
qualitatively different dynamics from their large $\lambda$
cousins. In particular they are no longer black.

\newsec{The Plasma-ball as a Stable Lump of Plasma Fluid}

In this elementary sub-section we will argue for the existence of a
meta-stable localized lump of plasma fluid (in the deconfined phase) -- a
plasma-ball -- in certain large $N$ theories that undergo first
order phase transitions. More details may be found in appendix A.

Consider an isolated spherical ball of static plasma fluid of radius
$R \gg {1 / \Lambda_{gap}}$ in $p$ spatial dimensions.  In
equilibrium its effective temperature $T$ and the pressure $P$ are
both uniform in the bulk of the ball\foot{In this section we ignore
energy loss by `radiation' from the surface of the plasma-ball. In
the next section we will argue that this approximation is justified
in the large $N$ limit. The energy loss by radiation to the bulk is
$\CO(1)$; as the energy density within the plasma ball is
$\CO(N^2)$, the temperature gradients induced by surface radiation
are negligible in the large $N$ limit.}. In order for the ball to be
static, the pressure of the plasma in the ball must precisely
balance the tension of the domain wall separating the two phases
(the ``surface tension'' of the plasma fluid; see appendix B.1 for a
precise definition). Denoting this tension by $\Sigma$, the forces
balance when
\eqn\fb{
\omega_{p-2} R^{p-2} \Sigma= {\omega_{p-2} \over {p-1}} R^{p-1} P
\qquad \Longrightarrow \qquad P = { (p-1)\Sigma \over R},}
where
$\omega_{p-2}$ is the surface area of the unit $(p-2)$-sphere, and
$\omega_{p-2}/(p-1)$ is the volume of the unit $(p-1)$-ball. Since the
bubble is in the deconfined phase, its pressure and surface tension
are both of order $N^2$ \foot{As we review in appendix A, the
pressure is simply minus the free energy density of the bubble. The
fact that the surface tension should scale as $N^2$ is predicted
from general arguments, since planar diagrams should contribute to
it, but it has not yet been directly verified by lattice
computations \teper.}, and are functions of the plasma-ball
temperature. According to \fb, static plasma-balls exist at
asymptotically large $R$ if and only if the pressure of the
deconfined phase vanishes at a finite temperature.

While the precise functional form of $P(T)$ (the pressure of the
deconfined phase as a function of temperature) depends on details,
this function is constrained by thermodynamic considerations. First, a
simple thermodynamical identity (see appendix A) ensures that the
pressure increases monotonically with temperature, provided that the
specific heat of the deconfined phase is positive. Second, as pressure
is continuous across a phase transition, the pressure of the
deconfined phase is $\CO(1)$ (rather than the generic $\CO(N^2)$) at
the deconfinement temperature $T_d$. Provided that the phase
transition is of first order (an assumption we will make in most of
the rest of this paper) the specific heat, and so $dP/dT$, are
positive and of order $N^2$ at $T=T_d$ (see appendix A) \foot{More
precisely this is true of the limit of the specific heat, as $T$
approaches $T_d$ from above. This limit determines the speed of sound
of the deconfined plasma fluid within the plasma-ball.}. It follows
that the pressure of the deconfined plasma vanishes at
$T=T_d-\CO(1/N^2)$. We conclude that uniform, asymptotically large
lumps of fluid plasma are static at the deconfinement temperature in
large $N$ gauge theories that undergo first order phase transitions.
Provided that $\Sigma$ is positive, large but finite static spherical
lumps of plasma fluid exist at a temperature slightly above the
deconfinement temperature, and have an energy density slightly above
the critical energy density.

These static lumps are also hydrodynamically stable \foot{They
resemble the bubbles of the deconfined phase which appear when the
energy density of the system is gradually raised through the first
order phase transition, except for the fact that they live in the
vacuum.}, provided that the effective  surface tension is positive.
As we have assumed the deconfinement transition to be of first
order, the deconfined phase continues to exist (as a meta-stable
phase) at temperatures below $T_d$. As we have explained in the
previous paragraph, the pressure of the deconfined phase is positive
for $T>T_d$ and negative for $T<T_d$; it follows that plasma-balls
are stable against homogeneous expansion or contraction. Local
stability of the plasma fluid against density fluctuations in the
bulk of the ball follows from the positivity of the speed of sound
at the phase transition temperature (see appendix A). Linearized
long wavelength fluctuations of the surface also obey a wave
equation; the squared speed of sound of these surface waves is
proportional to the effective surface tension; as a consequence the
plasma-ball is stable to fluctuations of the boundary provided that
this surface tension is positive and approaches a finite value at
the deconfinement temperature.

Note that according to \fb\ the pressure (hence temperature) of a
plasma-ball with positive surface tension is a decreasing function of
its radius (hence mass), so plasma-balls have negative specific heat
\foot{Note that the (negative) specific heat of the plasma-ball as an
object is distinct from the (positive) specific heat per unit volume
of the deconfined phase of which the plasma-ball is composed.}. Of
course, plasma-balls are not completely stable; they eventually
hadronize. In the next section we discuss this hadronization
process. Additional discussions of various aspects of plasma dynamics
may be found in appendix B.

In this section we have, so far, discussed plasma-balls that carry
no conserved charges besides their energy, and we will focus on such
configurations in the remainder of the paper. However the
hydrodynamical construction of stationary lumps of plasma
generalizes in a straightforward manner upon adding other conserved
charges. As a simple example consider a radially symmetric lump of
plasma rotating with angular velocity $\omega$ (about the origin) in
$p=2$ spatial dimensions. Force balance yields the equation
\eqn\fbrot{{d P \over d r}= \rho(r) \omega^2 r,} where $P(r)$ and
$\rho(r)$ are the pressure and density at radial distance $r$, and
we have used the non-relativistic approximation appropriate for
small rotational velocities. Substituting $\rho $ as a function of
$P$ using the equation of state, we find
\eqn\inteq{\int^{\Sigma/R}_{P(r)} {dP \over \rho(P)} = {\omega^2
\over 2} (R^2 -r^2).} This may be regarded as an equation for
$P(r)$, and potentially it can have two qualitatively distinct classes of
solutions. In solutions of the first class the variable $r$ has the
range $(0, R)$. Such a solution may be thought of as a rotating
plasma-ball. In the second class of solutions $r$ has the range $(R_1,
R)$, and \inteq\ is to be solved subject to the boundary condition
$P(R_1)=-\Sigma/R_1$ (we assume $R,R_1 \gg 1/\Lambda_{gap}$).
Such a solution is best thought of as a
rotating plasma-ring, filling an annulus in the plane.

\newsec{Hadronization of Plasma-Balls at Large $N$}

In this section we will compute the $N$ dependence of the rate at
which a plasma-ball hadronizes (decays into glueballs).
We model the plasma-ball as a gas of
initially thermally distributed uncorrelated gluons, whose collisions
sometimes produce glueballs. We then determine the $N$
dependence of all Feynman graphs (with arbitrary numbers of
interaction vertices) that contribute to glueball production. We
find that the rate at which any particular glueball is produced is
independent of $N$.

Of course, the gluonic constituents of a plasma-ball are not really
uncorrelated; however all correlations may be switched off without
encountering a phase transition (for instance by going to high
temperatures in an asymptotically free theory), so they should not affect
the $N$-scaling of the gluon production rates computed in this section.
The arguments of this subsection are valid at all orders
in perturbation theory; however, experience with counting powers of $N$
in the 't Hooft limit (e.g. in exactly
solvable matrix models) suggests that the powers of $N$ obtained from this
argument will be correct even non-perturbatively. In specific examples we
will provide evidence that this is indeed the case in sections 5-8.

\subsec{Counting Powers of $N$}

Consider a connected Feynman graph that describes $n$ initial
gluons scattering into $m$ final gluons and $k$ glueballs. The
graph in question has $n+m$ external gluon lines, and includes $k$
insertions of glueball creation operators such as  $\Tr F_{\mu\nu}
F^{\mu\nu}$.  These operators are normalized so that their two
point functions are of order $N^0$ to ensure that they create
glueballs with unit probability; this means that each such
operator insertion in a Feynman diagram
appears with an extra factor of $1/N$ compared
to an insertion of an interaction vertex. The contribution of this
graph to the inclusive probability for glueball production is
obtained by squaring the graph and summing over all initial and
final gluon states. We will now determine the $N$ dependence of
the result of this process.

\fig{A typical sewn graph contributing to the glueball production rate,
in double-line notation.}{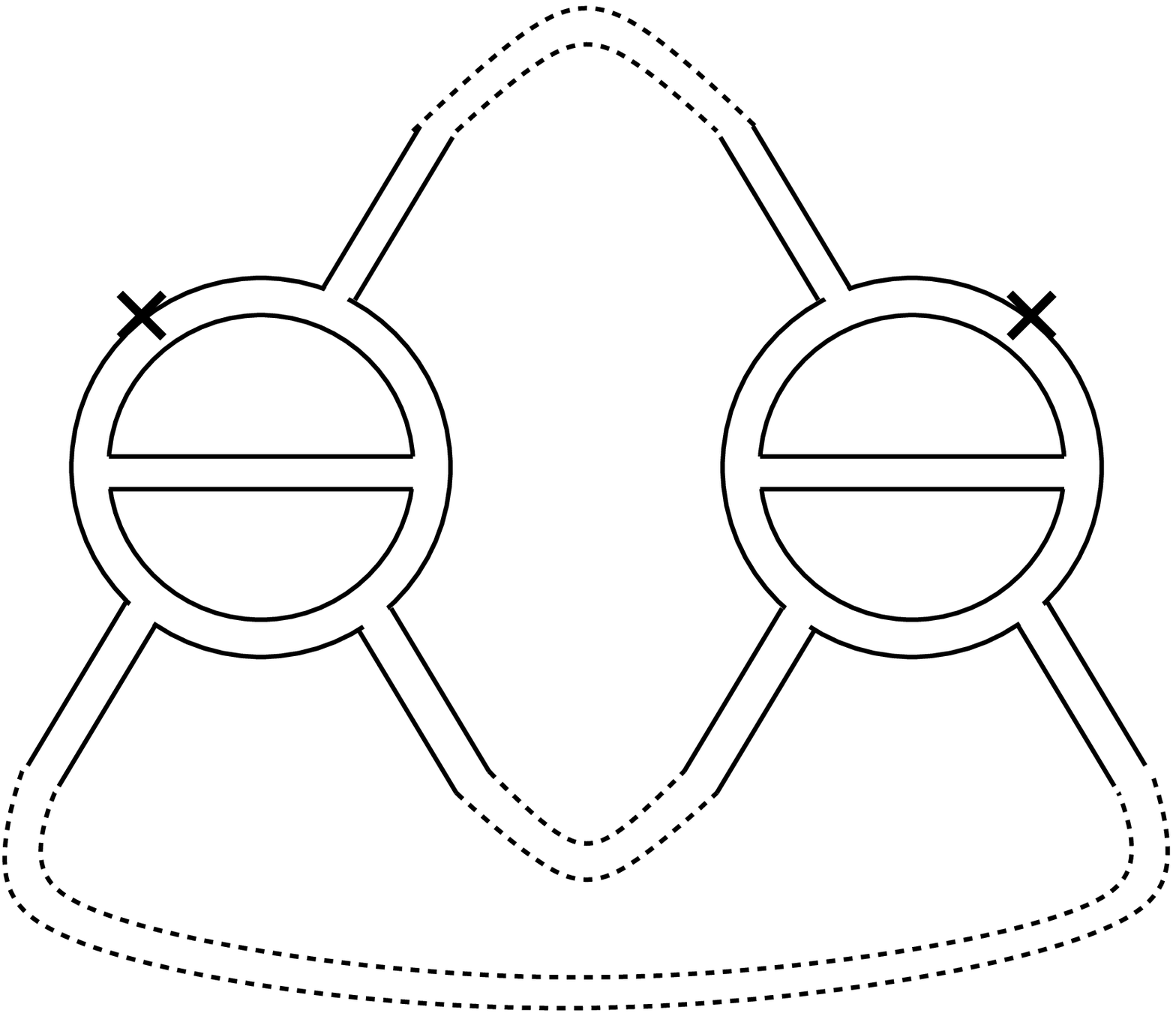}{3truein}
\figlabel{\loopgraph}

Consider the graph in question, drawn using the 't Hooft double line
notation, together with its CPT conjugate (a graph in which
fundamental and anti-fundamental indices are interchanged). Sew
these two graphs together, as in figure \loopgraph, by
attaching every free gluon line in the
first graph to the corresponding conjugate gluon line in the
conjugate graph; this gluing preserves the flow of all color indices. The
resulting graph has no free gluon lines, so the faces may be filled
in to form a genus $g$ Riemann surface in the usual manner.  The $N$
dependence of the contribution of this graph to inclusive glueball
production is obtained by freely summing over all indices of the
sewn graph\foot{This may be argued as follows. Color indices
attached to the $m$ final gluons must be summed over in summing
over all final states. Color indices attached to the $n$ initial
gluons must be summed over in summing over all initial gluons -- in
this step we use the fact that the plasma-ball has a finite density
of gluons of any given color variety. All internal color indices
must be summed over in squaring the original Feynman graph.}, and so
is proportional to $ N^{2-2g-2k}$ (using standard 't Hooft counting
and the normalization of the glueball creation operators). The
leading behavior at large $N$ is given by planar graphs ($g=0$) and
is proportional to \eqn\ndep{N^{2 - 2 k}.}

\subsec{k=0: Gluon Mean Free Time}

According to \ndep\ the effective `number' of gluon-gluon collisions
per unit time, in a large plasma-ball of volume $V$, is of order
$N^2$. It follows that the rate at which any given gluon
undergoes collisions is of order $N^0$ (recall that the plasma-ball
contains of order $N^2$ gluons). As a consequence, the relaxation
time scale of the plasma fluid is of order $N^0$. In appendix C we
verify this result at lowest order in perturbation theory.

\subsec{k=1: Glueball Production Rate}

The plasma-ball radiates glueballs by one of two mechanisms. Every
once in a while a gluon with indices $(i,j)$ collides with a gluon
with indices $(j,i)$
near the surface of the plasma-ball; this collision can produce a
glueball, which then escapes from the plasma-ball. Alternatively,
occasionally an energetic gluon shoots out of the surface of the
plasma-ball; the string that attaches it to the plasma-ball can then
snap, allowing it to escape into the bulk as a glueball. Glueball
production from each of these processes is governed by the Feynman
graphs studied in \S3.1 above. According to \ndep, in the
large $N$ limit glueball production is dominated by graphs with
$k=1$, and takes place at a rate that is independent of $N$.  In
appendix C we study a simple Feynman graph that contributes to
glueball production, to verify this conclusion at weak coupling (the
general analysis is valid at any value of the 't Hooft coupling).

In this subsection we have focused on the radiation of glueballs
from plasma-balls. However the arguments of this section apply
equally well to the time reversed process. Consider a glueball
incident upon the plasma ball from outside. Equation \ndep\ implies
that the interaction cross section for this glueball, per unit
length traversed through the plasma ball, is of order $N^0$. We
expect this interaction process to lead to the glueballs dissolving
into the plasma-ball. Similarly, glueballs formed in a gluon collision
far from the surface of a large plasma-ball will dissolve before they
escape, so the hadronization rate is proportional to the surface
area of the plasma-ball rather than its volume.

\subsec{k=2: Friction}

As we have discussed in the previous subsection, the cross section
for a glueball traversing the plasma-ball to dissolve into the
plasma is of unit order. Glueballs also interact with the plasma in
a more elementary manner; gluons incident on the glueball knock it
around, slowing down the glueball and causing it to jiggle around as
it traverses the plasma-ball. Such interactions are governed by
diagrams with one incoming and one outgoing glueball, have $k=2$ and
so are of order $1/N^2$.

\newsec{Localized Black Holes in Warped Backgrounds}

We now turn to a study of uncharged, static, finite energy (rather
than finite energy density) black holes in the backgrounds \backmet,
that are localized in the IR region of the radial direction
as well as in the $\IR^p$ directions,
in the range of parameters where gravity in the background
\backmet\ is a good approximation to the theory.
At low enough energies such solutions are well approximated
by ten dimensional Schwarzschild black holes of radius $R_s\ll
\sqrt{\alpha'} L$, centered at $u=u_0$ in \backmet. What is the
qualitative nature of the static black hole solutions at higher
energies? A consideration of the opposite, high energy limit gives us
a clue.  The confining backgrounds \backmet\ host infinite energy
translationally invariant (in the field theory $\IR^p$ directions)
black brane solutions, which undergo a first order phase transition at
some finite temperature\foot{Note that the deconfinement transition is
always of first order when the gravitational approximation is valid,
since it occurs far below the Hagedorn temperature which is related to
the string tension.}. The thermodynamic properties of these black
branes are clearly extensive (for instance, their energy is
proportional to their volume in $\IR^p$). Recall also that the
graviton (as well as all other fields) is massive in
\backmet; as a consequence all correlation functions decay
exponentially in this background.  Putting these facts together, it is
natural to expect that a black hole whose energy is very large, rather
than strictly infinite, is a large spherical lump in $\IR^p$ which in
its interior (meaning away from its edges in the $\IR^p$ directions -- not
the black hole interior behind the horizon) closely resembles a
translationally invariant black brane. Furthermore, taking the energy
and the radius of the lump to infinity and focusing on the edge of the
lump, we expect the solution to tend to a planar domain wall
interpolating between the confining gravity solution (which is just
\backmet) on one side and the deconfined homogeneous gravity solution
(which is the black brane in
\backmet) on the other side.

\fig{The localized black hole. The vertical axis is the radial
coordinate, and the horizontal axis $x$ is one of the spatial
$\IR^p$ coordinates. }{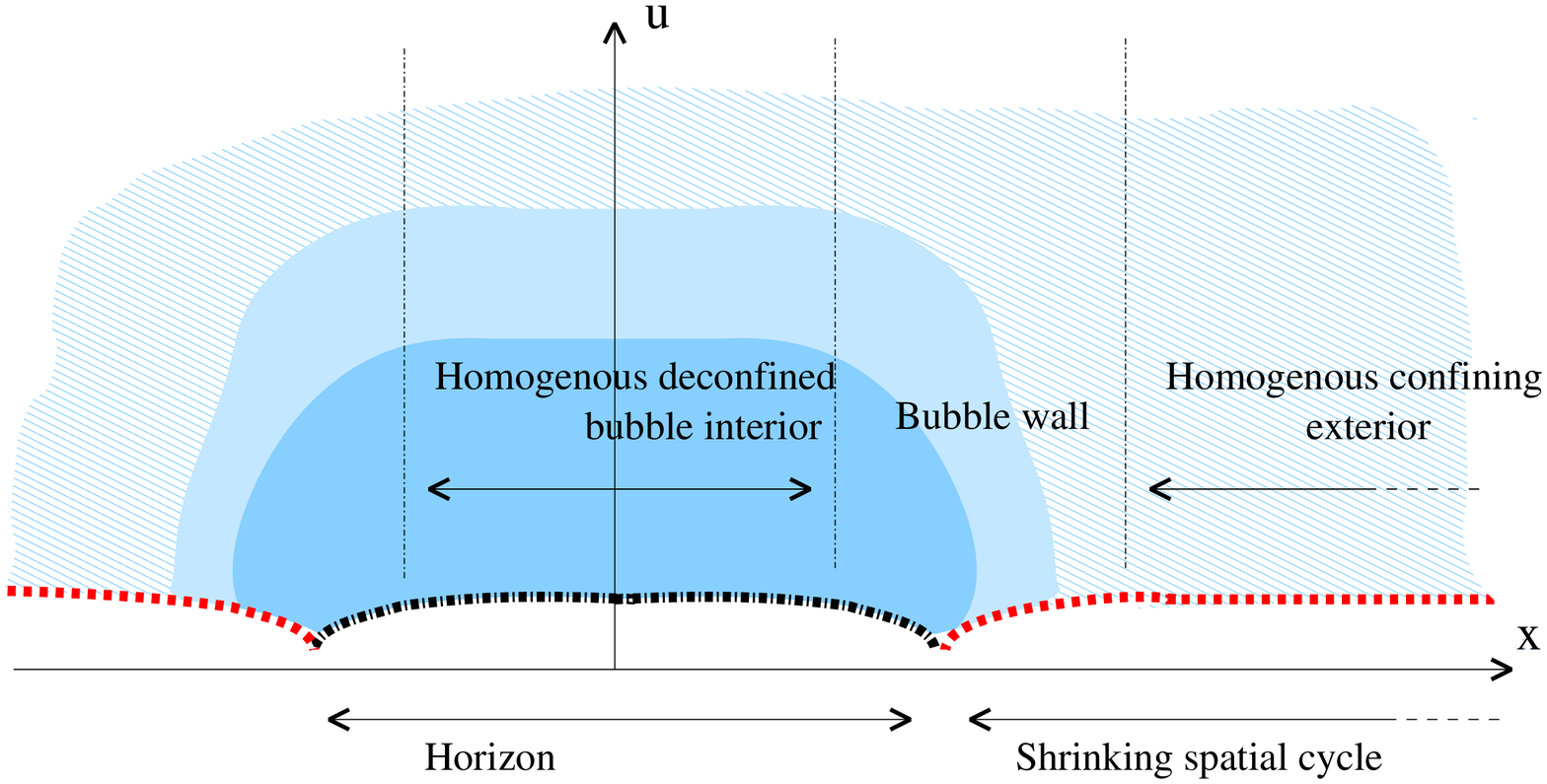}{8truein}
\figlabel{\locbhfig}

Of course, uncharged homogeneous black brane solutions exist over a
range of temperatures. However, the force balance arguments of
\S2, which apply to warped backgrounds whenever they possess a
well defined boundary stress tensor, tell us that such a large bubble
can be static only if the relevant black brane has a pressure that
vanishes in the large size limit. This is true for the black brane
precisely at the deconfinement temperature (see \S2 and appendix
A). Indeed, the force balance equation \fb\ is ensured by the
conservation of the (boundary) stress-energy tensor (see appendix B.1)
which in turn is a direct result of diffeomorphism invariance of the
bulk theory in warped gravitational backgrounds (see, for instance,
\refs{\hollands,\skend} and references therein, and see section 8 and
appendices E and F for how this works in a specific example). As a
consequence, in the classical gravity approximation (i.e. to leading
order in $1/N$), we expect to find large black holes whose interior
locally reduces to a homogeneous black brane at the deconfinement
temperature.  The width of the ``domain wall" separating the two
regions is expected to be of the order of the curvature scale (which
is also the scale of the mass gap in these theories).

Recall that any system undergoing a first order phase transition
goes through configurations that consist of bubbles of varying size
of one phase inside the other. The pancake-like solutions we have
described above are simply increasingly large bubbles of the
deconfined phase within the confined phase, except that the confined
phase is replaced by the vacuum. Within the classical supergravity
approximation the thermal confined phase is indistinguishable from
the vacuum (because its energy is $\CO(1)$ rather than $\CO(N^2)$),
so this last replacement changes nothing. As in section 2, we expect
these big black holes to be stable if their surface tension is
positive. \foot{The domain walls constructed in section 8 do,
indeed, have positive surface tension. We are not, however, aware of
a rigorous argument that guarantees this will always be the case. If
negative surface tension domain walls do exist, we would expect them
to possess Gregory-Laflamme like instabilities localized at the
surface of the black hole, in analogy with the Gubser-Mitra
conjecture \refs{\GregoryVY,\GregoryBJ,\gubsermitra}.}

Motivated by these considerations, we conjecture that warped
backgrounds of the form described in \S1.2 always host large
pancake-shaped black holes of the form described above (and in the
introduction) and depicted in figure \locbhfig. In section 8 we
will supply evidence for our conjecture by constructing the domain
wall that interpolates between the vacuum and the black brane at
the deconfinement temperature, in a particularly simple
background. We also show that in this specific background the surface
tension is positive, so we expect the corresponding black holes to be
stable.

Note that the black holes described in this subsection have a finite
temperature $T_d$ in the large mass limit. Moreover, assuming that
their surface tension is positive, they have negative specific heat
for a rather prosaic reason (see \fb); smaller black holes have to
be hotter in order to balance the surface tension. Note that this
behavior matches smoothly onto the negative specific heat of very small
black holes (which should behave like Schwarzschild black holes, see
\S7.1).

In this section we have focussed on a study of black holes with
vanishing angular momentum (and other conserved charges). However,
it seems likely that reasoning identical to that presented at the
end of section 2 could be used to predict the detailed properties of
rotating black holes and black rings (whose horizon contains an
$S^1$ component, similar to the ones of \EmparanWN), in the
backgrounds studied in this section. We will leave a detailed
analysis of such solutions to future work. Here we will merely note
that the rotating plasma-ring like solutions described in section 2
exist only when $p \geq 2$ for the excellent reason that when $p=1$
the plasma-fluid has no angular direction in which to rotate. As we
have explained, gravitational backgrounds with $p+1$ dimensional
Lorentz invariant sections, of the form considered in this section,
are at least $p+3$ dimensional (the two additional dimensions refer
to the radial $u$ coordinate plus the compact part of the internal
dimensions, which must be at least one dimensional since some
dimensions must shrink in the IR). \foot{For example, in the dual of
a $2+1$ dimensional confining theory which we will discuss in detail
in \S8, the ring would be a solution of a $4+1$ dimensional
gravitational theory (with negative cosmological constant), with
horizon topology $S^1\times S^2$. } It follows that an analysis of
the sort presented in section 2 predicts black ring solutions only
in $4+1$ and higher dimensions. This agrees nicely with the fact
that (at least in flat space) black rings do not exist in $3+1$
dimensions, and it predicts that (at least in some gravitational
theories) black rings should exist above $4+1$ dimensions.

Similarly, we expect to be able to generalize our solutions by adding
charge (global charge in the gauge theory, which maps to a local
charge in the gravitational dual). Again we will leave such
generalizations to future work. We also expect that there may be
phenomenological applications for these localized solutions
\Creminelli.

\newsec{Localized Black Holes as Plasma-balls}

We now turn to a dual gravitational description of plasma-ball
dynamics in bulk duals of confining large $N$ gauge theories at large
$\lambda$. As mentioned in the introduction, there exist several
examples of confining gauge theories that have purely gravitational
dual descriptions in the large $\lambda$ limit. Examples include (a
specific UV completion of) $4+1$ dimensional super-Yang-Mills theory
compactified on a circle with anti-periodic boundary conditions for
the fermions \wit, ``little string theories'' (coming from type IIB
NS5-branes) compactified on an $S^2$ \malnun, and cascading gauge
theories \refs{\kletsy, \klestr}. The geometries dual to these systems
differ qualitatively in several respects\foot{This is natural, as the
UV dynamics of the various systems differs qualitatively.}; however
they all share the general properties described in
\S1.2. According to the conjecture of the previous section,
these backgrounds all possess localized black hole solutions, that
closely resemble the deconfined black brane at their center.

A standard entry in the AdS/CFT dictionary identifies the
translationally-invariant black brane at any given temperature with the
deconfined phase of the dual gauge theory, at the same temperature \wit.
It follows that the localized black holes referred to above are dual
to localized lumps of plasma at approximately the deconfinement
temperature, which are plasma-balls.

In order to build a more detailed dictionary between bulk and
boundary degrees of freedom, it is useful to contemplate a Gedanken
experiment. Consider a black hole, of the form described in section
4, localized near the origin of $p$ dimensional space. Imagine
scattering a graviton in some mass eigenstate (about the original
confining background) off the black hole at a given $p$ dimensional
momentum. The wave-function that describes this scattering process
has four components; an incident wave, a reflected wave, a
transmitted wave (the part of the wave that
stays completely outside the horizon by going around the black hole,
including a part going `above' it in the radial direction), and a component 
of the wave
absorbed into the black hole. Each of these components has a natural
dual interpretation; a glueball incident on a plasma-ball may be
reflected from the ball, transmitted through the ball or may be
absorbed into the plasma-ball (thereby dissociating into gluons). We
thus conclude that gravitons outside the horizon `above' the black brane
region of the localized black hole
map to glueballs inside the plasma-ball \foot{Quantum mechanically
the momentum of such a graviton fluctuates. The lowest order
diagrams that contribute to such fluctuations are of order $g_{s}$,
leading to a fluctuation probability of order $g_{s}^2 \sim 1/N^2$,
in agreement with the analysis of \S3.4.} \foot{Gravitons located
far in the UV correspond to glueballs that are much smaller than
$\Lambda_{gap}^{-1}$. Of course, such special glueball
configurations have the propensity to expand, which is dual to the propensity
of a graviton at large $u$ to fall into the horizon.} . On the other
hand, anything near the horizon (or any possible degrees of freedom
behind the horizon) maps to gluon or deconfined degrees of freedom.

As we have already argued in \S3, lumps of the deconfined
phase have zero overlap with glueballs only at
infinite $N$. At finite $N$ lumps of gluon plasma overlap with
glueballs; this mixing explains the decay of plasma-balls by
hadronization. These statements have a direct gravitational
analogue; black holes at finite temperature mix with the gravitons
at finite $g_s$, a phenomenon that explains the decay of
non-extremal black holes by Hawking radiation.

\newsec{Plasma-Ball Dynamics from Black Holes}

\subsec{Plasma-Ball Hadronization as Hawking Radiation}

We have argued for the existence of stable pancake-like localized
black hole solutions in backgrounds of the form \backmet; we have
also argued that these black holes are dual to plasma-balls. Of
course, these black holes are stable only classically; quantum
mechanically they decay via Hawking radiation.

Hawking radiation from the `flat' surface of the pancake (see figure
\locbhfig), which goes out in the radial direction,
is simply reflected back into the black hole (by the
effective potential arising from the warp factor in the geometry of
\backmet, or, for special modes, by reflecting boundary conditions
in the large $u$ boundary of \backmet). This fact has a simple
interpretation in the dual gauge theory; as discussed in \S3,
glueballs created by
collisions between gluons in the bulk of the plasma-ball dissociate
before they are able to escape out of the plasma-ball. The black
hole loses energy only from Hawking radiation at its edge,
reflecting the fact that only glueballs produced by gluon-gluon
collisions near the surface of the plasma-ball can escape.

From a gauge theory point of view, the decay of a plasma-ball is a
non-perturbative process that is difficult to study even numerically
on the lattice. At large $\lambda$ the AdS/CFT correspondence maps
this hadronization to the quantitatively well understood process of
Hawking radiation, permitting a quantitative analysis of decay rates
and branching ratios in any model of interest. In particular, this
confirms the thermal nature of the decay of the plasma-ball,
at least at large $\lambda$.

In the field theory analysis adding a finite number of quark flavors
does not change the analysis, and the same is true also in the string
theory duals. Adding flavors in the duals is done by adding
D-branes, and in the 't Hooft large $N$ limit with a finite number
of D-branes, the D-branes have a very small back-reaction and do not
affect any of the qualitative discussion above. Note that the
Hawking radiation in this case is expected to be partly into closed string
modes (glueballs), such as gravitons,
and partly into open string modes (mesons).

\subsec{Plasma-Ball Production in Hadron-Hadron Collisions}

Consider the collision of two stable glueballs (with masses of order
$\Lambda_{gap}$) at center of mass energies large compared to $N^2
\Lambda_{gap}$. In the large $\lambda$ string theory dual these glueballs map
to light modes such as gravitons (or other light fields). Giddings
\giddings\ has emphasized that, at large $\lambda$,  black hole
production saturates the inclusive cross section for graviton
collisions at high enough energies (up to numbers of order unity);
furthermore, this cross section also saturates the Froissart
unitarity bound. We pause to review this argument (closely related
to an argument given by Heisenberg in 1952 \refs{\heisenberg, \nastase}
using pions instead of gravitons).  An upper bound for the inclusive scattering
cross section is easily estimated by assuming that gravitational forces
dominate at high enough energies.
Then, the force between two colliding particles separated by
impact parameter $b$ may be estimated to be of order $ {G_4 E^2
\over b^2} e^{-\Lambda_{gap} b}$, where we have used the fact that
gravity is massive in the relevant backgrounds. Consequently,
incident particles simply sail past each other when $b$ is larger
than a number of order $\ln(E) / \Lambda_{gap}$ (up to corrections
that are subleading at large energies), and the inclusive cross
section is bounded from above by \eqn\gravcros{\sigma \sim \ln^2(E)
/ \Lambda_{gap}^2.} However, one may independently use shock wave
metrics and singularity theorems to demonstrate that the cross
section for black hole formation is of the order of $\sigma$ in \gravcros\
\refs{\giddings, \nastase}.

As we have explained above, localized black holes map to localized
lumps of gluon plasma in the dual field theory. It follows as a
prediction of the AdS/CFT correspondence that, at large $N$ and
large $\lambda$, two glueballs shot at each other with an impact
parameter smaller than $({\ln(E)/ \Lambda_{gap}})$ will coalesce
into a lump of gluon plasma with a probability close to one when $E
\gg N^2 \Lambda_{gap}$. This lump of plasma, produced at a rate that saturates
the Froissart bound, will typically settle into a very long lived
plasma-ball, which proceeds to slowly hadronize over a time scale of
order $N^2$, in the manner described in the previous subsection.
Note that even though the black hole forms with a size of order
$\ln(E)$ in the spatial directions of the field theory, it quickly
expands to a size of order $E^{1/p}$ (when the dual field theory has
$p$ spatial dimensions) for which it can be meta-stable.

The reader may find the picture sketched in the last paragraph
clashing with her QCD-trained intuition, which might lead her to
expect the fast partonic constituents of the glueball to either pass
right through each other or to undergo a small number of hard
collisions rather than smoothly coalescing into a plasma-ball. Of
course the utility of partonic ideas is questionable at large
$\lambda$ (where partons interact strongly at all energies).
Nevertheless, to the extent that this notion is valid,
rapid gluon radiation ensures that the parton distribution function is
peaked at small values of $x$ \deepin\ so that the average parton
energy is of order \eqn\parten{E_{parton} \approx E ({\Lambda_{gap}
\over E})^{\lambda} } where $E$ is the center of mass energy of the
collision, ensuring that $E_{parton} \ll \Lambda_{gap}$ at large
$\lambda$, so that
glueballs simply do not contain fast partons. Thus, at large $N$ and
large $\lambda$, glueball-glueball collisions are conceptually
similar to heavy ion collisions, with the large center of mass
energy shared between a large number of constituents.

At small $\lambda$, as mentioned above, most of the energy of the
glueballs is carried by a small number of partons, each of which
carries a significant fraction (an energy of order $E/\ln(E)$)
of the energy of the glueball (at
high energies the partons tend to have smaller values of $x$ due to
asymptotic freedom, but this is a logarithmic effect that does not
affect our arguments). These very energetic partons interact weakly;
their interactions will not form a plasma. As a consequence we
expect a crossover in the dominant behavior of high energy
scattering at $\lambda$ of order one. \foot{Note that even at
small $\lambda$ the glueball still contains a large number of small
$x$ partons whose strong interactions might create a lump of gluon
plasma. However, this lump will not carry a finite fraction of the
center of mass collision energy. One might think that
even if only a small plasma-ball is formed, the fast partons would
be bound to it by a string (since they carry a color charge) so they
would eventually be pulled back into it, given that strings cannot
break in the large $N$ limit. However, since in order to form the
plasma we need an energy at least of order $N^2$, the fast partons
will also carry an energy of order $N^2$, allowing the relevant
strings to snap even in the large $N$ limit (since their length
would be of order $N^2$).}

It would be interesting (and may be possible) to quantitatively
verify the qualitative picture sketched in this subsection.

\newsec{Black Hole Physics from Gluon Plasmas}

As we have argued above, in confining gauge theories at large
$\lambda$, black holes in the IR of the background \backmet\ are
dual to plasma-balls. As discussed below, the decay process of these
plasma-balls goes through ten dimensional Schwarzschild black holes.
To our knowledge this is the first proposal for the dual of
classically stable small black holes (black holes whose size is
small compared to the curvature scale of the space that
hosts them). As a consequence, we are able to use the setup of this
paper to inquire how various mysterious phenomena involving black
holes (for instance, the complete absorption of any incident object)
manifest themselves from the dual viewpoint. The dual picture also
allows us, in principle, to investigate how $\alpha'$ and
$g_{s}$ corrections modify the classical properties of black
holes. In this section we will present a very preliminary discussion
of these extremely interesting questions.

\subsec{The Fate of Small Schwarzschild Black Holes}

As the pancake-shaped black holes (at large $\lambda$) lose energy
into radiation they shrink in size until their size approaches the
curvature length scale of the space \backmet\ that hosts them; they
then localize on the internal manifold (via a Gregory-Laflamme
\refs{\GregoryVY,\GregoryBJ} type transition)\foot{Note that such a
transition is not expected to occur at small $\lambda$, and this may
cause some differences between the details of the decay process at
small $\lambda$ and at large $\lambda$. Note also that this
localization transition may not be smooth, so part of the energy of
the black hole may be lost in the transition, but we expect the
end-point to be a localized black hole carrying a finite fraction of
the initial energy. See \HubenyXN\ for an initial study of the
localization transition on $AdS_5\times S^5$. Note that there is
still some question whether the Gregory-Laflamme instability leads,
in finite time, to an end state with localized horizon
\HorowitzMaeda. See \glg\ and references therein for a study of the
Gregory-Laflamme transition from a dual gauge viewpoint, in a
similar context.}, and then shrink further until their size is much
smaller than the length scale of the background curvature. At this
point the black holes resemble ten dimensional Schwarzschild black
holes\foot{Note that this is true for all known theories with large
$\lambda$, despite the very different UV physics of these theories.
It would be interesting to understand this universality better from
the field theory point of view.}. Thus, the subsequent evolution of
these black holes is identical to that of black holes in flat ten
dimensional space. As these black holes continue to lose energy by
Hawking radiation, they eventually shrink to the string scale; the
qualitative nature of their subsequent evolution is of great
interest (see appendix D for a discussion).

Of course the slow decay of warped black holes, described in the
paragraph above, has a dual description as the decay of
plasma-balls (by hadronization) in gauge theories at large
$\lambda$. The properties of small (string sized) black holes map to
the properties of small plasma-balls, a reformulation that may be
useful.

\subsec{Information Conservation in Hawking Radiation}

It has often been pointed out that the AdS/CFT correspondence ensures that
black hole evaporation is a unitary process. Our identification of
specific gauge theory configurations (occurring in the decay process
of plasma-balls) with `small' ten dimensional Schwarzschild black
holes makes this argument more specific. The production of a plasma
ball in a glueball-glueball collision, and its subsequent decay via
hadronization is clearly a unitary process; the end point of this
process is only approximately thermal.

Of course  a full resolution of the information paradox requires the
identification of the flaw in Hawking's argument that predicts a
breakdown of unitarity. While the dual description of Hawking
radiation is manifestly unitary, it has not yet proved possible  to
formulate Hawking's argument in gauge theory language in order to
identify its flaw. This is an important problem that deserves
attention.

\subsec{How Black is a Black Hole ? \foot{The results of this
subsection were obtained in collaboration with Nima Arkani-Hamed.}}

It is a striking feature of classical black holes in general
relativity (the feature responsible for their name) that a particle
squarely incident on the black hole is always absorbed, no matter
how large its energy.

In this paper we have identified a black hole with a plasma-ball,
a localized lump of gluon plasma at temperature $T_d$. It might,
at first, seem that a particle incident on the plasma-ball at an
energy $E$ much larger than $T_d$ would sail right through it
(undergoing a series of grazing collisions that hardly affect it),
in blatant contradiction with the expectations of black hole
physics. We have already explained in subsection 6.2 that this
naive expectation is incorrect. The only objects available to be
hurled at the plasma-ball are glueballs. At large $\lambda$
glueballs may be thought of as consisting of a very large number
of low energy partons; the average parton energy is $E_{parton}$
in
\parten. Note that $E_{part} \ll \Lambda_{gap}$ even in the limit
$E \to \infty$.  As a consequence, glueballs may
always be absorbed by plasma-balls, no matter how high the incidence energy.

As we have explained in subsection 6.2, however, glueballs at
small $\lambda$ are composed dominantly of a small number of high
energy partons. We expect that these partons should simply blast through the
plasma ball at high enough $E$ (note that
$E$ may have to be of order $N$ in order to permit
these partons to snap confining strings that may bind them to the
plasma-ball and to escape all the way to infinity, since the snapping
probability per unit time and unit length of the string is of order
$1/N^2$). Thus,
plasma-balls at small $\lambda$ (where string corrections are very
important) do not share all the properties of their large
$\lambda$ cousins (in particular they no longer efficiently absorb
glueballs).

It is tempting to translate the last paragraph into a statement
about a reduction in the absorptive properties of black holes in
warped backgrounds whose curvatures are of the order of the string scale. Of
course, this translation has an obvious pitfall; a glueball blasting
through a plasma-ball may have a more mundane dual description, as a
graviton that evades the corresponding black hole by going around it in
the radial direction\foot{This was discussed in a somewhat different
context in \GiddingsAY.} (needless to say, all these notions are a little fuzzy
when curvatures are stringy). Nonetheless, this suggestion is
intriguing and deserves further investigation.

\newsec{Numerical Solutions for Domain Walls in some Specific
Backgrounds}

\subsec{Scherk-Schwarz Compactifications of Anti-De Sitter Space}

In this section we turn to a detailed study of the specific warped background
\eqn\AdSSchIM{ ds^2 = L^2 \apm \left( e^{2 u} \left( -dt^2 + T_{2
\pi}(u) d\theta^2 + dw_i^2 \right) + \frac{1}{T_{2 \pi}(u)} du^2
\right),}
where $i= 1,\cdots,d-1$, $\theta\equiv \theta+2\pi$ and
\eqn\AdSSchII{ T_{x}(u) = 1 - \left(  \frac{x}{4 \pi} \, (d+1) \,
e^u \right)^{-(d+1)}. } This metric, known as the AdS soliton \HorowitzMyers,
is a solution to the $d+2$-dimensional Einstein equations with a
cosmological constant
\eqn\einst{R_{\mu \nu}= -{{d+1} \over {L^2
\alpha'}} g_{\mu \nu},}
and has a simple physical interpretation
\wit. It may roughly be thought of as a Scherk-Schwarz
compactification of $AdS_{d+2}$ on a circle. Indeed, at large $u$,
$T_x(u)\simeq 1$ and \AdSSchIM\ reduces to $AdS_{d+2}$ in Poincar\'e-patch
coordinates, with $u$ as the radial scale coordinate, and with one of the
spatial boundary coordinates, $\theta$, compactified on a circle
(the remaining boundary coordinates, $w_i$ and $t$, remain
non-compact). At smaller values of $u$, \AdSSchIM\ deviates from the
metric of periodically identified $AdS_{d+2}$. In particular the
$\theta$ circle shrinks to zero at a finite value of $u$, smoothly
cutting off the IR region of $AdS_{d+2}$.

In order to describe thermal physics about \AdSSchIM\ it will be
convenient to switch to Euclidean space. Compactifying time $\tau
\equiv \tau + \beta$ on the Euclidean continuation of \AdSSchIM\
\eqn\AdSSchI{ ds^2 = L^2 \apm \left( e^{2 u} \left( d\tau^2 + T_{2
\pi}(u) d\theta^2 + dw_i^2 \right) + \frac{1}{T_{2 \pi}(u)} du^2
\right),}
we obtain the Euclidean configuration for a thermal gas of
gravitons (at temperature $T=1/ \beta$) about \AdSSchIM.  The background
\eqn\AdSSchIb{ ds^2 = L^2 \apm \left( e^{2 u} \left( T_{\beta}(u)
d\tau^2 + d\theta^2 + dw_i^2 \right) + \frac{1}{T_{\beta}(u)} du^2
\right)} yields a second smooth Euclidean manifold with the same
asymptotics; upon continuing to Lorentzian space this solution has a
horizon, so we identify it with the black brane at temperature $T=1/
\beta$ about \AdSSchIM.

It is immediately evident that for $\beta = 2 \pi$ the solutions \AdSSchI\
and \AdSSchIb\ (the thermal gas and the black brane) are identical (they
differ only by a labeling of circles) and so have the same free
energy. In appendix E we compute the boundary stress tensor and free
energy for these two solutions at every temperature, and demonstrate
that the thermal gas has lower free energy for $\beta>2 \pi$, while
the black brane solution dominates the thermodynamics at all higher
temperatures. We also explicitly verify that the pressure of the
black brane vanishes at the phase transition temperature
$T_d=1/2\pi$, as anticipated by the general arguments of section 4.

Fluctuations around \AdSSchIM\ have a mass gap, and
the phase transition temperature is of the same order as
the mass gap (both are of order one in our conventions).
It is also easy to check that the thickness of the
brane at the phase transition temperature (the minimal value of $u$
in \AdSSchIb) and the confinement scale of massive particles (the
exponential decay in the radial direction
in solutions of $\partial ^2 \phi =0$ in the
background \AdSSchIM) are both of order one.

Therefore, the background
\AdSSchIM\ is a particularly simple warped background of the form
described in subsection 1.2; its simplicity stems partly from the
fact that it is a solution to Einstein's equations with a
cosmological constant but no additional matter fields. Moreover, the
background
\AdSSchIM\ also appears as a component of string compactifications
that are dual to interesting field theories, as we now review.

Some backgrounds of the form \AdSSchIM$\times M$ are solutions of string
theory (with appropriate fluxes) that are holographically dual to
$d+1$ dimensional conformal field theories with one spatial
dimension compactified  on a circle with anti-periodic
(Scherk-Schwarz) boundary conditions for the fermions. When $d=3$
the field theory dual to the simplest such compactification
(that with $M=S^5$; different choices of $M$
lead to different field theories) is $\CN=4$ supersymmetric Yang
Mills theory compactified on a Scherk-Schwarz spatial circle \wit.
At low energies (and for small $\lambda$) this theory reduces to
pure Yang Mills theory in $2+1$ dimensions, a confining gauge
theory, making contact with the gauge theory discussions in this
paper. When $d=4$ the field theories dual to such
backgrounds are five dimensional superconformal theories
compactified on a circle with anti-periodic boundary conditions for
the fermionic fields. Such theories are effectively `confining' in
four dimensions, and in some cases the corresponding five dimensional
conformal field theories may be viewed as the
strong coupling limit of a five dimensional gauge theory \FerraraGV.

For all these reasons, in the next subsection we will turn to a
detailed study of the background  \AdSSchIM.  In particular we
obtain a numerical construction of the domain wall that interpolates
between \AdSSchI\ and \AdSSchIb\ at $\beta=2\pi$.

\subsec{The Domain Wall: Asymptotic Behaviors}

As discussed in \S4, the black hole solutions we are interested in
should interpolate between the (Lorentzian versions of the)
confining vacuum \AdSSchI\ on their
outside and the black brane solution \AdSSchIb\ at the
deconfinement temperature on the inside.  In the limit that these
black hole solutions are very large, the solution in the
neighborhood of their edge should reduce to a domain wall (in one
of the field theory directions) interpolating between these two
solutions.

Since the metrics must depend explicitly on two coordinates (the bulk
radial coordinate and the field theory direction normal to the wall),
it is unclear whether any analytic solutions may be found. The elegant
Weyl metric, whilst extended to higher dimension, has not been
extended to include a cosmological constant
\refs{\weyl,\reall,\CharGregory}. Hence we will use numerical methods
to construct the metric, following those employed previously in
\refs{\wisemanA,\wisemanB,\kudoh} (see also
\refs{\kleihaus}). Note that whilst it seems possible that the domain
wall system we solve here may one day find an analytic solution,
similar equations with more bulk matter, such as that found in general
confining dual backgrounds \refs{\polstr, \klestr, \malnun}, would
likely not be solvable, whereas the numerical methods we set-up
here should easily extend to these cases. Furthermore, the finite
size black hole problem is likely to remain analytically
intractable.\foot{There has been interesting analytic progress in
lower dimensional ($d=2$) cases with just a negative cosmological
constant and the bulk ending at a thin brane
\refs{\horowitz,\zakout}. This progress follows from the AdS
C-metric \refs{\plebanski}, which unfortunately is not known in more
than 4 dimensions. We also emphasize that everything we say applies
only to warped backgrounds with a mass gap (and with no four dimensional
massless gravity). In particular the analysis
of this paper may not apply to `cut off' AdS geometries with a
massless `radion' field. } We focus our calculation on the domain wall
solution rather than on finite size solutions simply because the
equations and solution are most elegant in this case. Moving to finite
size is in principle a straight forward extension of the methods we
use, which we leave for future work.\foot{This will introduce new
terms in the Einstein equations as planar symmetry in the field theory
directions is replaced by spherical symmetry, and similarly the
boundary conditions will then include the origin of these
spherical-polar coordinates. However the same methods will apply, and
in particular the metric will still depend non-trivially on only two
coordinates.}

In this subsection we give an overview of our numerical construction
of this domain wall. The formal construction, detailing the equations
to be solved, the boundary conditions and the data, is presented in appendix
F. Numerical details are given in appendix G. We are mostly interested
in the cases of $d=3,4$ but we present our formulae for arbitrary
dimension. We will consider the problem in the Euclidean setting,
although we stress that the equations, and their boundary conditions,
are totally independent of the spacetime signature. The static
Lorentzian solution is obtained by trivial continuation of the
Euclidean one.

\fig{Domain of problem in $x,y$
plane.}{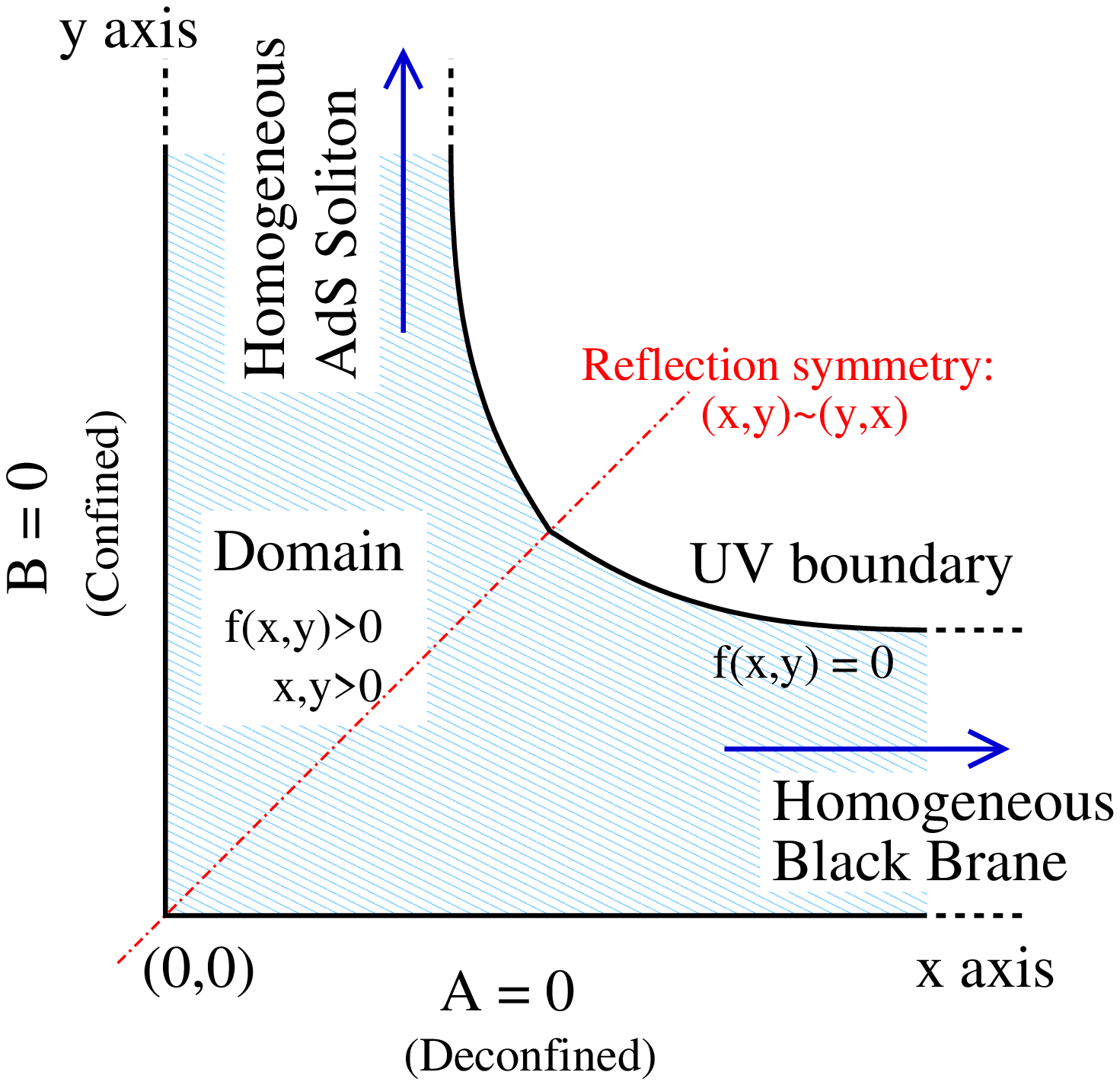}{5truein} \figlabel{\domain}

We will search for a solution to \einst\ that preserves rotational
and translational symmetry in $(d-2)$ of the $(d-1)$ spatial $w_i$
directions, which we call $r_a$ ($a = 1,\cdots, (d-2)$), and
translational symmetry in the $\theta$ and $\tau$ directions. The
background is also required to respect reflection symmetry in
$\tau$ and $\theta$. It follows that it is locally possible to
choose the metric for our background to take the form
\eqn\metricA{ ds^2 = A^2 d\tau^2 + B^2 d\theta^2 + e^{2 C} dr_a^2
+ e^{2 D} \left( dx^2 + dy^2 \right), }
where the metric functions depend on the coordinates $x,y$ which
represent some combinations of the field theory direction on which the
domain wall is localized and of the radial coordinate in the bulk
space (we choose specific combinations such that the metric takes the
form \metricA). We take the coordinate periodicity of the $\tau,
\theta$ circles to be $\beta, \tilde{\beta}$ respectively, although we
note that in our asymptotic AdS space-time (or, from the perspective
of the $(d+1)$-dimensional field theory, due to conformal invariance
in the UV) the actual value of $\beta$ or $\tilde{\beta}$ is
irrelevant. It is only the ratio of the proper size of the Euclidean
time circle in the UV to the proper size of the space circle in the UV
that is physical. As discussed above, these are equal exactly at the
deconfinement temperature and we only expect to find a Euclidean (and
hence static Lorentzian) solution in this case.

Note that the ansatz \metricA\ leaves a residual diffeomorphism
freedom of performing conformal coordinate transformations of
the $x, y$ plane which preserve the form of \metricA. Remembering
that (non-singular) two dimensional conformal transformations can
map any region into any other region (preserving angles at
corners), we will use this freedom to position the boundaries of
the domain of our coordinates $x$ and $y$ as in figure \domain. We
will take the IR boundaries to be at $x = 0$ and $y = 0$, and the
UV boundary will be defined by the vanishing of a function
$f(x,y)$. For large $x$ we take the function $f$ to vanish at $y =
c$, for some constant $c$. For large $y$, similarly we take $f$ to
vanish at $x = c$.  Furthermore we will choose $f$ so that the
function is invariant under the $\IZ_2$ symmetry $f(x,y) = f(y,x)$
(which will yield a $\IZ_2$ symmetry of our solution, to be
discussed below). We regard the locus of the zero of $f$, and thus
the particular value chosen for $c$, to be fixed throughout the
following discussion. Note that the particular value of $c$ chosen
is irrelevant, and may be changed simply by performing a global
scaling $(x,y) \rightarrow (\lambda x, \lambda y)$, which rescales
the UV boundary position and the metric functions $A,B,e^C$
together with $\beta,\tilde{\beta}$ but does not physically change
the solution.

The metric may always be written in the form \metricA\ locally. We
assume that there exists a solution for which the metric functions
$A,B,C,D$ in \metricA\ are finite everywhere in the coordinate
domain of the problem, except at the conformal boundary in the UV
where $f$ has a first order zero and $A,B,e^C,e^D \sim 1/f$. It is
certainly not obvious that such a solution exists, but we will
show that it does by constructing it numerically.  Note that once
we have made the choice of coordinate boundaries above, we have
completely fixed the residual conformal coordinate freedom (up to
pathological transformations which diverge at large $x, y$ and
thus are projected out under our finite metric component
assumption).

Let us discuss the boundary conditions in more detail. The bulk
space-time has a conformal boundary at $f(x,y) = 0$, and the
region near $f=0$ looks like anti-de Sitter space near its
conformal boundary. Thus the conformal boundary metric is
(conformally) flat. We require the bulk to close smoothly in the
IR, with either the $\tau$ or $\theta$ circles shrinking to zero
size. As indicated in figure \domain\ we choose the time ($\tau$)
circle to shrink for $x>0,y=0$ and the space ($\theta$) circle for
$y>0,x=0$.  At the origin $x = y = 0$, both circles smoothly vanish,
as discussed in more detail below.  At large $x$ and at large $y$
the geometry should tend to one of the homogeneous solutions
\AdSSchI\ and \AdSSchIb\ with the appropriate circle shrinking, as
detailed in the previous subsection.

The residual conformal transformations that preserve the form of
\metricA\ and maintain the finiteness of the metric functions must
preserve angles at corners of boundaries of the domain. Hence we
have made a crucial choice above, namely that the angle between
the $x$ and $y$ axes is a right angle. Let us now justify this
choice.
If we focus on a small enough region near the origin in $x$ and
$y$ we may ignore the bulk cosmological constant. The local
geometry here must be flat space, since we require the Euclidean
geometry to be smooth. Hence, the metric must take the form
\eqn\metricB{ ds^2 = \left[ A^2 d\tilde{\tau}^2 + B^2
d\tilde{\theta}^2  + \left( dA^2 + dB^2 \right) \right] + \left[
e^{2 C} dr_a^2 \right] + O(A^2, B^2) }
to leading order in $A, B$ around $A = B = 0$, where to leading
order $C$ is a finite constant, and $\tilde{\tau}, \tilde{\theta}$
are rescalings of $\tau, \theta$ with coordinate period $2 \pi$,
such that this Euclidean manifold is smooth. This is simply
$(d+2)$-dimensional flat space, written as a product of four
dimensional flat space, in double polar coordinates, and the flat
$(d-2)$-dimensional space parameterized by the $r_a$ directions.
In \metricB\ the lines defined by the shrinking circles, $A = 0$
and $B = 0$, are indeed at right angles, and the metric components
are finite, justifying our choice for the IR boundary.\foot{
Define $z = A + i B$, and consider a singular conformal
transformation that takes the form $w = z^p$ near $z = 0$, with $w
= \tilde{A} + i \tilde{B}$. Such a transformation with $p \ne 1$
will change the angle between the two axes. The new metric will
contain the terms $|w|^{p-1} ( d\tilde{A}^2 + d\tilde{B}^2 )$ from
the Jacobian factor of the transformation, so $e^{2 \tilde{D}} =
|w|^{p-1}$. For $p \ne 1$ this means $|\tilde{D}|$ is infinite at
$z = w = 0$.  Hence if we require a metric of the form \metricA\
with all the metric functions being finite, we {\it must} have a
right-angle between the axes defined by the shrinking of the
Euclidean time circle, and that of the space circle.  }

The analytic details of the construction are found in appendix F.
There we find that the boundary conditions have no remaining data once
we have specified our IR and UV boundary locations (and fixed the
trivial rescalings of the $\tau, \theta, r_a$ coordinates). The
equations and boundary conditions in the IR and UV (having chosen
appropriately the UV boundary location as described above) are
invariant under the $\IZ_2$ symmetry $(x,y) \leftrightarrow (y,x)$
providing we interchange $A$ and $B$, and the asymptotic solutions
\AdSSchI\ and \AdSSchIb\ we expect to find at large $x$ and at
large $y$ are also exchanged by the symmetry.  Hence we expect
that the solution for the domain wall will possess this $\IZ_2$
symmetry so that
\eqn\sym{ \eqalign{ A(x,y) & = B(y,x), \cr C(x,y) & = C(y,x), \cr
D(x,y) & = D(y,x). } }
For a smooth Euclidean geometry this implies $\beta = \tilde{\beta}$,
so the time and spatial circles have precisely the same proper size in
the UV, and so the Euclidean temperature is indeed exactly at the
deconfinement temperature, as discussed in the previous
subsection. Alternatively, in the Lorentzian setting, the surface
gravity of the horizon, as measured by an IR observer far from the
domain wall in the confining region, is directly related to the circle
size in that region.

Having fixed the coordinate boundary locations (and fixed the
trivial rescalings of the $\tau, \theta, r_a$ coordinates), we
obtain a unique solution for the metric functions $A, B, C, D$ in
both $d = 3,4$ with the above symmetry. We may then determine the
periodicity $\beta$ of the two circles from the proper gradient of
the vanishing of the $d\tau^2$ metric component in the asymptotic
black brane region (or equivalently, the vanishing of the
$d\theta^2$ component in the confining region). We see in appendix
F that the Einstein equations indeed guarantee that the circle
coordinate periods required to smoothly close the appropriate
shrinking circle in the homogeneous regions indeed smoothly close
the circles everywhere in the solution, in close analogy with the
zeroth law of black hole mechanics.

\fig{ Surface plot of $f$ in the $x,y$ plane (in fact we plot $(-f)$ for visual
convenience) for $d = 4$. The zero of $f$ gives the position of
the UV boundary. $f$ is not defined over our entire coordinate
domain, but rather in a finite region near the UV
boundary.}{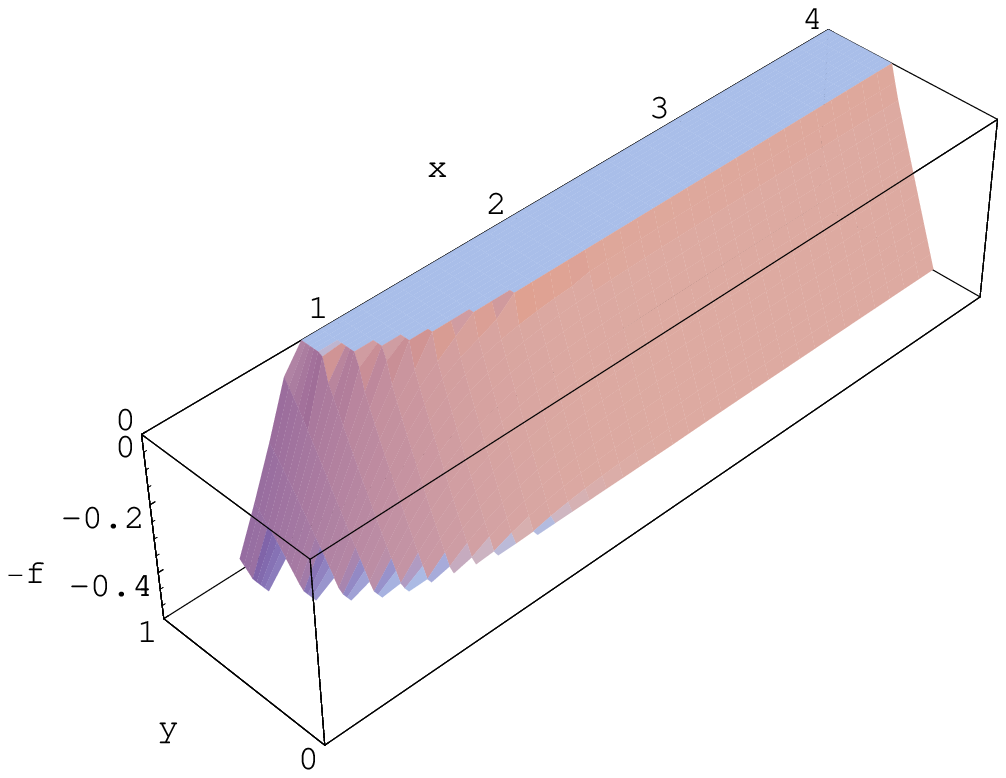}{3.5truein}\figlabel{\ffig}

\quadfig{Surface plots of $A$, $B$, $e^C$, $e^D$ in the $x,y$
plane for $d = 4$. The corresponding plots for $d = 3$ appear very
similar.}{3.5truein}{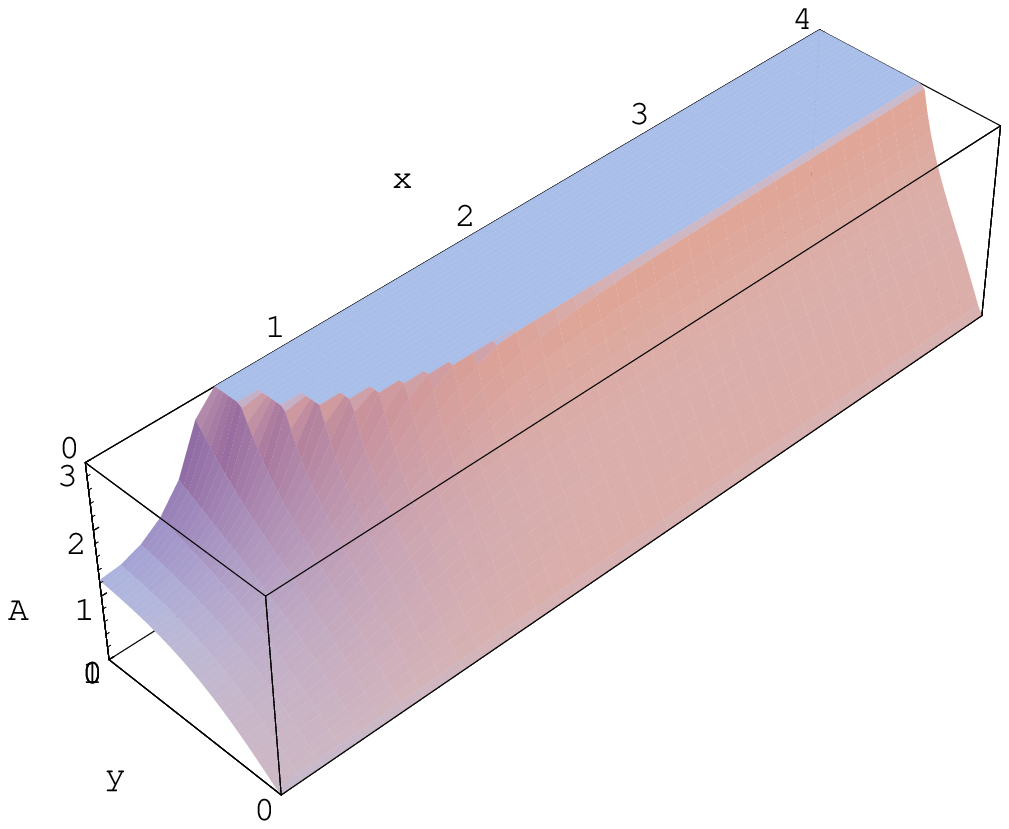}{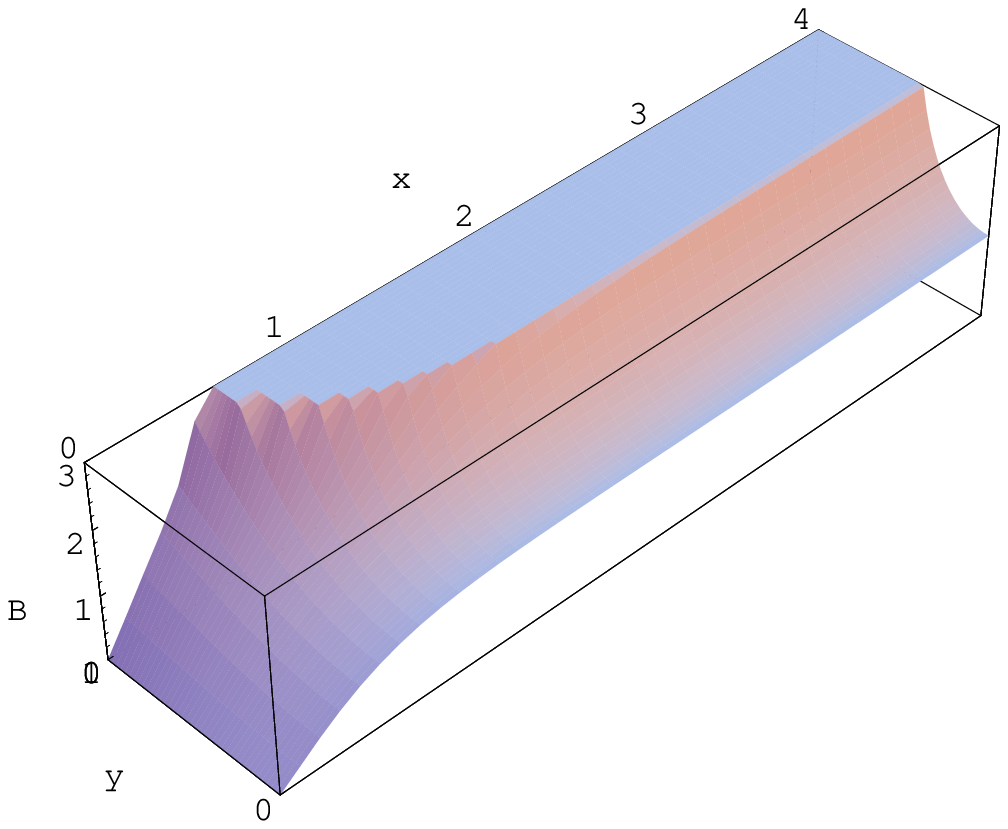}{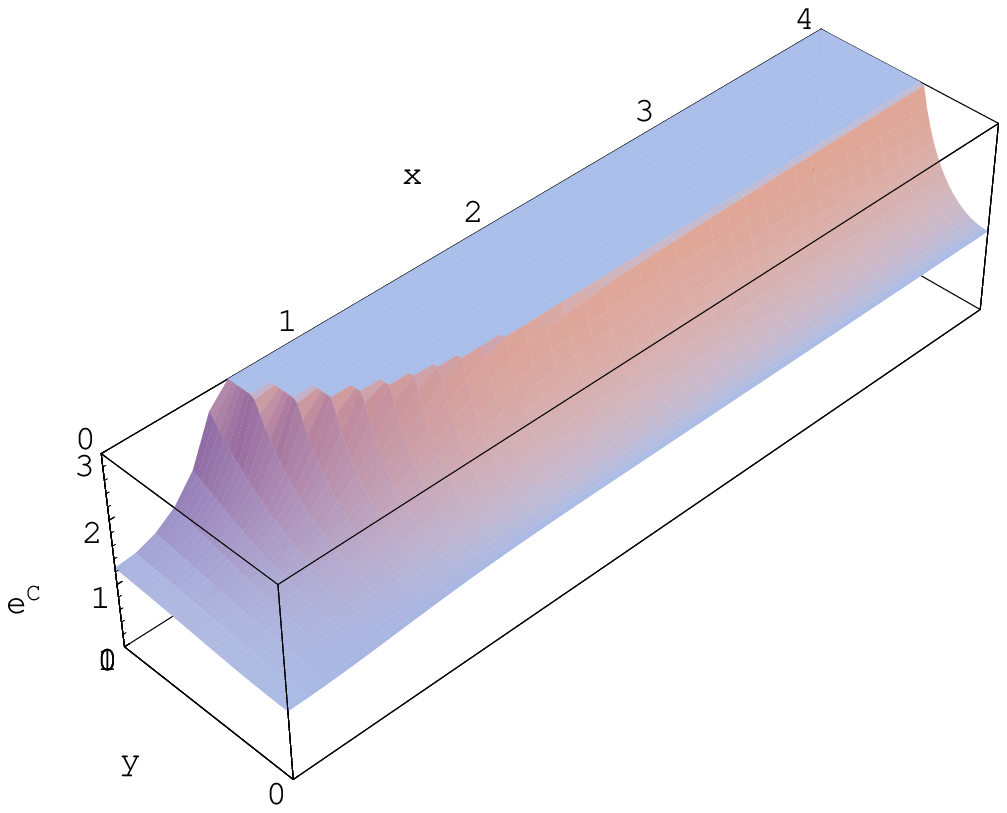}{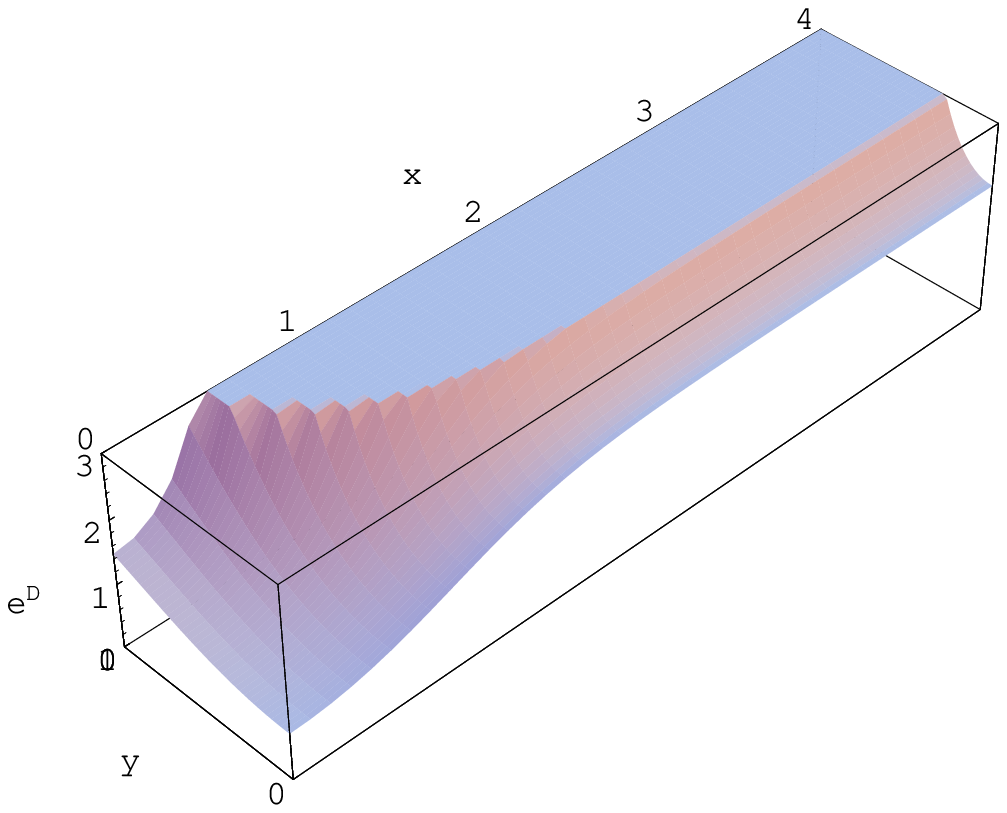}\figlabel{\TUSRfig}

In figures \ffig\ and \TUSRfig\ we present the form of the metric
functions we compute numerically for the case $d=4$. For $d=3$
their qualitative form is very similar. We work in units where
$L^2 \alpha' = 1$. In figure \ffig\ we show the function $f$,
whose zero gives the position of the UV boundary. Note that this
function is not defined over the entire domain, as it is only
required in a neighborhood of the UV boundary. The details
concerning the functional form of $f$ are discussed at length in
appendix F. Then, in figure \TUSRfig\ we show surface plots of the
metric functions $A, B, e^C, e^D$ as functions of $x, y$. We see
that these behave smoothly everywhere, except near the UV boundary
where they diverge as $\sim 1/f$ (where $f$ has a first order
zero). $A$ goes to zero at $y = 0$, and hence the time circle
shrinks there. By symmetry, $B$ goes to zero at $x = 0$ where the
space circle shrinks. We clearly see that at large $x$ (and by
symmetry large $y$) the solution does indeed quickly become
homogeneous as expected.

\twosidefig{The components of the boundary stress tensor (in units
where $T_{\tau\tau} \rightarrow 1$ in the deconfined phase) as a
function of a coordinate $g$ perpendicular to the domain wall
(negative is confined region, positive is deconfined region), for
$d=3$ (left) and $d=4$ (right): red (top) = $T_{\tau\tau}$, orange
(bottom) = $T_{\theta\theta}$, green (top middle) =
$T_{r_{a}r_{a}}$, blue (bottom middle) = $T_{gg}$.}
{3.5truein}{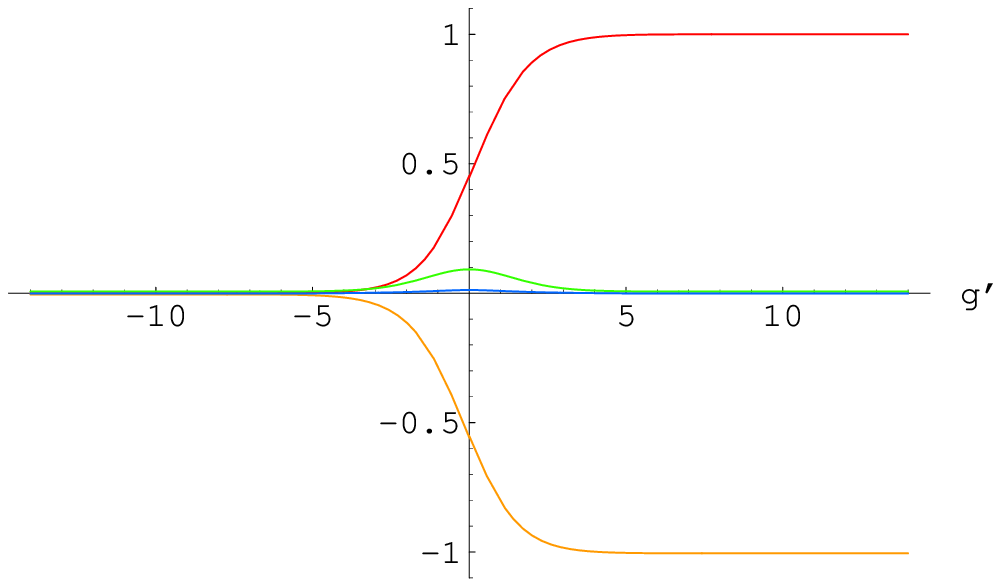}{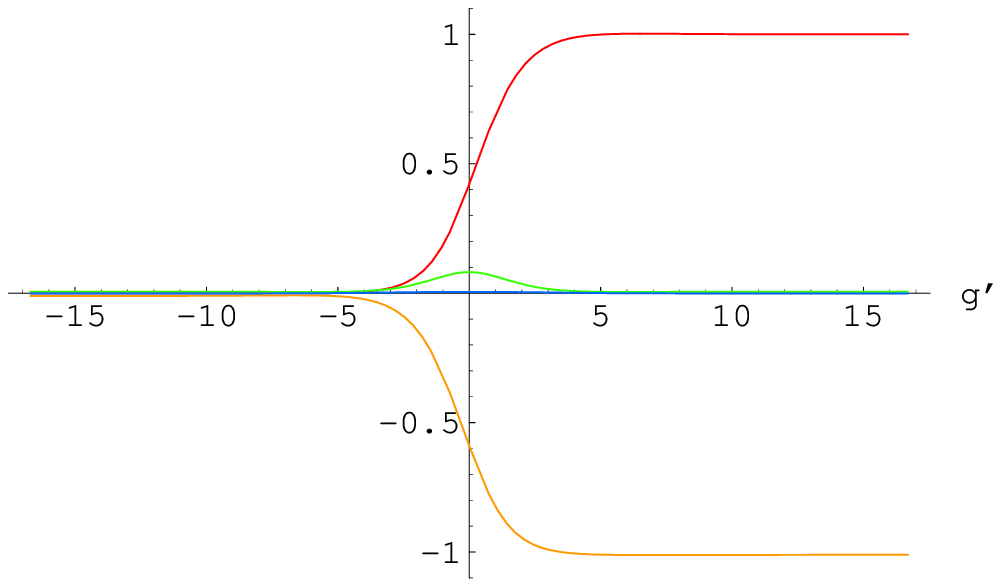}\figlabel{\stressfig}

We compute the boundary stress tensor for both the $d=3$ and $d=4$
solutions in figure \stressfig. The horizontal axis is the proper
distance along the boundary (up to a global scaling explained below),
which we identify with a coordinate $g$ in the field theory
perpendicular to the domain wall.  The vertical axis gives the various
independent components of the traceless stress tensor. A key point is
that $T_{gg}$ is constant (in fact it is zero), since the stress
tensor conservation yields $\partial_g T_{gg} = 0$ (we discuss this in
detail in appendix F).

As discussed in appendix B.1, the surface tension of the domain
wall is given by
\eqn\surften{ \Sigma = \int_{-\infty}^{\infty} d g T_{r_a r_a} }
(no sum over $a$). From our numerical results depicted in figure
\stressfig\ we find that the domain wall tension $\Sigma$, measured in
units (as in the figure) of the energy density $\rho_c$ of the
deconfined phase at the deconfinement temperature divided by the
deconfinement temperature $T_d$ (which was used to set the scale of
the $g$ coordinate in figure \stressfig), is given by $\Sigma = 2.0
\times \rho_c / T_d$ for $d=3$ and by $\Sigma = 1.7 \times \rho_c /
T_d$ for $d=4$. We estimate the systematic error in these numbers to
be no more than ten percent.

\fig{Surface plot of the scalar curvature invariant
$K=(R_{\mu\nu\gamma\rho} R^{\mu\nu\gamma\rho})^{1/4}$ $x,y$ plane
for $d =
4$.}{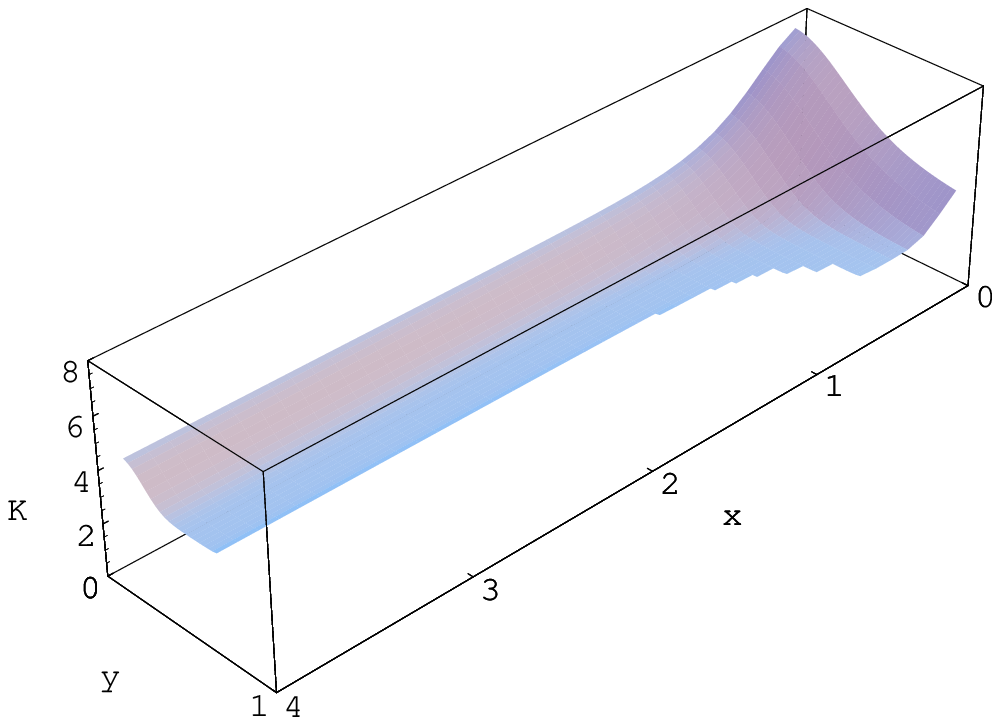}{4.5truein}\figlabel{\kretch}

Finally, in figure \kretch\ we plot the scalar curvature invariant
$(R_{\mu\nu\gamma\rho} R^{\mu\nu\gamma\rho})^{1/4}$ for the $d =
4$ solution. We see that near the UV boundary it
tends to a constant, namely that for AdS. In the homogeneous
regions it is a little larger in the IR, reflecting the fact the
solution is really Euclidean AdS-Schwarzschild, with greater
curvature localized where the circle shrinks. In the domain wall
region, we see the curvature is increased, but it is everywhere
smooth as we expect.

\bigskip
\bigskip
\centerline{\bf Acknowledgements}

We especially thank N. Arkani-Hamed for collaboration on some stages
of this project and for many useful discussions.  We thank T.
Friedman for suggesting the name plasma-balls. We would like to thank
M. Berkooz, R. Bhalerao, M. Douglas, R. Gavai, S. Giddings, R.
Gopakumar, D. Gross, S.  Gupta, G. Horowitz, V.  Hubeny, I.  Klebanov,
B. Kol, H. Kudoh, D. Kutasov, H. Liu, J. Maldacena, J.  Marsano, R.
Myers, H. Nastase, C. Nunez, K. Papadodimas, B.  Pioline,
J. Polchinski, E. Rabinovici, H. Reall, M. Rangamani, S.  Ross,
H. Schnitzer, N. Seiberg, S. Shenker, A. Starinets, A. Strominger,
M. Van Raamsdonk, C. Vafa, S. Wadia, and F. Wilczek for useful
discussions.  We especially thank K. Rajagopal for several very useful
comments and suggestions that have kept us honest. OA would like to
thank Harvard University and the Fields Institute for hospitality. SM
would like to thank the organizers of the Post-Strings workshop at
CERN and the ``QCD and string theory'' workshop in Santa Barbara for a
stimulating environment during the progress of this work. OA and SM
would like to thank the Einstein Symposium at the Library of
Alexandria, the Third Regional Conference on String Theory in Crete
and the Perimeter Institute for hospitality. SM and TW would also like
to thank the organizers of the ``String Theory'' program at the
Benasque Center for Science for a wonderfully relaxed atmosphere and
hospitality while this work was being completed. The work of OA was
supported in part by the Israel-U.S.  Binational Science Foundation,
by the Israel Science Foundation (grant number 1399/04), by the
Braun-Roger-Siegl foundation, by the European network
HPRN-CT-2000-00122, by a grant from the G.I.F., the German-Israeli
Foundation for Scientific Research and Development, and by
Minerva. The work of SM was supported in part by an NSF Career Grant
PHY-0239626, DOE grant DE-FG01-91ER40654, and a Sloan Fellowship. TW
is supported by NSF grant PHY-0244821.

\appendix{A}{Thermodynamics of Large $N$ Gauge Theories}

In this appendix we review the thermodynamic properties of large $N$
$SU(N)$ gauge theories that undergo first or second order
deconfinement phase transitions. We discuss explicitly theories having
only fields in the adjoint representation (such as pure Yang-Mills
theory and its supersymmetric generalizations), but the discussion
does not change if we add a finite number of flavors.

\subsec{Local Stability of Homogeneous Configurations}

We begin by reviewing the thermodynamic criteria for the stability of
a homogeneous system.

Consider any system (like a gauge theory) described by a local field
theory. In the micro-canonical ensemble the system is characterized by
the energy density $\rho \equiv E / V$, and the thermodynamical
behavior is determined by the entropy density $f(\rho)$, $S=V
f(\rho)$.  The effective temperature is given by $T = 1 / (\del S /
\del E) = 1 / f'(\rho)$.  Here we have assumed that the system is
homogeneous, but of course the homogeneous configuration does not
always maximize the entropy.  To see this, divide the system into two
pieces, of volume $\alpha V$ and $(1-\alpha) V$, respectively. Let the
energy in the two parts be $(\alpha+\delta
\alpha)E$ and $(1 - \alpha - \delta \alpha)E$, respectively. $\delta
\alpha=0$ corresponds to the homogeneous system; at small non-zero $\delta
\alpha$ the entropy of this system, above that of the homogeneous
configuration, is given by
\eqn\nonhom{{\delta S \over V} ={\delta \alpha^2 \rho^2
\over 2 \alpha (1-\alpha) } {d^2 f \over d \rho^2 }.}
Note that $\delta S$ has the same sign as ${d^2 f / d \rho^2}$. It
follows that the homogeneous phase is locally unstable when ${d^2 f
/ d \rho^2}$ is positive. Since ${d f / d \rho}$ is the inverse
temperature of the system, ${d^2 f / d \rho^2}$ has an opposite sign
from the specific heat of the system. Consequently, we have merely
rederived the well-known fact that homogeneous systems with negative specific
heat are unstable.

Note also that the pressure of a system is given by $p=- {\del E /
\del V}|_{fixed\ S}$. Using $d S = dV f(\rho) + f'(\rho) (dE-\rho
dV)$, we find that
\eqn\press{p(\rho)=-{f' \rho -f \over f'}= - F_{free}(\rho),}
where $F_{free}$ is the free energy density. The velocity of sound squared
is given by
\eqn\soundspeed{
v_{sound}^2 = {d p \over d \rho}= {-f f'' \over {(f')^{2}}}.}
It follows that
small pressure waves in this homogeneous system are tachyonic (so that the
system is dynamically unstable to small perturbations) if and only
if $f''>0$, reiterating the conclusion of this subsection.

\subsec{Density of States in Confining Gauge Theories}

We now specialize to the study of a confining large $N$ Yang-Mills
theory that undergoes a first order deconfining transition, and review
the behavior of the function $f(\rho)$ in this class of theories
(which may include the models of \refs{\wit, \polstr, \klestr,
\malnun} at various values of $\lambda$). Two interesting mass scales in
such theories are $T_H \sim M_s$, the Hagedorn temperature of the low
energy confining large $N$ gauge theory (related to the confining string
tension $M_s^2$), and the mass gap
$\Lambda_{gap}$. At large $\lambda$, $T_H / \Lambda_{gap}$ is
proportional to a positive power of $\lambda$, while for small
$\lambda$ this ratio is a number of order one.

We first study $f(\rho)$ at energy densities which are ${\cal O}
(N^0)$ (rather than ${\cal O}(N^2)$). When $\rho \ll M_s^4$  the
system is well-approximated by a gas of weakly interacting glueballs,
a system that has positive specific heat\foot{In theories with a
single mass scale, $M_s$ is of order $\Lambda_{gap}$ and this energy
range is pretty boring. However, at large $\lambda$, $M_s$ is
parametrically separated from $\Lambda_{gap}$ and this energy range
is more interesting -- typically the density of low mass glueball
states grows as a power of the mass (due to Kaluza-Klein modes in
extra dimensions), leading to $f(\rho) \propto \rho^{\alpha}$ with
$3/4 < \alpha < 1$.}. When the energy density reaches the string
scale, $\rho \gg T_H^4$, the system is well-described by a gas of
weakly interacting strings at approximately the Hagedorn temperature
$T_H$,  $f(\rho) \propto \rho / T_H$ (this is a consequence of the
exponential or Hagedorn growth in the high-energy density of states
$e^{M/T_H}$). Note that ${d^2f / d \rho^2}$ vanishes, indicating
marginal stability, when $\rho \gg T_H^4$ \foot{More precisely, the
behavior in this regime depends on the sign of the corrections to
the Hagedorn behavior, $f(\rho) \sim \rho / T_H + \alpha \ln(\rho)$,
but in any case the second derivative is very small for large
$\rho$.}.

Next, we turn to energy densities of order $N^2$. We first consider
the very high energy limit. At very high energy densities the system
is governed by its high-energy `fixed point'. This fixed point
could, for instance, be a free theory with a number of particle
species of order $N^2$, or a conformal field theory in four
dimensions. In either of these cases $f(\rho) \propto N^{1/2}
\rho^{3/4}$ for $\rho \gg N^2 \Lambda_{gap}^4$ \foot{Some of the
large $\lambda$ theories that we are interested in actually involve
more exotic UV `fixed points', which have a different form of
$f(\rho)$ but the same qualitative behavior.}. In this regime the
effective temperature grows indefinitely with increasing density and
the system has positive specific heat.

We now turn to intermediate energy densities of order $N^2$. When the
deconfinement transition is of first order, it must occur at a
temperature $T_d$ lower than $T_H$ (otherwise the deconfinement
transition would be a second order transition at $T_H$; clearly it
does not make sense for the confining phase to exist at temperatures
larger than $T_H$). Thus, there must exist an intermediate regime
between the Hagedorn regime and the high-energy-density regime, in
which the effective temperature decreases to a value below $T_d$.
Such a phase obviously has negative specific heat and is unstable
\foot{In many gravitational duals at large $\lambda$, this phase is
dominated by Schwarzschild-like black holes in ten dimensions, for
which $f(\rho) \propto N^{-2/7} \rho^{8/7}$.}. As $\rho$ increases
in this phase, we expect that ${d^2 f / d \rho^2}$ decreases
monotonically, passing through zero at some $\rho=\rho_0$ of order
$N^2 \Lambda_{gap}^4$, and eventually joining with the negative
values of $d^2 f / d\rho^2$ in the high-energy-density phase.

\fig{The (logarithm of) the effective temperature $T = 1 / (\del S / \del E)$
as a function of the (logarithm of) the energy density in large $N$
confining gauge theories which undergo a first order deconfinement
transition.}{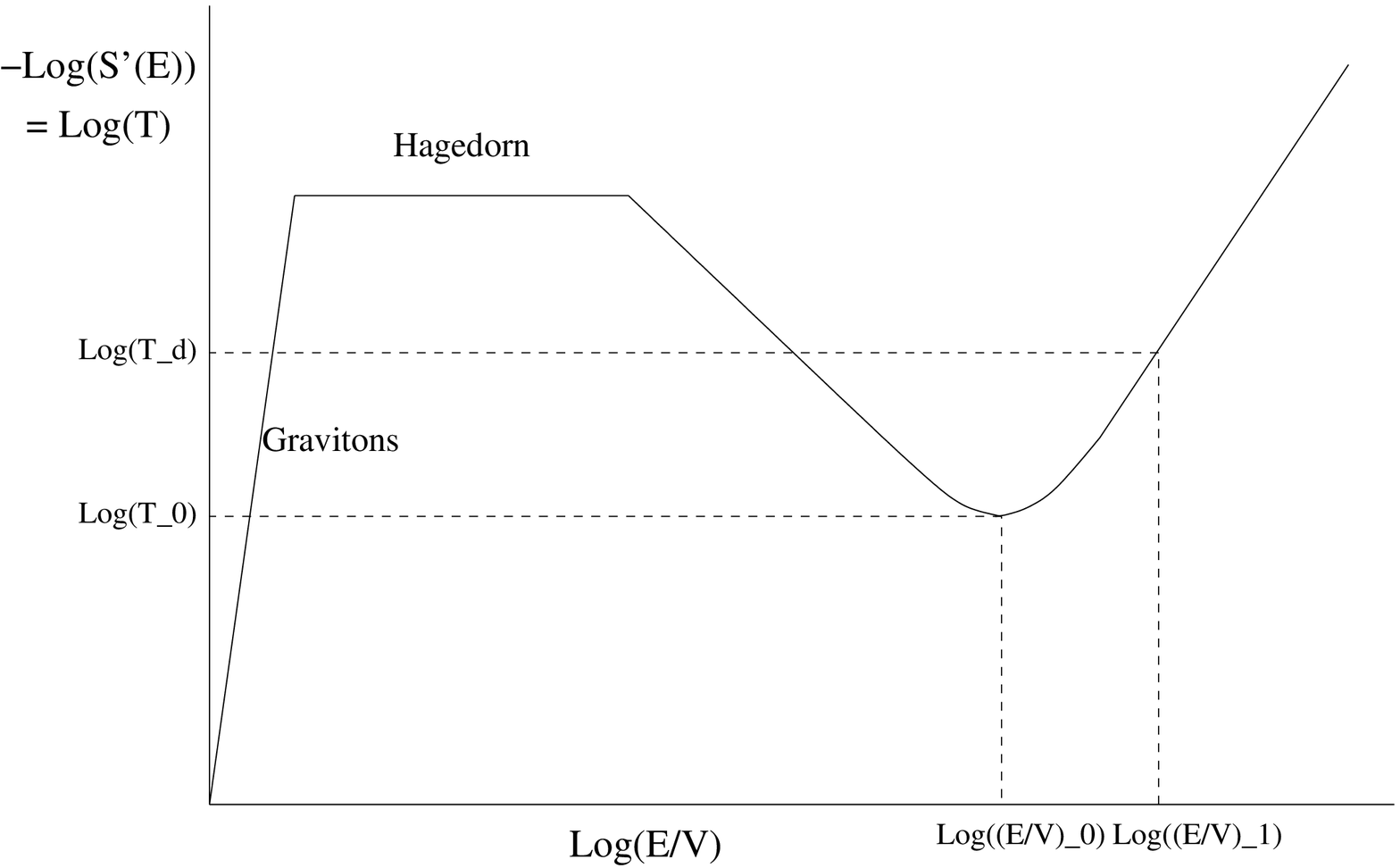}{4truein} \figlabel{\endens}

The behavior described above is depicted in figure \endens, which
plots the logarithm of the effective temperature as a function of the
logarithm of the energy\foot{A similar behavior appears also in the
gravitational dual of $(p+1)$-dimensional conformal field theories
compactified on $S^p$ \refs{\HorowitzBJ,\wit};
see \spenta\ for a recent discussion.}.
Stable homogeneous phases correspond to lines
with a positive slope in this figure.

The generic behavior of systems exhibiting a second order phase
transition is quite similar, except that the unstable
intermediate phase does not exist. In such systems the Hagedorn
phase joins smoothly with the high-energy-density phase and there
is a second order phase transition precisely at the Hagedorn
temperature \refs{\pisarski,\sphere}.

\subsec{Phase Separation in Confining Large $N$ gauge theories}

Consider the behavior of
a large $N$ gauge theory which has a first order deconfinement
transition, at an energy density $\rho$ which is of order $N^2$ but smaller
than $\rho_0$. As discussed above, the homogeneous phase is unstable
in this range of densities, so we expect the system to decay into
separate regions in two different phases, one with a density smaller
than $\rho$ and the other with a density larger than $\rho$. We expect
the two phases to be the stable phases in the discussion above, which
are the low-energy-density confined phase, at $\rho < T_H^4$, and the
high-energy-density deconfined phase, at $\rho > N^2
\Lambda_{gap}^4$. In the first phase the energy density, entropy
density and free energy density are all $\CO(1)$. We will find it
convenient to normalize the extensive quantities (such as $E$ and $S$)
by dividing them by a factor of $N^2$ as well as by the volume. All
normalized extensive quantities vanish in the first phase.

The second phase in this non-homogeneous mixture must have an energy
density larger than $\rho_0$ in order to be stable. We now determine
the properties of this phase. Let a fraction $\alpha$ of the net
volume of our system be occupied by this phase; since the energy
density of the confined phase is negligible, this means that the
energy density in the second phase should be $\rho/\alpha$. Now,
$\alpha$ may be determined by maximizing the entropy of the system,
which is done by by maximizing $\alpha f(\rho/\alpha)$. This is the
same as maximizing $S(\rho') / \rho'$, where $\rho'$ is the energy
density of the second phase. The energy density $\rho'=\rho/\alpha$ of
the second phase is determined by this maximization, which gives the
equation
\eqn\fdet{f'(\rho') = {f(\rho') \over \rho'}.}
It is not difficult to see that equation \fdet\ has a single solution $\rho_1$,
and that $\rho_1>\rho_0$. \foot{The argument goes as follows.
Recall that, for $\rho<\rho_0$ (but still of order $N^2$),
$f'$ is an increasing
function. It follows that, for $\rho<\rho_0$, $f'(\rho)$ is
more than its average value $(f(\rho)/ \rho)$ between
$0$ and $\rho$. Consequently, \fdet\ has no solutions for
$\rho<\rho_0$. Now, for $\rho$ in the high-energy-density phase, the
right-hand side of \fdet\ is larger than its left-hand side.
Consequently, \fdet\ must have
at least one solution for $\rho>\rho_0$. Let the smallest such
solution be $\rho_1$. Using the fact that $f'$ decreases
for $\rho>\rho_1$, it is easy to argue that no $\rho>\rho_1$ can
obey \fdet. It follows that $\rho_1$ is the unique solution to
\fdet. Note that we assume here that there is only one point where
the second derivative of $f$ vanishes (as in the discussion of the
previous subsection); this is true in all known cases, though more
complicated possibilities cannot be ruled out.}

Equation \fdet\ may be physically interpreted in several ways.
First, recall that the free energy density of our system is given by
$F_{free}(\rho)= (E - T S) / V = \rho-(f(\rho) / f'(\rho))$.
Comparing with \fdet, we conclude that our system has positive free
energy density for $\rho<\rho_1$ but negative free energy density
for $\rho>\rho_1$; the free energy density vanishes at
$\rho=\rho_1$. As the (normalized) free energy density also vanishes
in the first phase, the free energy densities of the two phases are
equal when the deconfined phase has $\rho=\rho_1$, so we conclude
that $\rho_1$ is the energy density of the gluon plasma just above
the deconfinement temperature. Second, recalling that the pressure
is simply equal to minus the free energy per unit volume \press,
it follows that $p(\rho)$ is negative for $\rho<\rho_1$ but is
positive when $\rho>\rho_1$, and the pressure vanishes precisely at
$\rho=\rho_1$.

The bubbles of gluon plasma at density $\rho_1$ which appear in this
discussion resemble the localized bubbles discussed in section 2; the
main difference between them is that the bubbles of section 2 had the
vacuum outside them rather than the finite-density confined phase, but
this difference is negligible in the large $N$ limit.

\appendix{B}{Properties of Plasma Dynamics}

\subsec{The Surface Tension of the Plasma-Ball}

The force balance equation \fb\ may be derived from stress energy
conservation applied to the plasma fluid in polar coordinates. In
$p$ spatial dimensions the equation  $\nabla^\mu
T_{\mu r}=0$, for static configurations,
reduces to
\eqn\stencon{\partial_r T_{rr}+ {1 \over
r} \left( p T_{rr} -g^{i j} T_{i j} \right) =0,}
where $i, j$ are
summed over all spatial coordinates. When the plasma fluid is
isotropic $p T_{rr} =g^{i j} T_{i j}$, and constant $T_{rr}$
solves \stencon. In the plasma-ball configuration, however, the
stress tensor will not be isotropic near the boundary of the
bubble. Integrating \stencon\ we find
\eqn\pres{T_{rr}(\infty)-
T_{rr}(0)= P(0)-P(\infty) = -\int_0^\infty {dr \over r}  \left( p
T_{rr} -g^{i j} T_{i j} \right),}
where $P(0)$ and $P(\infty)$ are, respectively, the pressure in the
interior and exterior of the bubble. In the plasma-ball configuration
the integral on the right-hand side of
\pres\ receives contributions only from the neighborhood of the
boundary of the bubble. Comparing with \fb\ we
conclude that the domain wall tension (for bubbles that are much
larger than the width of the domain wall, which is expected to be
of order $1/\Lambda_{gap}$, so that $r$ is
approximately constant in the region that contributes to the
integral in \pres) is given by \eqn\sig{\Sigma ={1 \over p-1}
\int_0^\infty dr \left( g^{i j} T_{i j} -p T_{rr}\right).} In the
limit of an infinite bubble whose domain wall is transverse to the
$x$ direction, stress energy conservation implies that $T_{xx}$ is
a constant (which vanishes since it is zero in the confined vacuum),
and the expression for the domain wall tension reduces to
\eqn\sigm{\Sigma ={1 \over p-1} \int_{-\infty}^\infty dx  g^{i j}
T_{i j}= \int_{-\infty}^\infty  dx T_{yy}, }
where $y$ is any spatial
dimension along the surface of the domain wall.

In this paper we have assumed that the effective dynamical surface
tension between the confined and deconfined phases is positive
near the phase transition temperature. This is true in the
specific example analyzed in section 8. However, it may turn out
that the effective surface tension is negative in some large $N$
gauge theories -- at least we are unaware of an argument that
rules out this possibility. Large plasma-balls will not be stable
in such theories -- they will be unstable to fragmentation into
small plasma-balls (whose size is presumably of order the inverse
mass gap). However note that, at least in gauge theories at large
$\lambda$, small plasma-balls (with $R \ll 1/\Lambda_{gap}$) are always
stable as they map to Schwarzschild black holes in the dual
gravitational theory.

\subsec{Dynamical Evolution of Lumps of Plasma}

Consider a lump of gluon plasma with large volume $V \Lambda_{gap}^3 \gg 1$,
at an
average energy density larger than the critical density. The initial
pressure of such a bubble is positive, driving the bubble to expand.
As the expansion proceeds, the pressure and temperature gradients
inside the bubble tend to even out.
In large $N$ systems that undergo first order transitions we have
mentioned in section 2 that the plasma-ball is an attractive fixed
point in the space of dissipative fluid flows, and so it will be the
equilibrium configuration if we start in its domain of attraction.
We believe that the only other stable configurations of plasma fluid
are disconnected plasma-balls moving away from each other. It
follows that, within fluid dynamics, hot lumps of gluon plasma
stabilize into a collection of plasma balls.\foot{Note that, under
some circumstances, it is rather natural for a single lump of gluon
plasma to break up into many components. This could happen if, for
example, during the process of evolution the local energy density of
some part of the fluid were to become low enough such that the local
speed of sound squared is negative (it should go below $\rho_0$, see
appendix A), triggering an instability that results in phase
separation.}

We will argue below that we do not expect stable plasma-fluid
configurations to exist in systems that undergo second order
deconfinement phase transitions. We expect the bubbles of plasma in
such systems to expand through the deconfinement transition and to
rapidly hadronize.

\subsec{Small Plasma-Balls}

As discussed in section 3, plasma-balls shrink as they slowly lose energy
by radiating glueballs. According to \fb\ the plasma-ball
temperature diverges as its radius goes to zero. However, \fb\ does
not apply when $R < 1/\Lambda_{gap}$, since the surface tension term does not
capture all surface terms for small bubbles (contributions to the
energy involving the curvature of the surface and inhomogeneities of
the plasma are as important or more important). A more detailed microscopic
analysis is needed to determine the properties of small
plasma-balls. The strongly coupled nature of the gluon plasma may
make this analysis rather difficult; however it could be worth the
effort, as small plasma-balls encode very interesting information
about black hole physics, as we have argued in \S6 and as discussed
further in appendix D.

\subsec{Plasma Lumps in Theories with Second Order Deconfinement
Transitions}

Stable plasma-balls do not exist in theories that undergo second
order phase transitions, at least when $\Sigma$ is positive. While
plasma-ball-like configurations of finite radius appear to exist as
stable solutions of fluid mechanics (using the arguments of section
2), their surface temperature is higher than the Hagedorn
temperature, so they cannot be in an approximate equilibrium in a
large $N$ gauge theory\foot{Recall that a hot ball above the
Hagedorn temperature immersed into a large $N$ gauge theory cools
instantaneously by radiating glueballs at a rate which diverges in
the large $N$ limit.}. Note also that in this case asymptotically
large plasma-ball-like configurations are unstable even within fluid
mechanics, as the specific heat diverges (and so the  speed of sound
vanishes) at the phase transition point in any second order
transition.

\subsec{Plasma-Balls in the Real World ?}

A large $N$ version of QCD may be obtained by coupling $SU(N)$ pure
Yang-Mills theory to (say) $N_f=2$ or $N_f=3$ light quarks (fermions
in the fundamental representation). As we have remarked above, there
is good evidence that this theory undergoes a first order
deconfinement transition and so possesses plasma-balls at large enough
$N$. If the phase transition stays of first order as we decrease $N$,
it seems possible that these meta-stable configurations will continue
to exist (albeit with decreasing finite lifetimes).
\foot{
Note, however, that in QCD-like theories with $N_f$ of order $N$,
plasma-balls would not have long lifetimes even if the phase
transition remains of first order, because the number of different
meson species that they can emit scales as $N_f^2$. See \schnitzer\
for a recent discussion of the deconfinement transition for large $N
\sim N_f$.}

The expected phase diagram of this theory at $N=3$, i.e. of real
world  QCD, may be found, for instance, in figure 1 of \krit.
The nature of the deconfinement `transition' in this diagram depends
crucially on the strength of the chemical potential that couples to
baryon number. When the chemical potential is small enough, QCD
undergoes a smooth crossover between confining and deconfined
behavior; this crossover becomes a strong first order phase
transition at large chemical potential. As a consequence, QCD could
possess meta-stable plasma-ball-like configurations at large enough
baryon density, if the phase boundary surface tension is positive
(see \refs{\krwi, \krnwi} for closely related discussions).
However, meta-stable plasma-ball-like configurations will not exist at low
baryon density, for instance at the baryon densities attained in the
central rapidity region of the fireball created at RHIC, see
\refs{\GyulassyZY,\LudlamRH} for reviews.

The rapidity tails of the RHIC fireball, as well as suitable regions
of the fireball (if one was produced) in the SPS experiment, or
plasmas produced in similar experiments where baryons are
heated\foot{See figure 1 of \HeinzAX\ for the location of various
experiments on the QCD phase diagram.} may have chemical
potentials large enough to be in the first order regime. It would
be interesting to analyze the relevant data from these experiments
for evidence of formation of meta-stable plasma-ball-like
configurations, that then decay into hadrons at approximately the
phase transition temperature (see \refs{\kro, \krt} and references
therein for discussions along these lines).

\appendix{C}{Counting Powers of $N$ at Lowest Order in Perturbation Theory}

In this brief appendix we illustrate the arguments of section 3 by
explicitly displaying and computing the $N$ dependence of sample
diagrams that contribute to the gluon mean free path and to the
glueball production rate at lowest order in perturbation theory.

\subsec{$k=0$ : Mean Free Path}

\fig{Tree diagrams for gluon-gluon scattering in double-line
notation.} {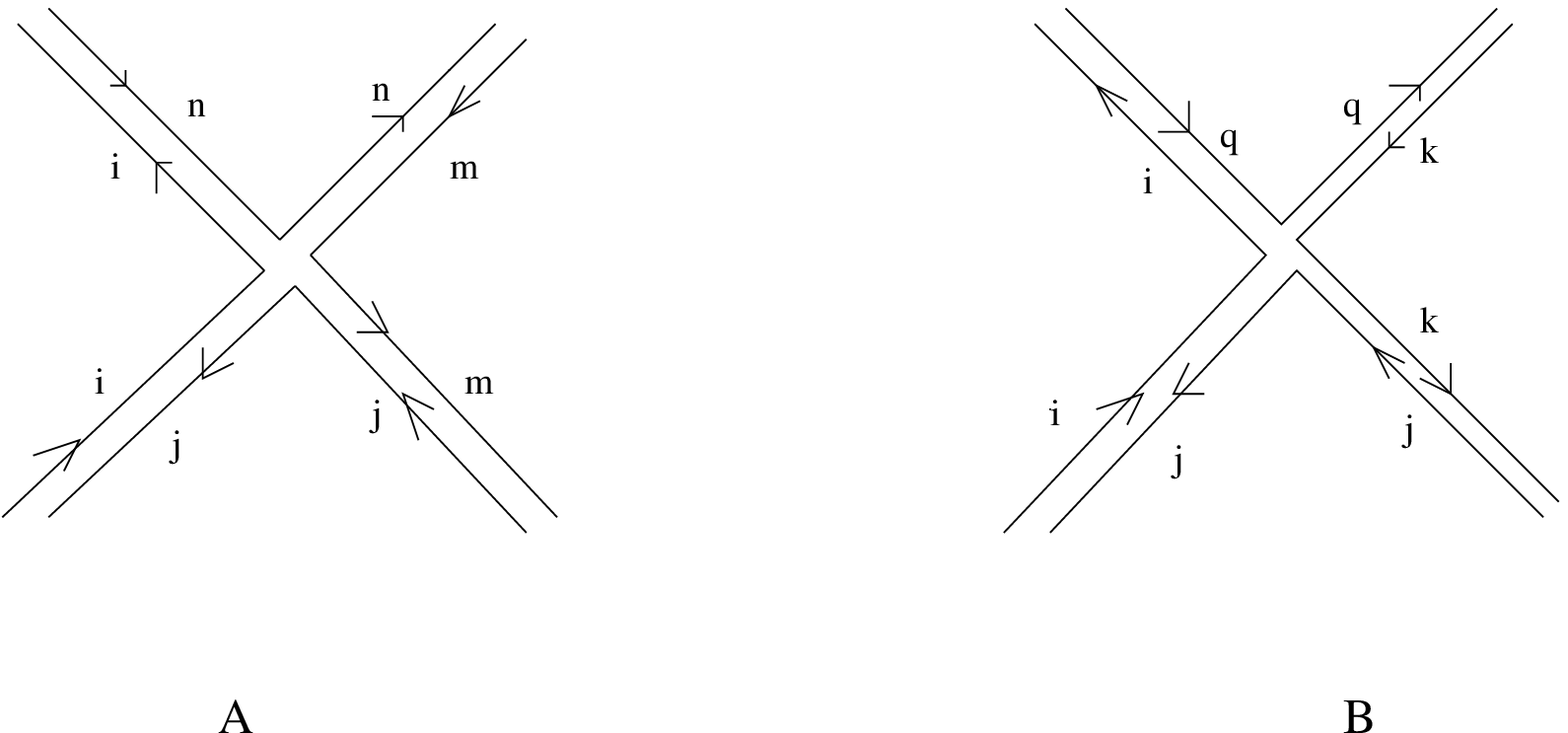}{5truein} \figlabel{\glueint}

We check the result of \S3.2 at weak coupling by
studying the leading gluon-gluon scattering graphs shown in figure
\glueint. First consider the graph in figure \glueint(A) for scattering
an $(i,j)$ gluon and an $(m,n)$ gluon, and let $m
\neq j$ and $n \neq i$. This diagram is proportional to $g^2 N^0 =
\lambda / N$, leading to a collision cross section proportional to
$\lambda^2/ N^2 $ and so a mean free collision time that is
independent of $N$ (the mean free time is obtained by multiplying
the collision cross-section by the density of potential collision
targets, whose number scales like $N^2$).

In the special case that $(m,n)=(j, k)$ for some $k$, the tree-level
diagram in figure \glueint(B) also contributes. The cross section
from this diagram is $\propto {\lambda^2 / N}$, where the additional
factor of $N$ results from the sum over the $q$ color index in the
final state. This graph too leads to a mean free collision time that
is independent of $N$ as, in this case, the density of potential
collision targets scales like $N$.

\subsec{$k=1$ : Rate of Glueball Production}

\fig{Tree diagrams for glueball creation by scattering
gluons.}{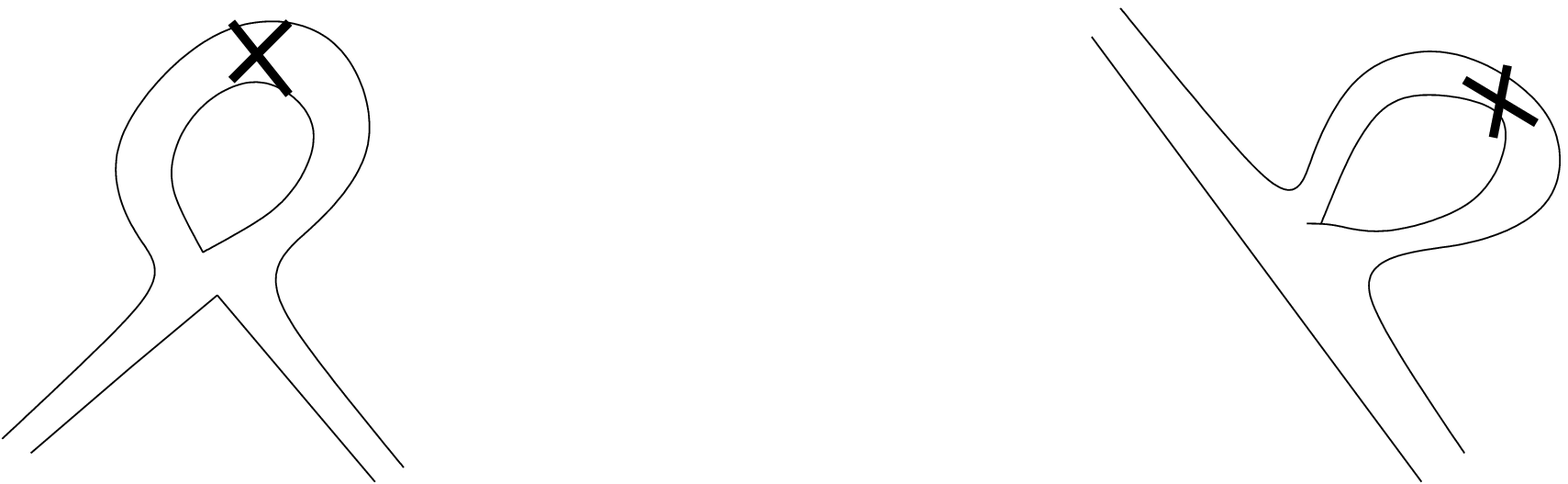} {4truein} \figlabel{\gluedecay}

Consider glueball production via the gluon-gluon collision depicted
on the left of figure \gluedecay. The vertex in this graph
contributes $g_{YM}^2$, the index loop contributes a factor of $N$,
and the overlap of the two-gluon state with the glueball
wavefunction is of order $1/N$ (this is the $1/N$ in the
normalization of the glueball creation operator), resulting in a
collision cross section of order $1/N^2$. Multiplying this by the
total number of gluons in the plasma-ball, we find a glueball
production rate independent of $N$.

The same Feynman diagram may be given a different interpretation
upon choosing the time slicing on the right of figure \gluedecay.
This diagram describes an energetic gluon escaping out of the
surface of the plasma, until the string that attaches it to the rest
of the plasma snaps by creation of a gluon-anti-gluon pair, allowing
the original energetic gluon and its anti-gluon partner to escape
away as a glueball. The analysis of the previous paragraph carries
through immediately for this process, leading, once again, to a
glueball creation rate independent of $N$.

\appendix{D}{The Final Decay of Small Plasma-Balls}

After the black hole/plasma-ball at large $\lambda$
shrinks such that it resembles a ten dimensional Schwarzschild black hole,
it continues to shrink as it evaporates until its proper
Hawking temperature is of the order of the Hagedorn temperature
$T_H$ (so that the proper size of the black hole is of order the
string scale). At this point the black hole is believed to resemble
a highly excited string state \refs{\SusskindWS,\HorowitzNW} \foot{See
\spenta\ for an intriguing discussion of a possible manifestation of 
the black-hole excited string transition in $\CN=4$ Yang Mills on $S^3$.} 
and there are several possibilities for the subsequent evolution. The
black hole could
\item{1.} Develop a tachyon (due to stringy $\alpha'$ corrections)
 at a temperature smaller than $T_H$ as a result of which the
 black hole decays classically, dominantly into low-lying string
 states (glueballs).
\item{2.} Reach the Hagedorn temperature when its mass is
of order ${1/ g_s^2} \sim N^2$, at which point it explodes
into a gas of dominantly highly excited string states (or transmutes
into a single excited long string, depending on details, see for
instance section 2 of \sphere)\foot{Note that the excited string
states are stable in the large $N$ limit, though for finite $N$ they
would eventually decay into lighter glueballs.}.
\item{3.} Continue to decay thermally (with temperatures less than $T_H$)
until its energy is of order one.

Which of the options 1-3 above is actually attained is a question of
obvious interest in the study of black hole physics. This question
could, in principle, be answered by an investigation of classical
string theory, but this has not yet proved possible. Our
identification of black holes with plasma-balls allows us to map
this question to the properties of small plasma-balls, a
reformulation that could lead to progress.

\appendix{E}{The Boundary Stress Tensor for Compactifications of $AdS_{d+2}$}

Asymptotically $AdS_{d+2}$ backgrounds possess a conserved $d+1$
dimensional stress tensor
\refs{\BrownBR,\BalasubramanianRE,\skena,\skenb,\skenc}.  The
stress tensor associated with
\AdSSchIM\ is (see equations 4 and 25 in \kraus)
\eqn\bst{T_{\mu \nu}=\Theta_{\mu \nu} - \gamma_{\mu \nu} \Theta
-(d-1) \gamma_{\mu \nu}={\alpha e^{-(d-1)u} \over 2} \left(
-\delta_{\mu \nu} + (d+1)\delta_{\mu \theta} \delta_{\nu \theta}
\right),}
where the indices $\mu$ and $\nu$ run over $i, t, \theta$,
$\gamma_{\mu\nu}$ is the restriction of \AdSSchIM\ to a constant $u$
slice, $\Theta_{\mu\nu}=-\half {1 \over \sqrt{g_{uu}}} {d
\gamma_{\mu\nu} \over d u }$ and $\alpha= \left( d+1
\right)^{-(d+1)}$. The boundary field theory stress tensor associated
with \AdSSchIM\ is obtained by multiplying \bst\ by $e^{(d-1)u}$, so
that
\eqn\ftst{T_{\mu \nu}^{confined\ vacuum}={\alpha  \over 2}\left( -\delta_{\mu
\nu} + (d+1)\delta_{\mu \theta} \delta_{\nu \theta} \right) . }
The boundary stress tensor for the black brane solution \AdSSchIb\ is
obtained similarly, giving,
\eqn\ftst{T_{\mu \nu}^{black\ brane}=\left( 2 \pi T \right) ^{d+1}{\alpha
\over 2}\left( -\delta_{\mu \nu} + (d+1)\delta_{\mu t} \delta_{\nu
t} \right) . }

The subtraction we used to define the stress tensor in \bst\ is the
standard subtraction for the $(d+1)$-dimensional field theory, which
gives a vanishing stress tensor in the vacuum of this theory. From the
point of view of the $d$-dimensional field theory which we obtain
after compactifying on the circle, it is more natural to use a
different subtraction in which the stress tensor in the vacuum
\AdSSchIM\ of this theory evaluates to zero.  Performing the necessary
subtraction we find
\eqn\ftst{T_{\mu\nu}^{black\ brane-sub}= {\alpha \over 2} \left[ -\left(
( 2 \pi T) ^{d+1} -1\right)\delta_{\mu \nu} + (d+1) \left( (2 \pi
T)^{d+1} \delta_{\mu t} \delta_{\nu t} -\delta_{\mu \theta}
\delta_{\nu \theta} \right) \right] . }
In particular, the energy density $\epsilon$, free energy density
$f$ and pressure $P$ of the black brane (the deconfined phase)
are given as functions of temperature by
\eqn\enp{\epsilon= {\alpha \over 2}\left( d (2 \pi T)^{d+1} +1
\right); ~~~P=-f= {\alpha \over 2} \left( (2 \pi T)^{d+1} -1
\right). }

\appendix{F}{Numerical Construction of the Domain Wall Solution}

\subsec{Equations and Constraints}

Working in units with $L^2 \alpha' = 1$, the metric we require is
Einstein and must solve
\eqn\Riccieq{ R_{\mu\nu} = - (d+1) g_{\mu\nu}. }
In our coordinate system \metricA\ the three equations from the
$\tau\tau$, $\theta\theta$ and $r_{a}r_{a}$ components manifest the
elliptic nature of the static Einstein equations, where the second
derivatives are simply the flat two dimensional
$(x,y)$-Laplacian, giving coupled
second order equations for $A, B, C$. The equation for $A$ is
\eqn\ellipticAB{ \nabla^2 A = (d+1) A e^{2 D} -
\left ( (d-2) \nabla C + \frac{\nabla B}{B} \right) \cdot \nabla A,
}
and the equation for $B$ is identical after the substitution $A
\leftrightarrow B$ above, since in the static context there is no
physical difference between the Euclidean time circle and the
compact spatial circle $\theta$. The equation for $C$ is of a
similar form,
\eqn\ellipticC{ \nabla^2 C = (d+1) e^{2 D} -
\left( (d-2) \nabla C + \frac{\nabla A}{A} + \frac{\nabla B}{B} \right)
\cdot \nabla C.
}
The last elliptic (i.e. Laplacian second derivatives) equation can be
found for the final function $D$ from the $xx+yy$ component of
\Riccieq,
\eqn\ellipticD{ \nabla^2 D = - \frac{(d+1)(d-2)}{2} e^{2 D} +
(d-2) \left( \frac{(d-3)}{2} \nabla C + \frac{\nabla A}{A} +
\frac{\nabla B}{B} \right) \cdot \nabla C +
\frac{1}{A B} \nabla A \cdot \nabla B. }
The remaining Einstein equations, namely the $xy$ and the $xx-yy$
components of \Riccieq, do not have elliptic second derivatives (but
rather $\partial_x\partial_y$ and $\partial_x^2 - \partial_y^2$
respectively) and should best be thought of as constraints associated
with the gauge-fixing performed in \metricA. Let us write them as
\eqn\Constraints{ \eqalign{ \alpha & \equiv \sqrt{ \det{ g_{\mu\nu}
} } \left( R^{x}_{~y} \right) = 0, \cr \beta & \equiv \frac{1}{2} \sqrt{ \det{
g_{\mu\nu} } } \left( R^{x}_{~x} - R^{y}_{~y} \right) = 0.} }
The two non-trivial contracted Bianchi identities then relate $\alpha$
and $\beta$ as a very elegant Cauchy-Riemann problem,
\eqn\Bianchi{ \eqalign{
\partial_x \alpha + \partial_y \beta & = 0, \cr
\partial_y \alpha - \partial_x \beta & = 0,
} }
provided that the above elliptic equations, \ellipticAB, \ellipticC\
and \ellipticD, are satisfied.

Our strategy in finding solutions will be to solve the elliptic
equations, subject to boundary data that ensures that the
Cauchy-Riemann problem \Bianchi\ only has the trivial solution
$\alpha=\beta=0$, and hence the constraint equations \Constraints\
will be satisfied. The natural data to take for these elliptic
equations is to specify a Dirichlet or Neumann (or mixed) condition on
each boundary for each metric function.

\subsec{The IR Boundary Conditions}

We now consider the boundary conditions on the IR coordinate boundary
where the $\tau$ or $\theta$ circle shrinks. Since these two circles
are identical from the point of view of the equations, let us choose
to discuss the shrinking $\tau$ circle at $y = 0$ (the other case
being identical with $x \leftrightarrow y$ and $A \leftrightarrow B$).

There are two sources of data for the functions $A,B,C,D$. Firstly,
the shrinking cycles imply regular singular behavior in the elliptic
equations, and our choice that these metric components be finite
singles out certain boundary conditions. Secondly, the constraint
equations $\alpha=0$, $\beta=0$ also lead to boundary conditions.

Firstly, consider the boundary conditions from the elliptic equations
at $y = 0$. By construction we impose the Dirichlet data $A = 0$. When
$A \rightarrow 0$, some terms in the `sources' for these elliptic
equations \ellipticAB, \ellipticC\ and \ellipticD\ are potentially
singular, as they go as $1/A$. Regularity therefore imposes Neumann
boundary conditions on $B$ and $C$,
\eqn\elliptbc{ \eqalign{
 \partial_y B & = 0, \cr
 \partial_y C & = 0,
} }
at $y = 0$. Secondly, consider the constraints at $y = 0$. These give,
\eqn\constbc{
\eqalign{
\alpha: \quad & \partial_x \left( e^{-D} \partial_y A \right) = 0, \cr
\beta: \quad & \partial_y D = 0.
}
}
We may immediately recognize the $\alpha=0$ constraint as the zeroth law
of black hole mechanics.  Similarly, for the spatial circle shrinking
on $x = 0$, it ensures that the geometry will close smoothly for one
value of the circle coordinate period.

Note that these boundary conditions are derived from the equations of
motion, and are totally independent of whether we work in the
Euclidean or Lorentzian setting.

Solving the elliptic equations in an elliptic manner we require one
piece of data on each boundary for each metric function. Thus we see
that we now have an over-complete set of boundary conditions for
$A,B,C,D$ in the IR. We therefore choose to impose only the $\beta=0$
constraint, and hence obtain the rather simple Neumann and Dirichlet
data,
\eqn\ellipticbc{ \eqalign{
 A & = 0, \cr
 \partial_y B & = 0, \cr
 \partial_y C & = 0, \cr
 \partial_y D & = 0,
}
}
at $y = 0$, and similarly for the coordinate boundary at $x = 0$ the
same with $x \leftrightarrow y$ and $A \leftrightarrow B$. We leave
the $\alpha=0$ constraint to be imposed via the constraint system, by
choosing data appropriately on the other boundaries.

An important point is that at the phase interface $x = y = 0$ both
constraints are satisfied automatically due to the factor of $A B$ occurring
in $\sqrt{\det{g_{\mu\nu}}}$. The two constraints take the form,
\eqn\fullconst{
\eqalign{
\alpha & = A \; \alpha_{x=0} + B \; \alpha_{y=0} + A \, B \; \alpha_{remainder} ,\cr
\beta  & = A \; \beta_{x=0} + B \; \beta_{y=0} + A \, B \; \beta_{remainder} ,
}
}
where $\alpha_i, \beta_i$ are linear functions in derivatives of $A$
and $B$, but are not explicit functions of $A$ and $B$ themselves. The
terms $\alpha_{y=0}, \beta_{y=0}$ give the constraints \constbc\ on
the $y = 0$ shrinking time circle boundary, but do not contribute to
the shrinking space circle boundary $x = 0$ due to the prefactor of
$B$ (omitted in \constbc). Likewise $\alpha_{x=0}, \beta_{x=0}$ give
the constraints for the shrinking circle boundary $x = 0$, and don't
contribute to the $y=0$ boundary. The final terms $\alpha_{remainder},
\beta_{remainder}$ contribute to neither boundary as $A B$ is zero on both,
and also at the interface.  Hence no additional data is required to
satisfy the constraint system at the interface, beyond \constbc\ (and
its symmetric counterpart at $x = 0$).

\subsec{The UV Boundary Conditions}

The UV boundary resides at the locus $f = 0$ where we demand that the
metric looks like the boundary of AdS space, with $A,B,e^C,e^D \sim
1/f$ (assuming that $f$ has a first order zero). In order to impose
this boundary condition, near the UV boundary we write
\metricA\ in the form
\eqn\metricC{ ds^2 = \frac{1}{f^2} \left( T^2 d\tau^2 + U^2 d\theta^2 +
e^{2 S} dr_a^2 + e^{2 R} | \nabla f |^2 \left( dx^2 + dy^2 \right) \right) }
and we take the new metric variables $T,U,S,R$ (which are trivially
related to $A,B,C,D$ via $f$) to be finite there. Coupled with the
form above, we also make the further choice that the function $f$
obeys the Laplace equation
\eqn\feq{ (\partial_x^2 + \partial_y^2) f = 0,}
and is defined in a finite neighborhood of the UV boundary. Note that
this is consistent with the fact that for large $x$ we expect $f
\propto (y-c)$, and $f \propto (x-c)$
for large $y$. The reason for choosing $f$ to be
harmonic is that then we can easily map \metricC\ to
\eqn\metricD{ ds^2 = \frac{1}{f^2} \left( T^2 d\tau^2 + U^2 d\theta^2 +
e^{2 S} dr_a^2 + e^{2 R} \left( df^2 + dg^2 \right) \right) }
where the metric components are now written as
functions of $f(x,y)$ and $g(x,y)$, with
the coordinate $g$ defined by the Cauchy-Riemann relations,
$\partial_x f = \partial_y g$ and $\partial_y f = - \partial_x
g$, so that the new coordinates $f, g$ are conformally related to
$x,y$. Since $f = 0$ at the UV boundary, it can be thought of as the
asymptotic bulk radial coordinate (like $u$ earlier), and $g$ is
naturally then the coordinate along the boundary at this zero of $f$.

In the coordinates \metricD\ it is easy to see that AdS space is
simply given by $T, U, e^S= $constant and $e^R = 1$, since then the
metric \metricD\ is just AdS written in the usual Poincar\'e
coordinates. Hence, $T, U, e^S$ going to constants and $e^R
\rightarrow 1$ near $f = 0$ will reproduce asymptotically AdS space.
Reversing the argument, starting with AdS in Poincar\'e coordinates near
the boundary and choosing our coordinate system \metricA\ {\it and}
the position of the UV boundary to be fixed on some curve $f(x,y)=0$,
then the most general form that the asymptotically AdS region can take
is \metricC, with $T, U, e^S= $constant and $e^R = 1$, for some
harmonic function $f$ with the correct zero locus.

We now make the further choice that
\eqn\asym{
T, U, e^S, e^R \rightarrow 1
}
as $f\rightarrow 0$, for some given $f$, so that the metric of the
dual field theory, in the conformal frame defined by $f$, is just
\eqn\boundarymetric{
ds^2_{d+1} = d\tau^2 + d\theta^2 + dr_a^2 + dg^2.  }
This choice is always possible, as we may independently rescale the
$\tau$, $\theta$ and $r_a$ coordinates without effecting the Einstein
equations. It is this choice (together with the choice of UV
coordinate boundary location) that will set the actual values of
$\beta,\tilde{\beta}$ required to smoothly close the manifold in the
IR. Note that we have chosen the asymptotic values of $T, U$ to be the
same in order to preserve the natural $\IZ_2$ symmetry of the
solution.

Suppose that for some metric $A,B,C,D$ we choose a particular $f$
satisfying \feq\ such that the behavior \asym\ is true. Now let us
consider a different $f$, say $f'$, which shares the same locus for
its zero, and also satisfies
\feq. Then the induced $T',U',S',R'$ using $f'$ will not satisfy
\asym, but rather will go as,
\eqn\asymfprime{
\eqalign{
T', U', e^{S'} & \rightarrow \phi(g) = \frac{ | \nabla{f'} | }{ | \nabla{f} | } , \cr
e^{R'} & \rightarrow 1 .
}
}
This gives a boundary metric in the conformal frame defined by $f'$
which is
\eqn\boundarymetricprime{
ds'^2_{d+1} = \phi(h)^2 \left( d\tau^2 + d\theta^2 + dr_a^2 + dh^2
\right) }
with $dg' = \phi \; dh$. Hence, different choices of $f$, together with
the condition \asym, pick out conformally related boundary metrics,
where the normal gradient of $f$ to its zero determines the conformal
scaling.

Thus we could pick a particular $f$ and implement the freedom of the
conformal factor in the asymptotic behavior directly (as in
\asymfprime). However this considerably complicates the asymptotic
behavior of the metric functions. Instead, we do not pre-determine
$f$, but impose the choice \asym\ and then this scalar degree of
freedom is just the normal gradient of $f$ to its zero.

For any $f$ satisfying \feq, the `elliptic' Einstein equations for $T,
U, e^S, e^R$ given by \ellipticAB, \ellipticC\ and \ellipticD, then
yield the following simple behavior near $f=0$ :
\eqn\UVbc{
\eqalign{
T & = 1 + f^{d+1} \delta(g) + f^{d+2} \tau_{d+2}(g) + f^{d+3} \tau_{d+3}(g) +
\cdots ,\cr
U & = 1 + f^{d+1} \gamma(g) + f^{d+2} u_{d+2}(g) + f^{d+3} u_{d+3}(g) +
\cdots ,\cr
e^S & = 1 - \frac{1}{(d-2)} f^{d+1} (\delta(g) + \gamma(g) + 2 \rho(g)) +
f^{d+2} s_{d+2}(g) + f^{d+3} s_{d+3}(g) + \cdots ,\cr
e^R & = 1 + f^{d+1} \rho(g) + f^{d+2} \sigma(g)  + f^{d+3} r_{d+3}(g) +
\cdots ,
}
}
where again $g$ is a coordinate normal to $f$ which measures the
position along the locus of the zero of $f$. The four leading ones
in these expansions should be likened to Dirichlet data, and as
discussed above, guarantee that this geometry is asymptotic to AdS.
\foot{
Choosing the three leading (Dirichlet) terms in the asymptotic
expansions of $T, U, e^S$ to be finite functions of $g$ (rather than
one as in \UVbc) represents a `non-normalizable' deformation of the
asymptotic region of the space away from AdS, and will give a non-flat
boundary metric. The fourth piece of Dirichlet data is implicit in the
finiteness of $R$ at the UV boundary. Whilst $e^R
\rightarrow 1$ if $R$ is finite, there is an additional possible
non-normalizable behavior, seen most easily in the linear theory
where $R \rightarrow r(g) \log(f)$ is also possible. This can simply
be understood as arising from a conformal coordinate transformation
moving the coordinate position of the UV boundary, and hence being
singular in $R$. This function $r(g)$ which we have implicitly set to
zero in \UVbc\ is the fourth and last piece of Dirichlet data for the
functions $T,U,S,R$ at the UV boundary.  }
The four undetermined finite functions $\delta,\gamma,\rho,\sigma$
should be regarded as Neumann data for the elliptic equations. Since
we fix the Dirichlet data, we are not at liberty to also fix these
functions when solving the system in an elliptic manner, and these
will be determined implicitly by the IR boundary conditions. The
functions $\{\tau,u,r,s\}_i$ for $i \ge d+2$ are terms in the
power-series expansion of the metric in $f$ which are determined in
terms of these leading `Dirichlet' and `Neumann' coefficients as in
the usual Fefferman-Graham expansion
\refs{\fefferman,\henningson}.

Using these expansions we may compute the boundary stress tensor as in
appendix E, with components given by
\eqn\stresscomponents{
\eqalign{
T_{\tau\tau} & = (d+1) \delta(g) + \rho(g) , \cr
T_{\theta\theta} & = (d+1) \gamma(g) + \rho(g) , \cr
T_{r_ir_i} & = - \frac{1}{(d-2)} \left( (d+1) ( \delta(g) +
\gamma(g) ) + (d+4) \rho(g) \right) , \cr
T_{gg} & = (d+2) \rho(g) \cr
}
}
in an appropriate normalization. Note that, as required, this is
traceless. Recall that we should subtract the confining vacuum
stress tensor in order to obtain the standard dual field theory
stress tensor. In the flat boundary metric, the conservation
equation of the stress-energy tensor gives $\partial_g T_{gg} =
0$, and this is crucial in order to understand the pressure
balance argument which singles out a particular temperature
thermal bubble interior. In the gravity context this arises as a
result of the equation of motion of $G^{f}_{~g}$, as we now
discuss.

So far we have only discussed the behavior of the `elliptic'
equations. Now we evaluate the constraint equations
$\tilde{\alpha}=0$, $\tilde{\beta}=0$, in the new coordinates $f, g$ :
\eqn\Constraintstilde{ \eqalign{ \tilde{\alpha} & \equiv \sqrt{
\det{ \tilde{g}_{\mu\nu}
} } \left( R^{f}_{~g} \right) = 0, \cr \tilde{\beta} & \equiv \frac{1}{2}
\sqrt{ \det{
\tilde{g}_{\mu\nu} } } \left( R^{f}_{~f} - R^{g}_{~g} \right) = 0.} }
We note that $\tilde{\alpha}, \tilde{\beta}$ are simply linearly
related to $\alpha, \beta$, the coefficients depending on the
(non-singular) coordinate transformation between the $f,g$ and $x,y$
coordinates.  Using the expansion \UVbc\ we find that both these
constraints are finite near $f = 0$, although neither is automatically
zero, with
\eqn\UValphaA{
\eqalign{
\tilde{\alpha} & = \frac{d (d+3)}{(d-1)} \; \partial_g \rho(g)  + O(f) , \cr
\tilde{\beta} & = (d+2) \; \sigma(g)  + O(f) .
}
}
We focus on the $\tilde{\alpha}=0$ constraint, which is particularly
elegant as it implies that $\rho(g)$ is a constant, say $k$, and hence
contours of $f$ and $R$ are aligned asymptotically. Thus it implies
that near the zero of $f$,
\eqn\UValphaB{
R(f,g) = k f^{d+1} + O(f^{d+2}).
}
Furthermore, this obviously guarantees that $T_{gg} = $constant in the
boundary stress tensor (and will be zero with a suitable vacuum
subtraction). Note that the actual value of $k$ is unphysical since in
the definition of $f$ via \metricC\ we see the physical metric is
invariant under rescalings of $f$, while appropriately rescaling the
$\tau$,$\theta$ and $r_a$ coordinates too. Therefore we may choose any
value for $k$ that is convenient.

\subsec{Solving the Constraint System }

At large $x,y$ we expect our solution to return to the homogeneous
black brane \AdSSchIb\ or to the AdS soliton \AdSSchI\ (this should
occur exponentially fast in proper distance, since the effective dual
theory has a mass gap). Since the $\alpha=0$ constraint is satisfied for a
homogeneous solution (depending only on $x$ or $y$), at large $x,y$
where the solution should become homogeneous it will quickly go to
zero. By the Cauchy-Riemann constraint relations \Bianchi\ $\beta$
must then go to a constant, and since we are actually imposing $\beta=0$
on the IR coordinate boundary, it will also vanish.  Hence, the
constraints will be satisfied in these asymptotic regions.  In the IR
we have explicitly imposed $\beta=0$ on the axes $x = 0$ and $y = 0$.
This is still not sufficient data to solve the constraint problem.

In order to solve the system we could try to impose $\alpha$ in the
IR, but as we discussed above, the elliptic system is already
completely determined there. Instead, we may impose the vanishing of
some linear combination of $\alpha$ and $\beta$ in the UV. With $\beta
= 0$ in the IR, and a linear combination being zero in the UV,
combined with the correct large $x,y$ asymptotics, this is sufficient
to guarantee that the solution of the Cauchy-Riemann system \Bianchi\
is such that both $\alpha$ and $\beta$ vanish everywhere.

From the discussion above, the natural combination of $\alpha$ and
$\beta$ to impose is exactly $\tilde{\alpha}$, coming from the $fg$
component of the Einstein equations in the UV, which implies the
relation \UValphaB. Now we already have Dirichlet data for $T, U, S,
R$ in the UV, determined by leading behavior in \UVbc, and we cannot
also fix the Neumann terms $\delta,\gamma,\rho,\sigma$. The only
undetermined quantity is the choice of our harmonic function
$f$. Thus, we impose the constraint $\tilde{\alpha}=0$ in the UV by
choosing $f$ to have the correct normal gradient solving \UValphaB\ on
its zero locus, so
\eqn\fbc{
| \nabla (f^{d+1}) | = \frac{1}{k} | \nabla R |
}
when $f = 0$. Note that since the Laplace equation is elliptic, this
data does not determine $f$ globally in our coordinate domain, but it
is sufficient to determine it exponentially well near the UV boundary,
and we only require $f$ in the immediate vicinity of this boundary
(i.e. we really only require its normal derivative) to define the
asymptotic behavior of $A,B,C,D$.  The behavior \UVbc\ will now ensure
that the constraint $\tilde{\alpha}=0$ is satisfied. As mentioned
above, the constant of proportionality, $k$, is unphysical and can be
chosen arbitrarily.

\subsec{Summary}

Our complete system, in either Euclidean or Lorentzian signature, is
then the elliptic equations \ellipticAB,\ellipticC,\ellipticD\ and
\feq, together with their boundary conditions \ellipticbc, \asym\ and
\fbc. We see explicitly that, as indicated in section 8, the $\IZ_2$ symmetry
\sym\ is manifest in the system. Furthermore, for fixed boundary
locations, and once we have fixed the arbitrary scaling symmetry of
the $\tau$, $\theta$ and $r_a$ coordinates (through the combined
choices of $c$, $k$ and the choice \asym), there is no further data to
be specified. Whilst we have not rigourously demonstrated that this system
is elliptic, this analysis strongly suggests that the solution is
unique, having no additional parameters.

\appendix{G}{Numerical Details}

We define our coordinate domain using an auxiliary function $h$. We
set $h=1$ on the lines $x = 0$ and $y = 0$, and $h = -0.6$ at $x \ge
1, y = 1$ and $y \ge 1, x=1$. We solve a Laplace equation for $h$ in
the interior of this region. Using this solution we then determine the
coordinate position of the UV boundary at $h = 0$ where we will demand
$f = 0$, the IR boundaries at $x = 0$ and $y = 0$, and hence the
coordinate domain for our problem. Note this gives the required
$\IZ_2$ symmetry of the UV boundary location, and for these choices,
at large $x$ the UV boundary tends to $y = 5/8$.

Furthermore we also determine an interior `boundary' at $h(x,y) =
h_0$, which satisfies $0 < h_0^{d+1} << 1$ for computational
purposes. To the IR of this interior boundary ($h>h_0$) we use the
metric variables $A,B,C,D$ when solving the elliptic equations, and to
the UV ($h<h_0$) we use $T,U,S,R$ which allows us to encode the
correct UV behavior, without requiring large gradients that reduce
accuracy and stability. The function $f$ is now only defined in this
UV region $0 \le h \le h_0$, where we must be able to translate
between $A,B,C,D$ and $T,U,S,R$. The initial guess for $f$ is then
taken to be $h$, and sensible initial guesses for the metric functions
$A,B,C,D$ and $T,U,S,R$ are then made (using $h$).

Due to the symmetry \sym, we only implement the region $y \le x$ and
use the symmetry to impose boundary conditions at $x = y$. In practice
we truncate the asymptotic region at large $x=L_{trunc}$ and simply
require that normal gradients of all functions vanish there
(i.e. Neumann boundary conditions). We also truncate the UV boundary
to be at $f = \epsilon$, rather than $f = 0$, since the behavior of
various terms in the elliptic equations are potentially rather
singular there. We check that when the position of the large $x,y$
boundaries is large compared to the bulk curvature length this
truncation does not influence the solution. Similarly, when $\epsilon$
is small, the difference between the solution for finite $\epsilon$
and the extrapolated $\epsilon = 0$ solution is also negligible.

The elliptic equations are second order finite differenced. The metric
functions are relaxed using Gauss-Seidel iteration with the boundary
conditions discussed earlier.  Some under-relaxation is required
depending on the resolution used. For some number of iterations $f$ is
not updated from its initial guess, to allow the metric components to
settle to have sensible asymptotic behavior.  Once this has occurred,
$f$ is also relaxed in the small UV region by the same method.

The value of $f$ at the interior boundary $h=h_0$ is required to solve
$R = k f^{d+1}$ there for some value of $k$. We solve this condition
by updating $f$ on this interior boundary by $\delta f = \omega (R / k
f^{d+1} - 1)$, where $\delta f$ is the update for $f$ after each
Gauss-Seidel step and $\omega$ is a small number (say $\sim 10^{-6}$
for the resolution used to produce the data shown here). This very
slow update is required as the UV boundary is rather unstable
numerically due to the singular nature of the elliptic equations
there.

Ideally we only require the normal gradient behavior of $f$ and thus
really require only a tiny $h_0$, so the UV region where $f$ is
defined is very small. In practice, the singular behavior of the
equations near the UV boundary mean $h_0$ cannot be taken too small at
a given resolution. However, we checked that for the values and
resolutions used, the solutions were insensitive to the choice of
$h_0$, provided it is small enough.

Results shown in the paper use the resolution $320 \times 80$ in $x,
y$ (recall we only use the region $y \le x$) with $L_{trunc} = 4$,
$\epsilon = 0.05$ and $h_0 = 0.6$ for both $d = 3 , 4$. The values of
$k$ were chosen for convenience to be $0.35$ and $0.3$ for $d = 3, 4$
respectively.

We remind the reader that the explicit program code used to generate
the solutions may be downloaded at {\tt
http://schwinger.harvard.edu/$\sim$wiseman/IRblackholes/}.

\quadvertfig{Contour plots of $A, B, e^C, e^D$ in the $x,y$
plane for $d = 4$.}{6truein}{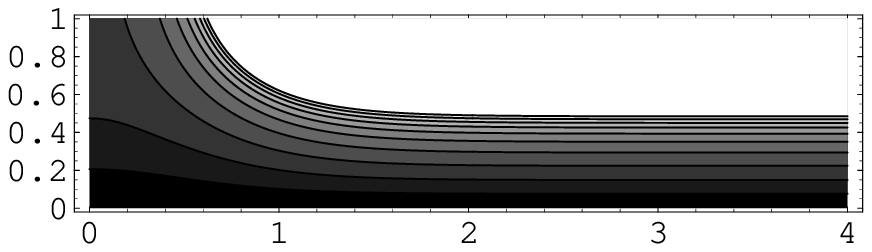}{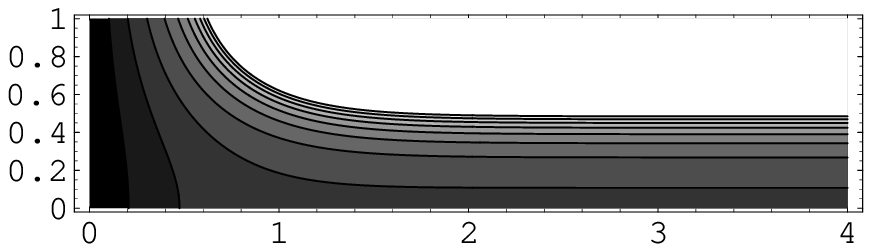}{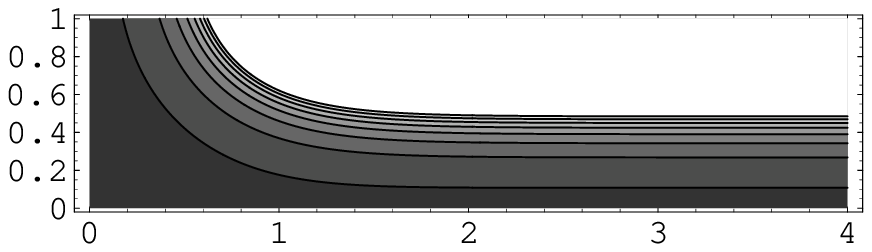}{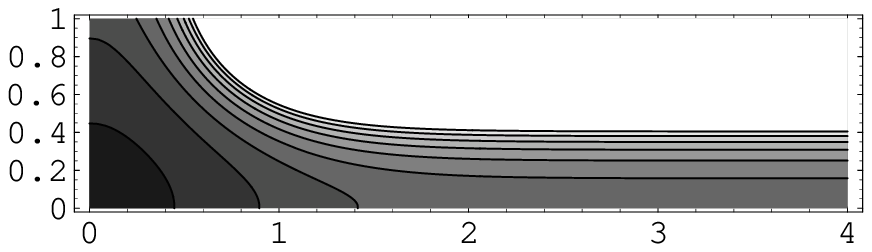}\figlabel{\TUSRfigB}

\fig{Plot showing contours of constant $f$ (red, dashed) and
$R$ (black, solid) for $d = 4$ near the UV boundary where $f$ is
defined. These are aligned, indicating that the constraint
$\tilde{\alpha}=0$ is satisfied near the UV
boundary.}{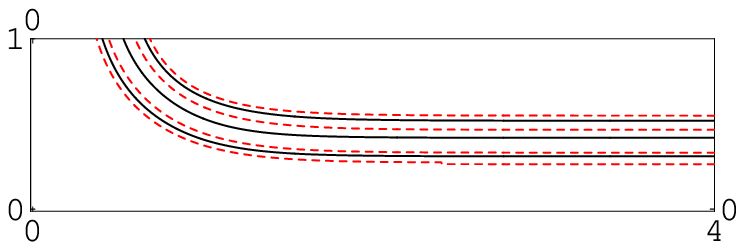}{6truein}\figlabel{\contours}

\twosidefig{
Plot showing violation of the $\alpha=0$ constraint in the IR. The proper
gradient of the vanishing circle size, $\kappa$, is plotted against
position on the boundary $x$ for $d=3$ (left) and $d=4$ (right),
normalized to its value as $x \rightarrow \infty$. The $\alpha=0$
constraint implies that this should be independent of $x$. Recall that
we impose
the constraints $\beta=0$ in the IR and $\tilde{\alpha}=0$ in the UV. Here
we see that $\alpha=0$ is true to high accuracy in the IR, implying we have
indeed solved the Cauchy-Riemann constraint system.
}{3.5truein}{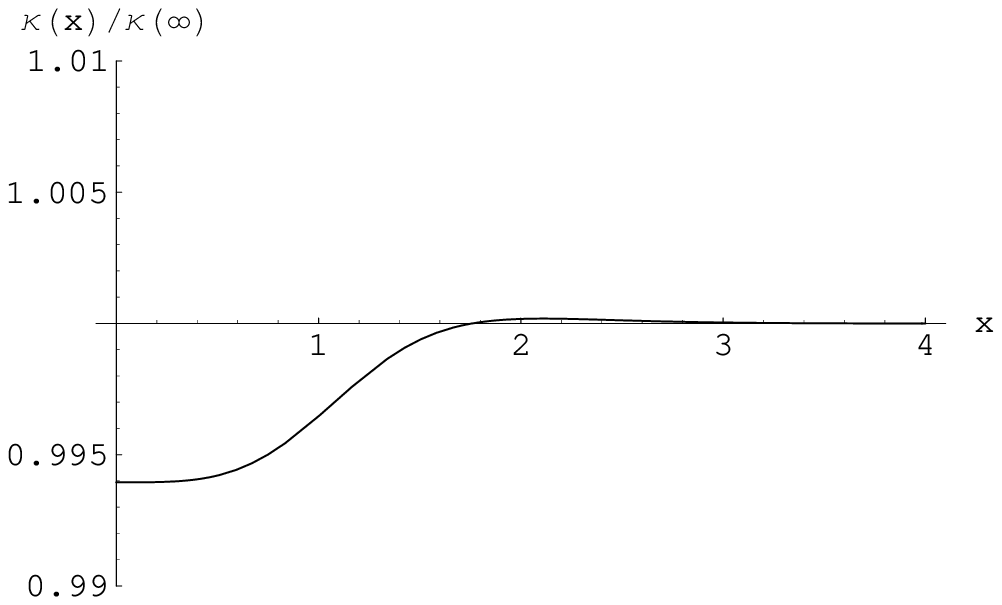}{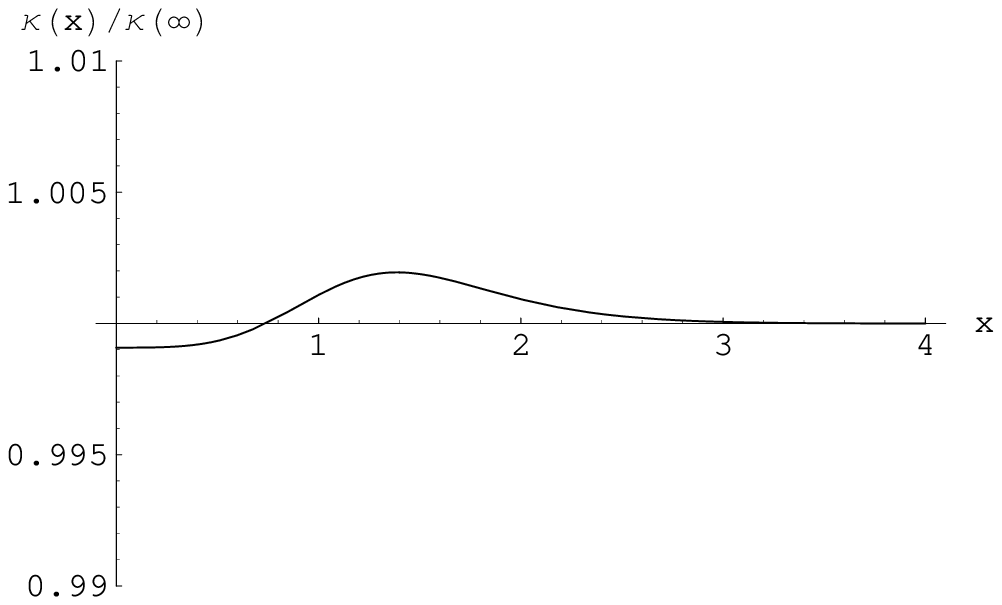}\figlabel{\constfig}

The procedure works very well, converging to a unique solution
independent of initial guesses. We now give the results in more
detail.  The earlier figures \ffig\ and \TUSRfig\ showed surface plots
of the functions $f$ and $A,B,e^C,e^D$ for $d=4$. In figure \TUSRfigB\
we show contour plots of the functions $A,B,e^C,e^D$ for $d = 4$. The
functions appear qualitatively similar for $d=3$, and can not be
distinguished from these $d=4$ contour plots by eye.

Let us now check that the $\tilde{\alpha}=0$ constraint is correctly
imposed in the UV. In figure \contours\ we plot, for $d = 4$, contours of
$R$ and $f$ near the UV boundary, and indeed they agree very well. One
obtains a very similar plot for $d = 3$.

We directly impose the $\beta=0$ constraint in the IR and the
$\tilde{\alpha}=0$ constraint in the UV. Thus we do not impose the
$\alpha=0$ constraint in the IR and this gives a rather physical test of
how well we solve the constraint system \Bianchi\ by plotting the
proper gradient of $A$ at the IR boundary $y = 0$, i.e. $\kappa(x) =
e^{-D} \partial_y A |_{y=0}$. We recall from \constbc\ that the
$\alpha=0$ constraint in the IR implies that $\kappa(x)$ should in fact be a
constant, and $2 \pi/\kappa$ gives the coordinate period of the two
circles $\beta, \tilde{\beta}$. In figure \constfig\ we plot
$\kappa(x)$ for the $d = 3$ and $d = 4$ solutions, and we see that they
are both beautifully constant indeed, reflecting the fact that we have
solved the Cauchy-Riemann constraint system to a high accuracy for the
resolution used.

\listrefs

\end